\documentclass[10pt]{article}
\usepackage[utf8]{inputenc}
\usepackage[T1]{fontenc}
\usepackage{microtype}
\usepackage{amsmath}
\usepackage{graphicx}
\usepackage{geometry}
\usepackage{floatrow}
\usepackage{layout}
\usepackage{mathtools}
\usepackage{amssymb} 
\usepackage{multirow}
\usepackage{multicol}
\usepackage{caption}
\usepackage{subcaption}
\usepackage{authblk} % authors
\usepackage[square,numbers]{natbib} % https://www.overleaf.com/learn/latex/Bibliography_management_with_natbib
\usepackage{indentfirst}
\usepackage{titlesec} % Change title spacing http://ftp.sun.ac.za/ftp/CTAN/macros/latex/contrib/titlesec/titlesec.pdf
\titlespacing*{\section}{0pt}{5ex plus 1ex minus .2ex}{3ex plus .2ex}
\titlespacing*{\subsection}{0pt}{2.5ex plus 1ex minus .2ex}{1.5ex plus .2ex}
\titlespacing*{\subsubsection}{0pt}{1.5ex plus 1ex minus .2ex}{1ex plus .2ex}
\usepackage{wrapfig}
\usepackage{url}
\usepackage{amsmath, dsfont}
\usepackage{lscape}
\usepackage[modulo]{lineno} % line numbers
\usepackage{tabularx} % adaptive tables
\usepackage{hyperref} % hyperlinks
\hypersetup{
    colorlinks,
    linktoc=all, % Hyperlinks for sections
    citecolor=black,
    filecolor=black,
    linkcolor=black,
    urlcolor=blue
}
\usepackage{titlesec} 
\usepackage{tocloft} % Table of contents

 \fontencoding{T1}
  \fontfamily{garamond}
  \fontseries{m}
  \fontshape{it}
  \fontsize{12}{15}
  \selectfont
 
\usepackage{fancyhdr} % Set nice headers
\begin{document}

\title{\textbf{Quantified Sleep}\\Machine learning techniques for observational n-of-1 studies}
\author{Gianluca Truda}
\affil{Vrije Universiteit Amsterdam}
\date{\today}

\maketitle
\begin{abstract}
\noindent This paper applies statistical learning techniques to an observational Quantified-Self (QS) study to build a descriptive model of sleep quality. A total of 472 days of my sleep data was collected with an Oura ring. This was combined with a variety of lifestyle, environmental, and psychological data, harvested from multiple sensors and manual logs.
Such n-of-1 QS projects pose a number of specific challenges: heterogeneous data sources with many missing values;  few observations and many features; dynamic feedback loops; and human biases. This paper directly addresses these challenges with an end-to-end QS pipeline for observational studies that combines techniques from statistics and machine learning to produce robust descriptive models. 
Sleep quality is one of the most difficult modelling targets in QS research, due to high noise and a large number of weakly-contributing factors. Sleep quality was selected so that approaches from this paper would generalise to most other n-of-1 QS projects. 
Techniques are presented for combining and engineering features for the different classes of data types, sample frequencies, and schema. This includes manually-tracked event logs and automatically-sampled weather and geo-spatial data. Relevant statistical analyses for outliers, normality, (auto)correlation, stationarity, and missing data are detailed, along with a proposed method for hierarchical clustering to identify correlated groups of features.
The missing data was overcome using a combination of knowledge-based and statistical techniques, including several multivariate imputation algorithms. ``Markov unfolding'' is presented for collapsing the time series into a collection of independent observations, whilst incorporating historical information. The final model was interpreted in two key ways: by inspecting the internal $\beta$-parameters, and using the SHAP framework, which can explain any ``black box'' model. These two interpretation techniques were combined to produce a list of the 16 most-predictive features, demonstrating that an observational study can greatly narrow down the number of features that need to be considered when designing interventional QS studies.
\end{abstract}

\begin{center}

\textbf{Source code}: \href{https://github.com/gianlucatruda/quantified-sleep}{github.com/gianlucatruda/quantified-sleep}
\end{center}

\newpage
\tableofcontents
\newpage

\section{Introduction}

Connected wearable devices are making personal data collection ubiquitous. These advances have made Quantified-Self (QS) projects an increasingly-interesting avenue of research into personalised healthcare and life extension \cite{swan2013quantified}. Whilst there has been considerable progress made on the challenges of collecting, storing, and summarising such data, there remains a gap between basic insights (e.g. sleep duration) and the deeper kinds of analysis that allow for effective interventions \cite{choe2014understanding} --- e.g. taking precisely 0.35 mg of melatonin 1-2 hours before bedtime to increase deep sleep by 22\%. There are also major challenges like missing values, few observations, feedback loops, and biological complexity \cite{Hoogendoorn2018}. 

\begin{figure}[h!]
    \centering
    \includegraphics[width=\textwidth]{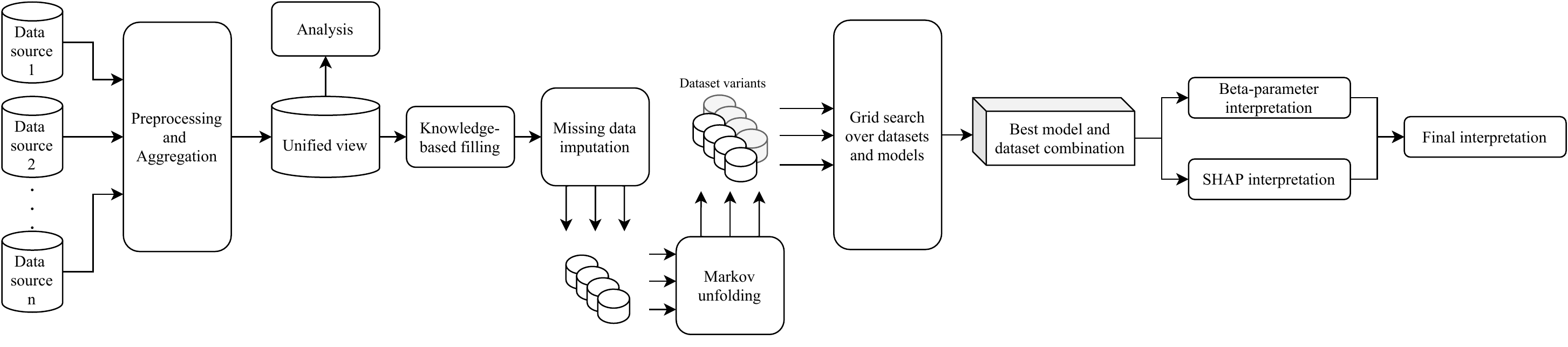}
    \caption{An overview of the data pipeline for this study, from disparate data sources to a final interpretation of the effects on sleep quality.}
    \label{fig:QuantifiedSleepOverview}
\end{figure}

This QS study utilised 15 months of my personal data to find useful relationships between sleep quality and hundreds of lifestyle and environmental factors. Multiple heterogeneous data sources from both active and passive tracking systems were combined and preprocessed into a day-level timeseries (\S \ref{sec:data_sources}). The study combines various techniques for feature engineering (\S \ref{sec:data_wrangling}), dataset analysis (\S \ref{sec:analysis_of_dataset_properties}), missing value imputation (\S \ref{sec:overcoming_missing_data}), temporal representation (\S \ref{sec:collapsing_markov_unfolding}), and model interpretation (\S \ref{sec:model_interpretation}). These methods were evaluated for various learning algorithms through a series of experiments (\S \ref{sec:experiments}). The context of sleep quality is a good case study in observational n-of-1 research, as its challenges generalise to other areas of QS research. Fig. \ref{fig:QuantifiedSleepOverview} gives a graphical overview of the components of this study.  

\subsection{Specific challenges} \label{ssec:challenges}

\begin{figure}[]
    \centering
    \includegraphics[width=0.4\textwidth]{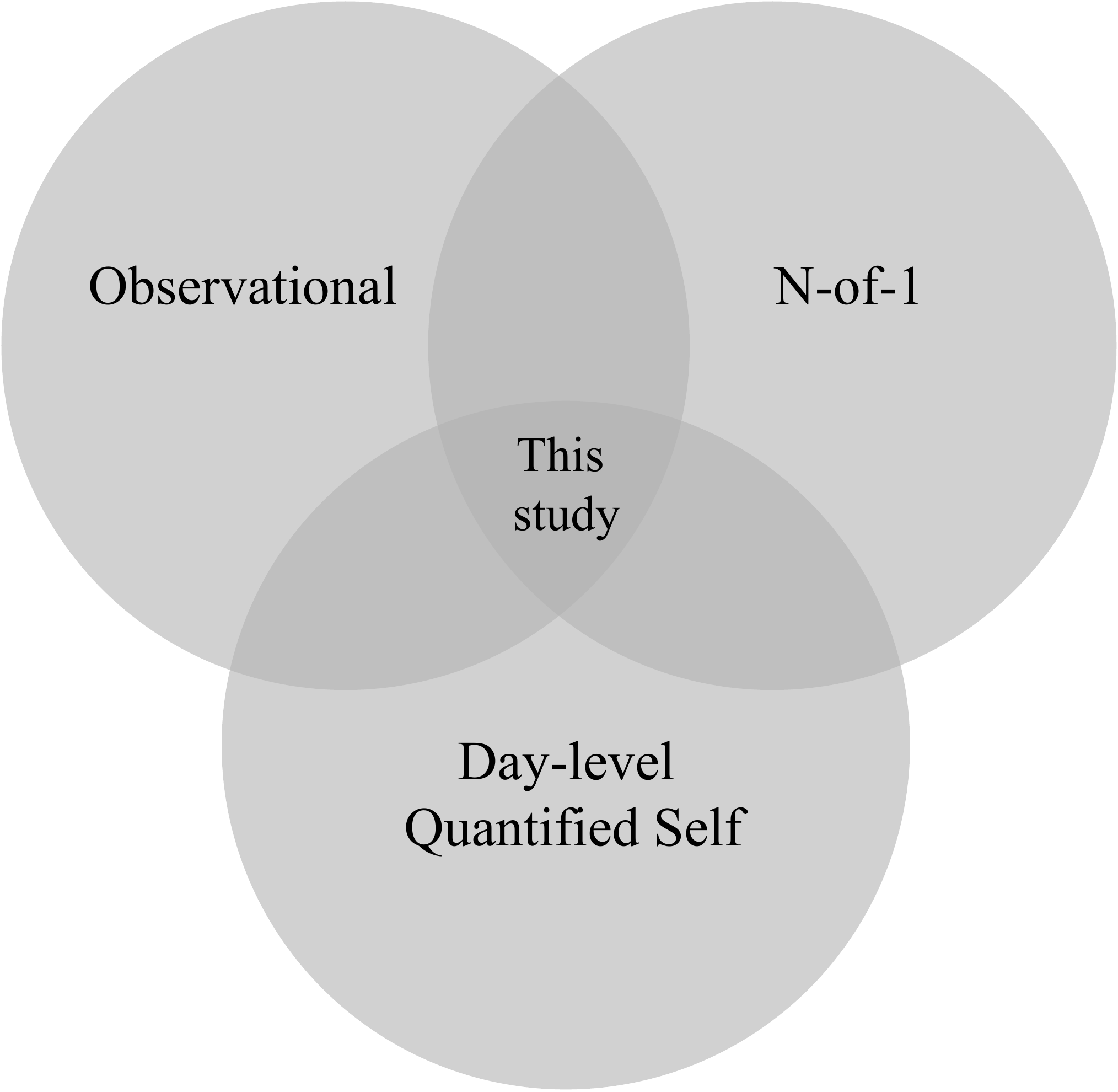}
    \caption{Illustrative diagram of how this study sits at the intersection of observational research, n-of-1 experiments, and day-level Quantified-Self (QS) research.}
    \label{fig:QS_venn}
\end{figure}

N-of-1 studies pose a number of unique challenges, but so do QS projects and observational studies. At the intersection of all of these (Fig. \ref{fig:QS_venn}) is the set of attributes that make this study uniquely challenging. We begin by exploring these challenges.

\subsubsection{Establishing causal relationships}
The gold standard for experiments involving human subjects is the double-blind, randomised, controlled trial (RCT). Such a study is \textit{interventional}, typically having a single variable that is manipulated, such as whether a subject receives a specific drug therapy. All other variables are controlled by strict laboratory conditions, selective recruiting of participants, and rigorous protocols. All of these measures help minimise various statistical and human biases that might affect the outcome. This allows researchers to establish causal links between variables. 

\newpage
In total contrast, we have the n-of-1 QS study. This involves a single individual who designs, administers, and is the subject of the experiments \cite{swan2013quantified}. This makes the study very difficult to blind. Confounding factors, ordering effects, and human biases are thus exacerbated when studying single individuals. Moreover, it is often unclear at the outset what variables are relevant to study in this way. This paper addresses these challenges by using a wide-spanning \textit{observational} study to identify interesting factors that can then be examined further in controlled studies. Techniques for effectively executing these initial observational studies are the focus of this paper.

Whilst there are many challenges in observational n-of-1 research, it is also important to acknowledge the many advantages. Firstly, n-of-1 studies offer something that RCTs cannot. Namely, direct applicability to the subject in question. The cohort of an RCT is very carefully selected. The results of these studies may, therefore, not fully generalise to the genetic, environmental, and psychological attributes of other groups. For complex systems, n-of-1 studies allow data to be collected about the specific individual in question \cite{lillie2011n}. Secondly, by beginning with a wide-ranging observational study, n-of-1 research allows the subject to fit the study around their daily life. This further increases the relevance of results, because they fit within the specific context that we are seeking to optimise. The observational approach can also capture unexpected interactions between variables, which helps identify the most relevant factors for future controlled studies.

\subsubsection{Many heterogeneous data sources}
The QS domain often requires combining multiple bespoke systems to collect both active and passive data \cite{choe2014understanding}. For instance, I log my caffeine intake actively using the \textit{Nomie} app on my phone, but my screen time is monitored passively by the \textit{RescueTime} desktop software. For simple analytics, it is sufficient to work with these sources in isolation. For more valuable multivariate analysis, however, it is essential to unify the data sources and produce a combined dataset. This is challenging for two reasons. Firstly, each source has its own schema for storing, processing, and exporting data. These schema are usually specific to each system. Many tools often limit the level at which data can be exported or analysed. Secondly, the nature and structure of the data differs from source to source. For example, I keep track of caffeine intake by actively logging the date-time-stamp whenever I drink coffee. This follows an event-based scheme, where each date-time-stamp is a record in a relational database. Conversely, my sleep quality is passively measured and logged by my \textit{Oura} ring and stored (primarily) as a daily summary. To analyse the interactions between my caffeine intake and my sleep, I would need to reconcile these two schema by aggregating the caffeine logs to daily summaries and then aligning the dates with the sleep summaries. Instead of events, I now have daily features like average number of coffees, time of last coffee, etc. As more sources are included in the dataset, the complexity of the processing and alignment increases. 

This study addresses these challenge by parsing each data source into a dataframe \cite{mckinney2011pandas}, then aligning these into a unified view. In order to do so, a number of feature aggregation and engineering techniques were developed for this context (\S \ref{sec:data_wrangling}). 

\subsubsection{Human in the dynamic feedback loops}
The very aspect of n-of-1 QS that makes it interesting also imbues it with challenges. Having a human ``in the loop'' of the experiment introduces a number of biases and errors \cite{tversky1974judgment} that affect the results \cite{swan2013quantified}. It is typically too impractical to use any blinding methods in QS because we mostly study multiple variables at once and cannot control for all the other factors. Self-blinding is quite challenging even for a single independent variable \cite{kravitz2014design}. For problems like sleep modelling, blinding the subject is outright impossible, as almost all of the interesting variables include at least some degree of subjectivity. For instance, it may be that tracking my sleep quality makes me more anxious about my sleep, which in turn keeps me up longer and degrades my sleep efficiency. Or, I may be less effective at estimating my mood and energy levels when I am sleep deprived. Or, the act of logging caffeinated drinks or melatonin tablets may have a stronger placebo effect than the actual substance. This means that even an accurate and explainable model will only be able to describe the system it was based on --- biases, feedback loops, and placebos included. It also means that we need to take great care when interpreting our engineered features and the results we produce.

Moreover, modelling \textit{dynamic} systems with feedback loops requires time series techniques. Unfortunately, these place constraints on the choice of models and the interpretations we can perform. In this study, a technique called ``Markov unfolding'' (\S \ref{sec:collapsing_markov_unfolding}) is used to collapse the time series into independent observations, allowing for historical data to be captured by non-temporal models. 

\subsubsection{Wide datasets}
A fundamental challenge for n-of-1 QS projects is having insufficient data. Whilst many sensors collect large quantities of data at high sample rates \cite{swan2013quantified, Hoogendoorn2018}, much of this is aggregated down to summaries. This is because we are often interested in longer time periods, like hours or days \cite{choe2014understanding}. In this study, for instance, the focus is sleep quality. Because this is quantified daily, all data sources must be aggregated to match this day-long window size of observations. A 200 Hz accelerometer ultimately becomes dozens of engineered features summarising daily motion and activity. This has the effect of collapsing low-dimensional, high-frequency data into high-dimensional, low-frequency data. In other words, our dataset becomes wider than it is long. Naïve modelling of such a dataset results in overparameterised models that are high in variance \cite{fortmann2012understanding} and do not generalise \cite{hawkins2004overfitting, vapnik2013nature}. 

This challenge is addressed through the use of multiple feature selection techniques (\S \ref{sssec:rfe}), which reduce dimensionality \cite{Hoogendoorn2018}. This is complemented by analysing the distributions of cross-validated results to detect the ones that are robust across subsets of the data (\S \ref{sec:experiments}).

\newpage
\subsubsection{Missing values}
Missing data is a major problem for modelling, because most techniques assume complete data \cite{beaulieu2018machine}. The missing values either have to be filled in (imputed or interpolated) or discarded (dropped) along with all other data for the observation. Imputation maintains the size and shape of the dataset, but reduces the quality by introducing noise. Dropping introduces no additional noise, but reduces the number of observations --- making the dataset relatively shorter and wider --- which increases model variance. 

N-of-1 QS studies exacerbate the problem of missing data dramatically. Because the study is about a single individual, there are already fewer observations at the outset. Moreover, there is a great deal of noise and complexity in the observations, as they are specific to the individual being studied \cite{kravitz2014design}. Additionally, QS projects involve data from multiple heterogeneous sources, and so often require some experimentation with different pieces of software and hardware \cite{choe2014understanding}. This results in a dataset that has a great deal of missing values in one of a few characteristic types (\S \ref{ssec:theory_of_missing_data}). Additionally, sensors and systems can fail, resulting in missing values scattered throughout the data. Active-tracking sources are prone to poor adherence. 

Fortunately, the unique nature of n-of-1 QS studies also allows us to utilise a collection of tricks and tools that can overcome some of these missing data issues. This paper organises missing values into distinct types (\S \ref{ssec:theory_of_missing_data}), inferring some from domain knowledge (\S \ref{ssec:knowledge_based_filling}), whilst others are imputed using various sophisticated techniques (\S \ref{ssec:imputation_strategies}).

\subsubsection{Complexities of sleep}
This study's target variable (sleep quality) posed some additional challenges. Not only is sleep a complex and little-understood process \cite{walker2017we}, but it is one for which the subject is necessarily not fully conscious, making measurement far more difficult. Many lifestyle factors have been shown to affect sleep quality and quantity \cite{st2016effects, bihari2012factors, lee2008sleep}. Poor sleep also impairs a subject's ability to accurately assess their sleep quality and quantity \cite{kaplan2017gold, walker2017we}. Consumer-grade hardware for sleep tracking is increasingly available due to wearable technology, but must infer sleep stages from other physiological markers like heart rate, movement, heart rate variability (HRV), and body temperature \cite{TheAccuracyofTheOuraRingOuraH}. Unlike exercise tracking, the ground truth for sleep tracking is uncertain. This adds additional noise to the target variable. 

Sleep exists within a complex feedback loop with hundreds of other factors (including itself) \cite{lee2008sleep, walker2017we}, so a simple univariate analysis is clearly insufficient. Instead, this study is framed as a modelling problem in which we wish to construct an explanatory model of sleep. To build the best possible model, we convert the task to an optimisation problem under the framework of supervised learning: models with a low prediction error on unseen (out-of-sample) data are more likely to have captured the relevant variable interactions \cite{vapnik2013nature}. However, our end goal is not to make a good \textit{predictive} model. That is just an intermediate step to finding a good \textit{descriptive} model. By interpreting the model (\S \ref{sec:model_interpretation}), it is possible to find which variables are most influential in determining sleep quality, highlighting potential avenues for further studies. For instance, the model may reveal that melatonin consumption and timing of intense exercise are, together, two of the biggest predictors of sleep quality. This information could then be used to design interventional n-of-1 studies that specifically determine the effect size or optimal ``dose'' of each variable in isolation. 

\newpage
\subsection{Terminology and notation}\label{ssec:terminology}
Because machine learning sits at the intersection of a number of fields --- statistics, computer science, and software engineering --- terminology is varied and overlapping. The goal of this paper is to model a dependent variable (sleep quality) in terms of the independent variables that influence it, such as caffeine consumption, exercise, weather, and previous sleep quality. 

In this paper, the term \textit{feature} will be used to refer to a preprocessed independent variable, whilst \textit{target} (feature) will be used for the dependent variable (sleep quality). For instance, caffeine consumption is an \textit{input variable} but, after preprocessing and aggregating, the hour at which caffeine was consumed is a \textit{feature}\footnote{Because this paper may be of interest to readers from varying backgrounds, it should be noted that the term \textit{feature} is synonymous with terms like \textit{predictor}, \textit{regressor}, \textit{covariate}, and \textit{risk factor}; whilst the \textit{target} variable might be known to others as a \textit{response} variable, \textit{regressand}, \textit{outcome}, or \textit{label}.}. Input variables are often non-numeric, but all features are numeric. Individual points in time (days) will be called \textit{observations}\footnote{In machine learning, \textit{observations} are sometimes called \textit{examples} or \textit{instances}, but that is avoided in this paper to prevent confusion.}. 
When we dig beneath this modelling view, datasets are organised as dataframes --- $m \times n$ matrices with labels for each column and row. Row $i$ corresponds to an \textit{observation} on a particular day. Column $j$ correspond to a timeseries for some \textit{feature}. So \textit{columns} are how we practically store our \textit{features} whilst \textit{rows} are how we practically store our \textit{observations}. This means that value $x_{i,j}$ is the single value in row $i$ and column $j$ of our data matrix $\boldsymbol{X} \in \mathbb{R}^{m \times n}$. So on day $i$, the $j$th feature had a value of $x_{i,j}$. 

\section{Data sources}\label{sec:data_sources}
An overview of the data sources is found in Table \ref{tab:sources}. A detailed explanation follows.

\subsection{Sleep data}
The target variable for this study was sleep quality. This is a function of several sleep-related variables that were captured using a second-generation Oura ring. The Oura ring is a wearable device with sensors that measure movement, heartbeats, respiration, and temperature changes. Being located on the finger instead of the wrist or chest, it can measure pulse and temperature with greater sensitivity and accuracy, resulting in measurements suitable for sleep analysis \cite{TheAccuracyofTheOuraRingOuraH}. 

Studies have found that the 250 Hz sensors of the Oura ring are extremely accurate for resting heart rate and heart rate variability (HRV) measurement when compared to medical-grade ECG devices \cite{kinnunen2020feasible}. This is likely due to the use of dual-source infrared sensors instead of the more common single-source green light sensors when performing photoplethysmography \cite{TheAccuracyofTheOuraRingOuraH}.  Respiratory rate was found to be accurate to within 1 breath per minute of electrocardiogram-derived measures by an external study \cite{HowAccurateIsOurasRespiratoryRat}. Internal studies \cite{TheAccuracyofTheOuraRingOuraH} found the temperature sensor to be highly correlated with leading consumer hardware, but uncorrelated to environmental temperature.

Combining the sensor data, the Oura ring has been found by a number of studies \cite{de2019sleep, chee2021multi, YouTubeOuraRingSleepTest} to produce reasonable estimates of sleep behaviour when compared to medical-grade polysomnography equipment. This is remarkable given the significantly lower cost and invasiveness of the Oura ring. Whilst all of the studies report that sleep detection has high sensitivity (and reasonable specificity), the classification of different sleep stages diverges considerably from the polysomnography reference. This, combined with the underlying opaqueness of sleep, made the target variable of this study noisy and uncertain. Despite this, the low costs and simplicity of the Oura ring make it an invaluable tool for QS research in the domain of sleep.   

The \href{https://cloud.ouraring.com/docs/sleep}{Oura API} gives access to daily summaries generated from the raw sensor data. For this study, the \texttt{oura\_score} variable was of most interest, as it is intended to represent the overall sleep quality during a sleep period. It is a weighted average of sleep duration (0.35), REM duration (0.1), deep sleep duration (0.1), sleep efficiency (0.1), latency when falling asleep (0.1), alignment with ideal sleep window (0.1), and 3 kinds of sleep disturbances: waking up (0.05), getting up (0.05), and restless motion (0.05). 

\subsection{Supporting data}
\begin{itemize}
\item My electronic activities and screen time were tracked with the \textit{RescueTime} application and exported as daily summaries of how much time was spent in each class of activity (e.g. 3h42m on software development). 
\item My GPS coordinates, local weather conditions, and phonecall metadata were logged using the \textit{AWARE} application for iOS and queried from the database using SQL. Local weather conditions and location information were logged automatically at 15-minute intervals (when possible). Phonecall metadata was logged whenever a call was attempted, received, or made. 
\item Full timestamped logs of all caffeine and alcohol consumption were collected with the \textit{Nomie} app for iOS. Logs were made within 5 minutes of beginning to consume the beverage. Caffeine was measured in approximate units of 100mg and alcohol was measured in approximations of standard international alcohol units.
\item Logs of activity levels and exercise measured on my phone were exported from Apple's \textit{HealthKit} using the \textit{QS Export} app in the form of non-resting kilocalories burned at hourly intervals and activities (running, walking, etc.) logged as they began and ended.  
\item Heart rate was recorded approximately once per minute\footnote{Higher frequencies were used during workout tracking and lower frequencies were used when the device was not worn.} on a \textit{Mi Band 4} and synchronised with \textit{AWARE} via \textit{HealthKit}. 
\item My daily habits -- meditation, practising guitar, reading, etc. -- were captured (as Boolean values) daily before bedtime using the \textit{Way of Life} iOS app.
\item My daily eating window (start and end times) was logged in the \textit{Zero} app as daily summaries. 
\item My mood and energy levels were logged at multiple (irregular) times a day in the \textit{Sitrus} app. 
\item A collection of spreadsheets were used to log the timestamps and quantities of sleep-affecting substances like melatonin and CBD oil.
\end{itemize}

\begin{landscape}
\begin{table}[]
\caption{The data sources used to build the dataset for this study. The prefix column indicates the string that the feature names in the dataset inherit from their source. These prefixes help associate features with their sources and allow easier grouping of features. $\star$ Other prefixes: location, city, country, travelling.}
\label{tab:sources}
\footnotesize
\begin{tabularx}{\textwidth}{XlXllllX}
\hline
\textbf{Kind}            & \textbf{Prefix} & \textbf{Hardware}              & \textbf{Software} & \textbf{Format} & \textbf{Frequency} & \textbf{Active/Passive} & \textbf{Hoogendoorn-Funk \cite{Hoogendoorn2018} categories}            \\ \hline
Sleep                    & oura            & Oura ring, 2nd gen.            & Oura API          & JSON            & Daily summaries    & Passive                 & Physical                                        \\
Readiness                & oura            & Oura ring, 2nd gen.            & Oura API          & JSON            & Daily summaries    & Passive                 & Physical                                        \\
Computer activity        & rescue          & Personal computer              & RescueTime        & CSV             & Daily summaries    & Passive                 & Mental \& Cognitive                             \\
GPS coordinates          & aw\_loc         & iPhone 8                       & AWARE v2          & SQL             & \textasciitilde15 mins            & Passive                 & Environmental                                   \\
Barometric pressure      & aw\_bar         & iPhone 8                       & AWARE v2          & SQL             & \textasciitilde15 mins            & Passive                 & Environmental                                   \\
Local weather conditions & aw\_weather     & iPhone 8                       & AWARE v2          & SQL             & \textasciitilde15 mins            & Passive                 & Environmental                                   \\
Phonecall metadata       & aw\_call        & iPhone 8                       & AWARE v2          & SQL             & \textasciitilde15 mins            & Passive                 & Social, Environmental                           \\
Activity / Exercise      & hk              & iPhone 8, Mi Band 3, Oura ring & Apple HealthKit   & CSV             & Hourly summaries   & Passive                 & Physical                                        \\
Heart rate               & aw\_hr          & Mi Band 3                      & AWARE v2          & SQL             & \textasciitilde1 min              & Passive                 & Physical                                        \\
Caffeine and Alcohol     & nomie           & iPhone 8                       & Nomie app         & CSV             & Timestamped logs   & Active                  & Diet                                            \\
Habits                   & wol             & iPhone 8                       & Way of Life app   & CSV             & Daily              & Active                  & Mental \& Cognitive, Psychological, Situational \\
Eating / Fasting periods & zero            & iPhone 8                       & Zero app          & CSV             & Daily summaries    & Active                  & Physical, Diet                                  \\
Mood and Energy          & mood            & iPhone 8                       & Sitrus app        & CSV             & Timestamped logs   & Active                  & Psychological                                   \\
Melatonin use            & melatonin       & N/A                            & Spreadsheet       & CSV             & Timestamped logs   & Active                  & Diet                                            \\
CBD use                  & cbd             & N/A                            & Spreadsheet       & CSV             & Timestamped logs   & Active                  & Diet                                            \\
Daily metrics            & daily $\star$           & N/A                            & Spreadsheet       & CSV             & Daily logs         & Active                  & Environmental, Situational                      \\ \hline
\end{tabularx}
\end{table}
\end{landscape}

\section{Data wrangling}\label{sec:data_wrangling}
The sampling intervals of the data sources ranged from below 1 minute to a full $24$ hours. Much of the data was necessarily at irregular intervals because it was an event log (e.g. having coffee or taking melatonin). The data sources were all structured, but were a mix of temporal and numeric types that required different feature engineering and aggregation techniques. Because the target  (sleep quality) was calculated at a daily interval, all the data needed to be up- or down-sampled accordingly. 

This section details how the heterogeneous data sources were ingested (\S \ref{ssec:data_ingestion}), how custom feature engineering was used to align (\S \ref{ssec:midnight_unwrapping}) and aggregate (\S \ref{ssec:transformations}) the different classes of time series, and how the sources were concatenated into a unified dataset (\S \ref{ssec:dataset_concatenation}).

\subsection{Data ingestion}\label{ssec:data_ingestion}
Each data source had its own bespoke data ingester that ultimately fed into a single unified view from which features could be engineered to produce a dataset (Fig. \ref{fig:QuantifiedSleepOverview}). Each ingester is a function responsible for reading a data source in its source format, transforming it into a dataframe, renaming the columns with appropriate conventions and a descriptive prefix, then returning the dataframe. This modularity allowed for iterative development during this study. It also allows the downstream code to generalise to future studies on different data sources. 

\subsection{Midnight unwrapping}\label{ssec:midnight_unwrapping}
\begin{figure}[H]
    \centering
    \includegraphics[width=0.5\textwidth]{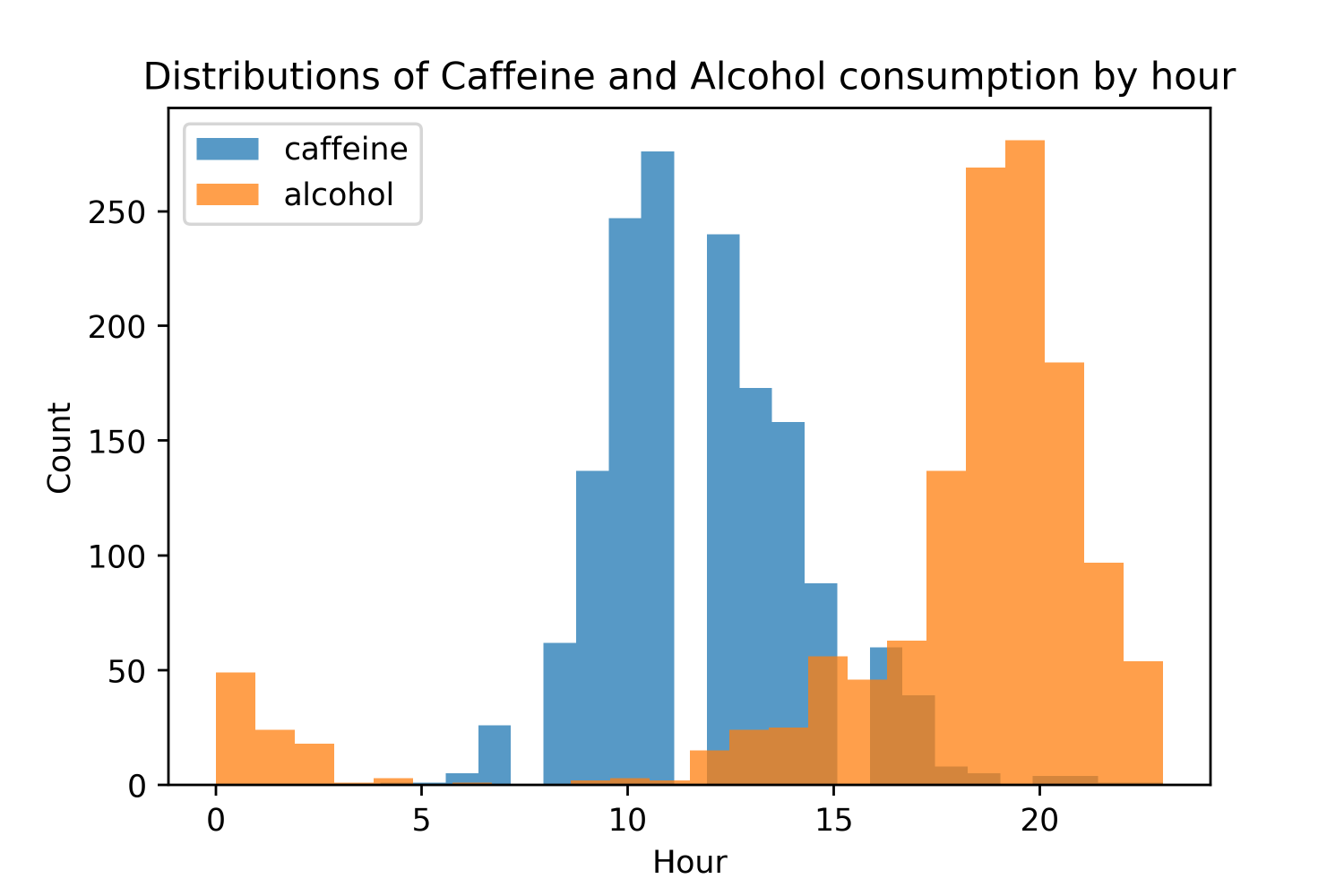}
    \caption{Temporal histograms for for alcohol and caffeine consumption, illustrating how a reference time of 05:00 was selected for midnight unwrapping.}
    \label{fig:01b_caffeine_alcohol_distributions}
\end{figure}

Because the data typically showed one sleep pattern per night, the window size for observations in this study was necessarily one sleep-wake cycle (i.e. one day). Because sleep runs over midnight, it was essential to select another time as the point around which each ``day'' was defined. To determine this, temporal histograms of important activities like sleep, food consumption, and exercise were plotted (e.g. Fig. \ref{fig:01b_caffeine_alcohol_distributions}). From this, a time of 05:00 was selected as the offset point. A window size of 24 hours was then applied from that reference. For instance, alcohol and caffeine consumed between midnight and 05:00 count towards aggregates for the \textit{previous} day. This simple technique preserves causal relationships between input features and the target, as the order of events is preserved.  

The summary date (based on 05:00 offset) was generated for each timestamp in each data source. For instance, if alcohol was logged at \texttt{2020-04-07 01:03:41}, the summary date was set to \texttt{2020-04-06}. All records were then grouped on that summary date attribute, with specific time-domain aggregations applied over the attributes to produce numerical features. This was informed by intuition and domain knowledge and thus the aggregations varied across data sources. 

\subsection{Transformations}\label{ssec:transformations}
The unified dataset required all sources to be sparse daily summaries. This required aggregations and interpolations. The data sources fell into three groups: 
\begin{enumerate}
    \item \textbf{Daily summaries}: those sources that were already daily summaries (e.g. Oura, Zero, RescueTime). These required no further adjustment, provided their datestamp format was correct. 
    \item \textbf{Event logs}: those that were records of the date and time that events occurred (e.g. caffeine, alcohol, melatonin, calls). There sources needed to be \textit{pivoted} into wide format. The dates were then interpolated to daily summaries. Events occurring on the same day were aggregated. 
    \item \textbf{Intra-day samples}: higher-frequency records (e.g. weather, location, heart rate). These sources needed aggregation to produce daily summaries. 
\end{enumerate}

\begin{figure}[H]
    \centering
    \includegraphics[width=0.9\textwidth]{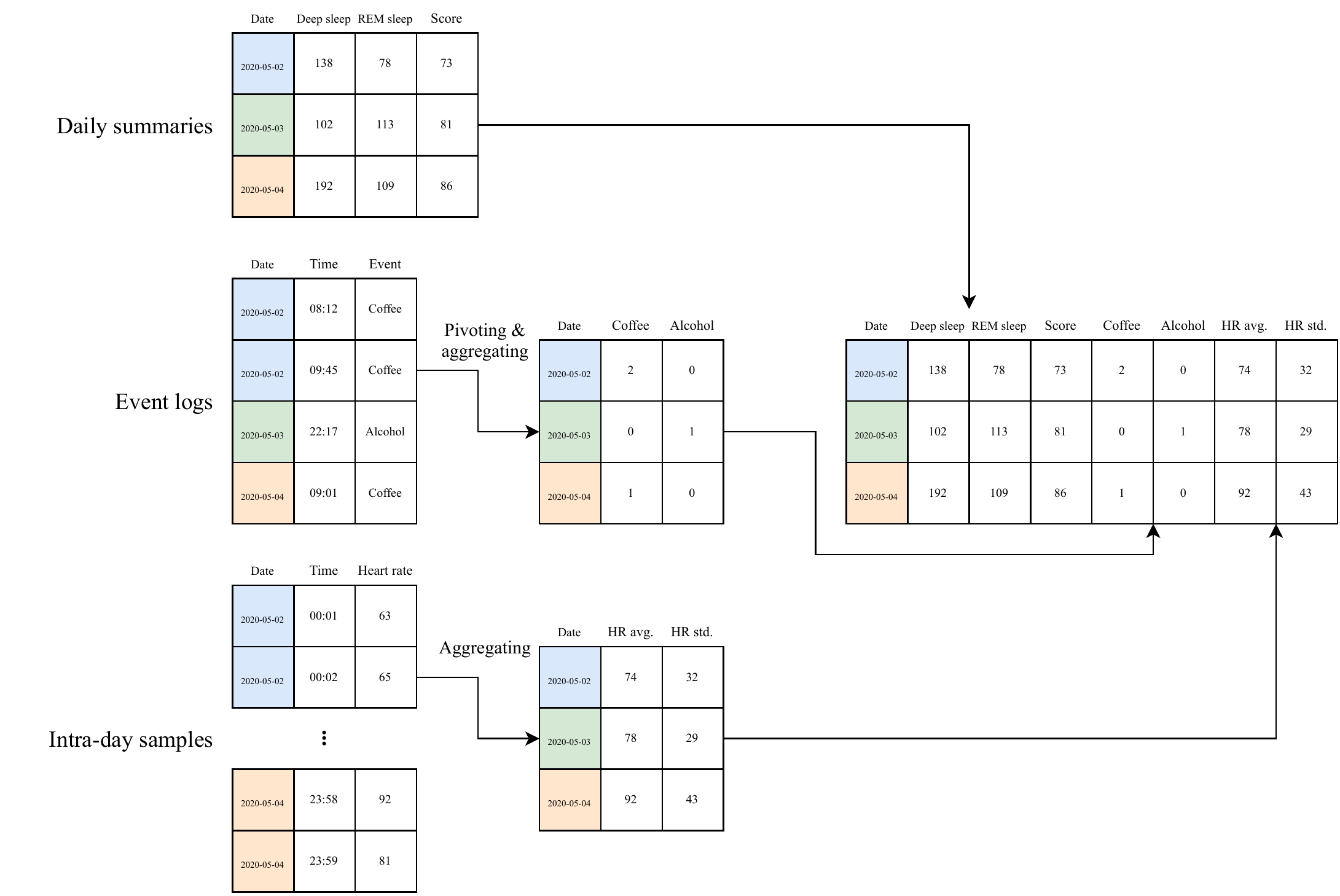}
    \caption{Toy examples of how the three kinds of data source in this study were transformed prior to concatenation into the unified dataset.}
    \label{fig:data_transformations}
\end{figure}

Fig. \ref{fig:data_transformations} illustrates how such transformations would take place using highly-simplified scenarios.

\subsection{Aggregation techniques}\label{ssec:aggregation_techniques}
When collapsing event logs and intra-day samples into daily summaries, aggregation functions summarise the distribution. The choice of aggregation functions needs to consider the original data source and the downstream learning pipeline. 

For continuous variables (e.g. heart rate), standard summary statistics --- minimum, mean, maximum, standard deviation --- capture the shape of the data well. This is particularly true when the variable is close to a normal distribution, which many QS sources are. For categorical variables, one-hot encoding and summation were used to aggregate data. For event logs (e.g. coffee and alcohol consumption), counts are the most intuitive aggregation. Because the times that the events took place is also important, specialised temporal aggregation techniques were needed.

\subsubsection{Aggregating temporal data}
For event logs, the \textit{hour} of occurrence was encoded as a numeric feature along with the quantity, allowing aggregation with \texttt{min}, \texttt{max}, and \texttt{range} functions. Hourly resolution is a reasonable level given the inherent noise in the data.

Caffeine data is shown by way of an example:
\begin{center}
\begin{tabular}{|c|c|c|c|c|}
\hline
\textbf{Summary date} & \textbf{Value sum} & \textbf{Hour min} & \textbf{Hour max} & \textbf{Hour range} \\ \hline
2020-06-12 & 2.0 & 12 & 14 & 2 \\
2020-06-13 & 1.0 & 13 & 13 & 0 \\
2020-06-15 & 2.0 & 11 & 13 & 2 \\
\hline
\end{tabular}
\end{center}

Because midnight unwrapping had been applied, the \textit{net} hour of occurrence was used. So having a last beer at 1AM on Saturday would result in the \texttt{net\_hour\_max} feature on \textit{Friday} having a value of 25.

\subsubsection{Aggregating location data with geohashes}
\begin{figure}[H]
    \centering
    \includegraphics[height=4cm]{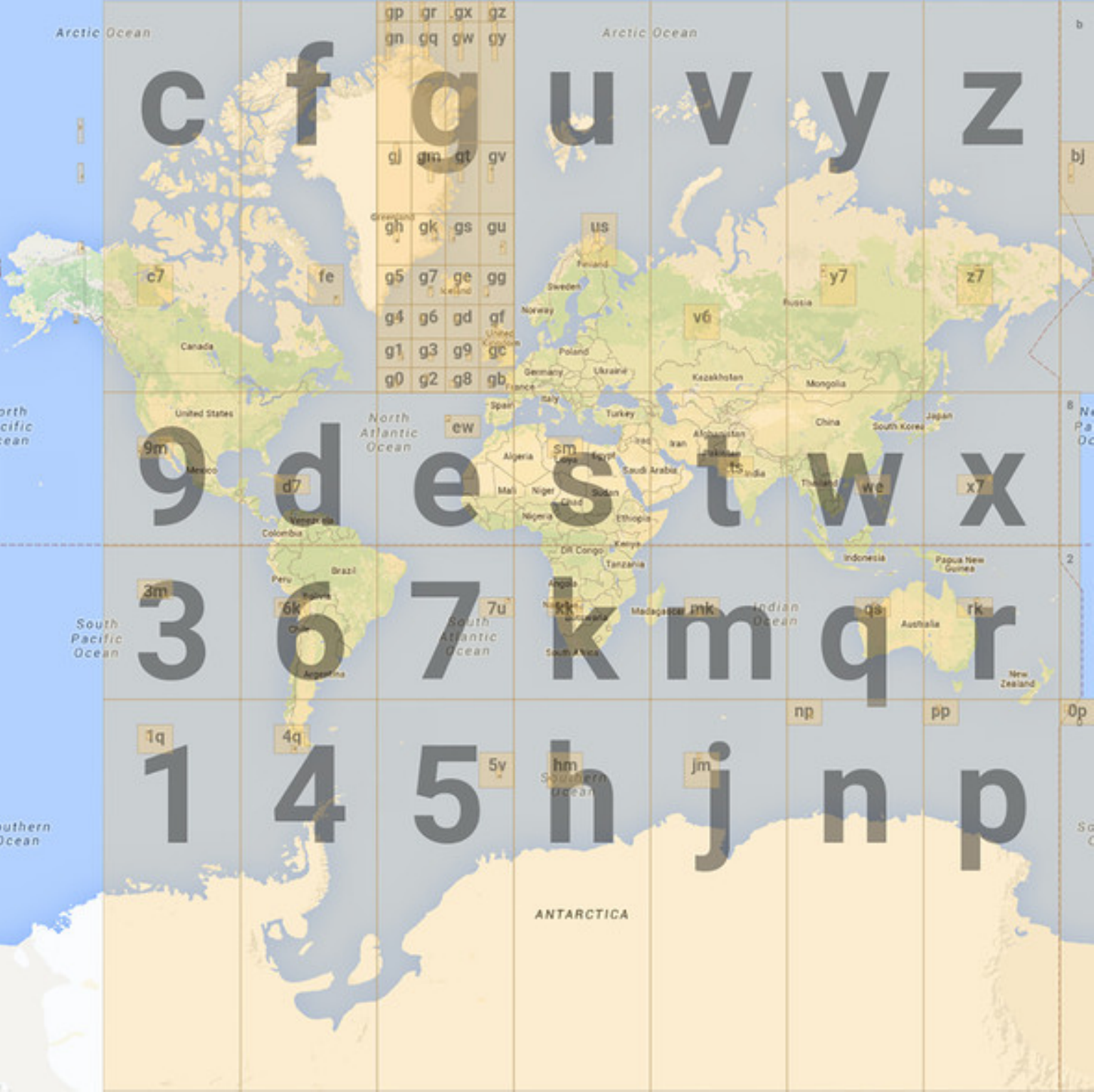}
    \includegraphics[height=4cm]{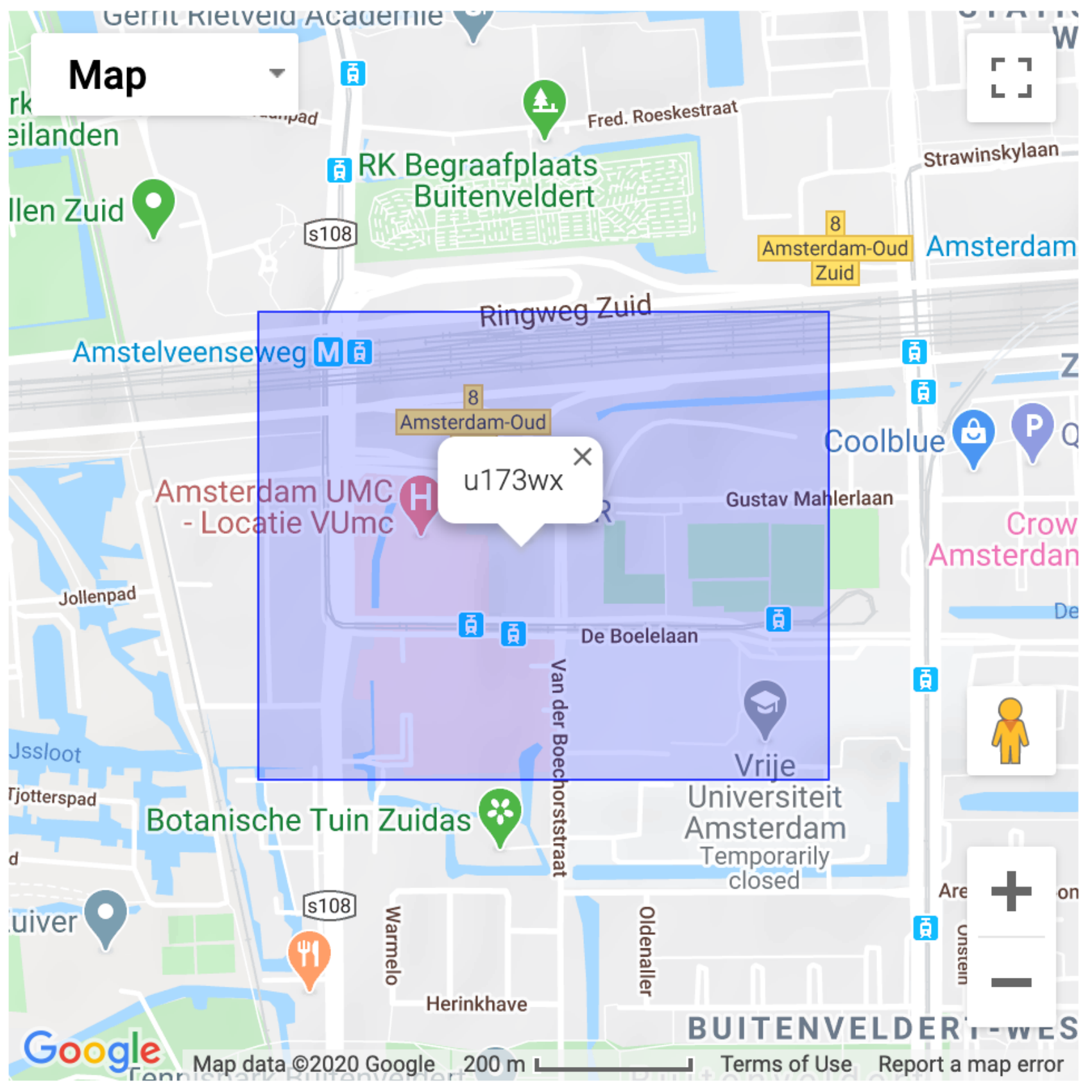}
    \caption{Illustration of the geohash system and example of the level-6 geohash for the Vrije Universiteit \cite{geohashscript}.}
    \label{fig:geohashing}
\end{figure}

High-frequency GPS data has huge potential for building information-rich features, but poses two major challenges. Firstly, not all GPS measurements have the same level of accuracy, due to signal availability and power-saving measures. Secondly, comparing or aggregating coordinates is difficult and ill-defined. One solution is to manually define regions as places of interest and calculate which GPS coordinate pairs fall within those regions. This proves to be very computationally expensive and time consuming. Instead, this study used the \textit{geohash} system \cite{Niemeyer} (Fig. \ref{fig:geohashing}). 

Geohashing recursively divides the earth into 32-cell grids that are each codified with an alphanumeric sequence. Because geohashes are based on the mathematics of z-order curves, they offer some useful properties like arbitrary precision and fast encoding. For example, the GPS coordinates of the Vrije Universiteit can be mapped to a level-6 geohash: $(52.3361, 4.8633) \to \text{u173wx}$, which corresponds to a rectangle with an area of $0.7$ km. But by simply truncating the last 3 characters, we can ``zoom out'' to \textit{u17}, which covers the northern half of the Netherlands. 

The variable precision of geohashes helps overcome the GPS accuracy problem. Reducing pairs of GPS co-ordinates to single alphanumeric strings with well-defined neighbourhood properties solves the aggregation problems of GPS data.

The geohash \texttt{encode} function was mapped over all 61 000 GPS values in the dataset to generate the level-12 geohashes in seconds. By simply truncating digits from the ends of these level-12 geohashes, features were generated for levels 5 through 9. These levels correspond to blocks ranging from $25m^2$ to $25km^2$, capturing location information at a variety of resolutions. 

With this in place, daily aggregation was performed in two ways: (1) by counting the number of unique geohashes from each level (5-9) for that day, (2) by finding the 10 most common level-5 geohashes over the entire dataset and counting the proportion of logs that matched each in that day. This produced 15 features that summarised the locations and movements of the day in an efficient format. 

\subsection{Dataset concatenation}\label{ssec:dataset_concatenation}
All the preprocessed data sources were sequentially concatenated on the summary date column using a \textit{left join} operation. The sleep dataframe was used as the starting object. This served two purposes. Firstly, it prevented the need for interpolating dates, as the sleep data was complete. Secondly, it automatically resulted in all rows being trimmed to match the start and end of the target feature (contained in the sleep data). This produced a unified dataset with 789 observations (rows) and 309 columns, of which 271 were numeric features.

Because the data from the Oura ring contained numerous linear components of the target feature, there was a risk of data leaks. For instance, \texttt{oura\_yesterday\_total} alone contributes $35\%$ of the \texttt{oura\_score} target. But these variables were relevant to predicting future nights of sleep. To remedy this, all variables with the \texttt{oura\_} prefix were copied, shifted one day later, and re-prefixed with \texttt{sleep\_yesterday\_}. Before fitting the model, the features with the \texttt{oura\_} prefix were always dropped. This way, each observation had no features leaking information about the target, yet still included useful information about prior sleep behaviour. 

\section{Analysis of dataset properties}\label{sec:analysis_of_dataset_properties}

\subsection{Time period}
The subset of the data used in this study ran for 472 days from mid-October 2019 to mid-January 2021. Some $65\%$ of this time was under various restrictions due to the Covid-19 pandemic. This meant much less variety in location and much more consistent patterns of behaviour, due to the stay-at-home orders. This formed a natural experiment by keeping many factors consistent from March 2020 to December 2020.

On one hand, this produced a more ``controlled'' experiment, with fewer free variables. On the other hand, the unique circumstances mean that the results may not generalise as well. Moreover, a lack of variability in features makes them less useful to a predictive model \cite{vapnik2013nature}. This can result in highly-relevant variables being absent in the final model because they remained consistent during the course of the lockdown. To mitigate this, data from 5 months of pre-pandemic conditions was retained. 

\subsection{Outliers}

It is essential to differentiate \textit{variational} outliers from \textit{measurement-error} outliers \cite{Hoogendoorn2018}. The former are a legitimate result of natural variation in a system and must be retained in order to build a fully-descriptive model. The latter are a result of failed sensor readings, corrupted data, or erroneous data entry. These measurement errors add noise to the dataset that makes it more challenging to fit a model to the underlying signal. They should therefore be removed. Unfortunately, it is often difficult to differentiate the two types of outlier. To err on the side of caution, minimal outlier removal was used in this study. 

The focus of the study was sleep quality, so the most relevant sleep features were assessed to detect outliers. By inspecting the distributions and linear relationships in the sleep data\footnote{See Fig. \ref{fig:01b_oura_sleep_pairplots} in the Appendices.}, the presence of some outliers was apparent. Both the sleep efficiency and overall sleep score were negatively skewed --- with potential outliers in the left tail. 

Chauvenet's criterion is a technique to find observations that have a probability of occurring lower than $\frac{1}{cN}$, where $N$ is the number of observations and $c$ is a strictness factor \cite{Hoogendoorn2018}. Chauvenet's criterion ($c=2$) identified a total of 6 outliers across the 4 key sleep features: score, total, efficiency, duration. All but one of these outliers pre-dated the intended timespan of the study, and that outlier was removed.

For the non-target features, distribution plots and 5-number summaries were inspected to detect erroneous measurements. For instance, a heart rate of 400 would have been clearly erroneous. No values were deemed obvious errors\footnote{It is, of course, possible that some data-entry errors or measurement errors made it past this conservative filter, adding further noise to the data.}. 

\newpage
\subsection{Normality}
Many learning algorithms assume that the target feature is normally distributed \cite{burkov2019hundred, bishop2006pattern}. If it significantly differs from a normal distribution, we can either (1) normalise the target feature, or (2) apply a transformation to the model predictions to map them onto the same distribution as the target feature. Neither of these are ideal. Normalising the target feature makes our model predictions harder to interpret \cite{molnar2019}. Transforming the predictions can introduce a number of errors and points of confusion. 

\begin{figure}[H]
    \centering
    \includegraphics[width=0.5\textwidth]{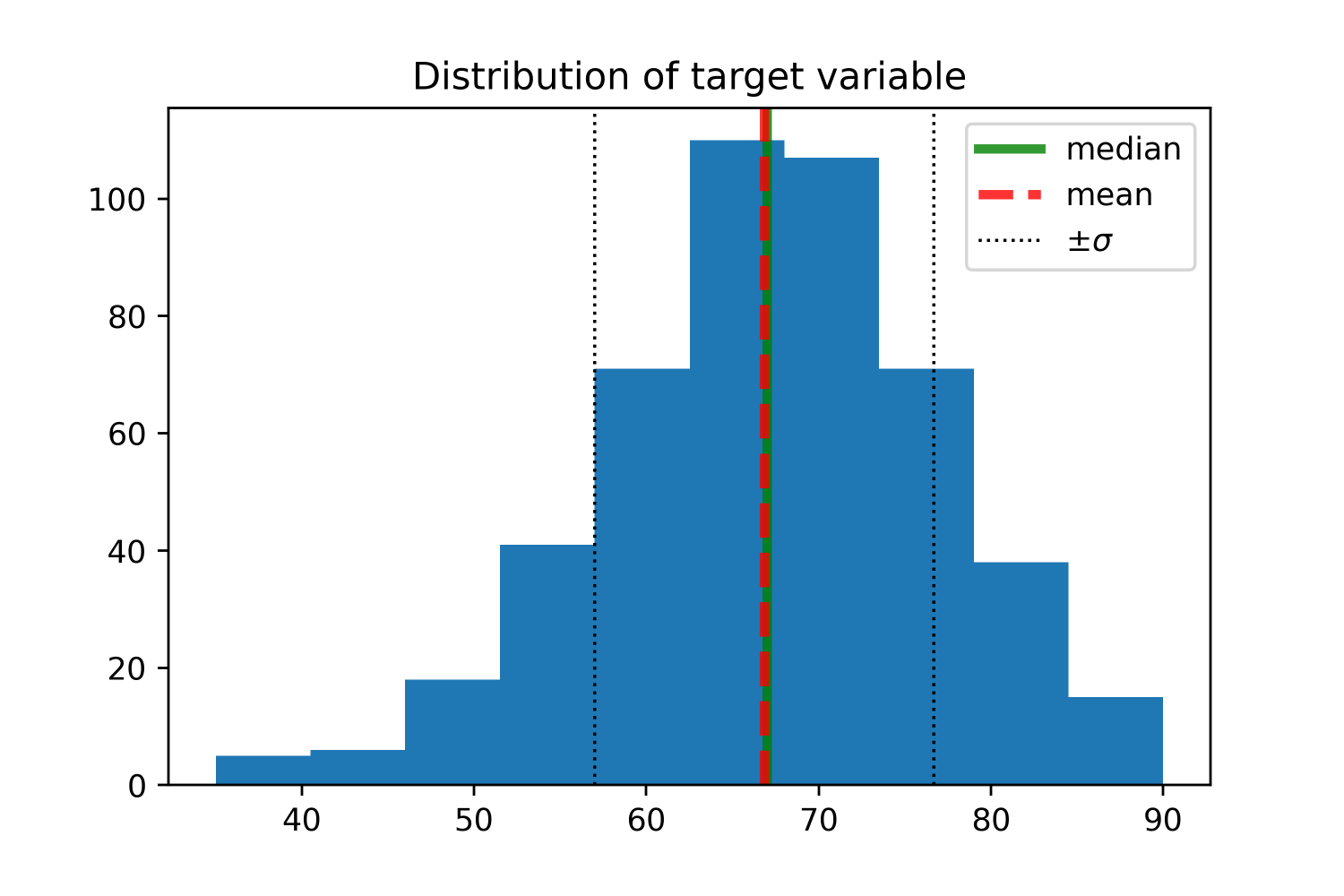}
    \caption{Histogram showing the distribution of the target feature for the time period of the study: \texttt{oura\_score}. The median is indicated with a green vertical line. The mean is indicated with a dashed red line. We can see that they are almost identical. Dotted black lines indicate one standard deviation ($\sigma$) in either direction of the mean.}
    \label{fig:02_target_distribution}
\end{figure}

The target feature (\texttt{oura\_score}) followed the general shape of a normal distribution, but with a skewness of $-0.325$ and an excess kurtosis of $0.154$, indicating thin tails and a negative skew (Fig. \ref{fig:02_target_distribution}). We know that the target is bounded by $[0, 100]$ and sleep behaviour generally regresses to the mean \cite{lee2008sleep}, so there is little chance that the population distribution is extreme \cite{taleb2020statistical}, even if it is slightly skewed. These heuristics, along with the desire to keep the model interpretable, informed the decision not to transform the target feature.

\subsection{Correlation}
It is important to understand how features linearly relate with one another (correlation) and with themselves at different points in time (autocorrelation). If features that share a source are highly correlated, we may want to combine them or discard one of them, as this maintains most of the same information (and interpretability), whilst reducing the number of features our model needs to process \cite{burkov2019hundred}. More importantly, features that are correlated with the target are strong candidate features for our final model. 

\newpage
\subsubsection{Pairwise correlation with target}
\begin{figure}[H]
    \centering
    \begin{subfigure}[b]{0.49\textwidth}
        \includegraphics[width=\textwidth]{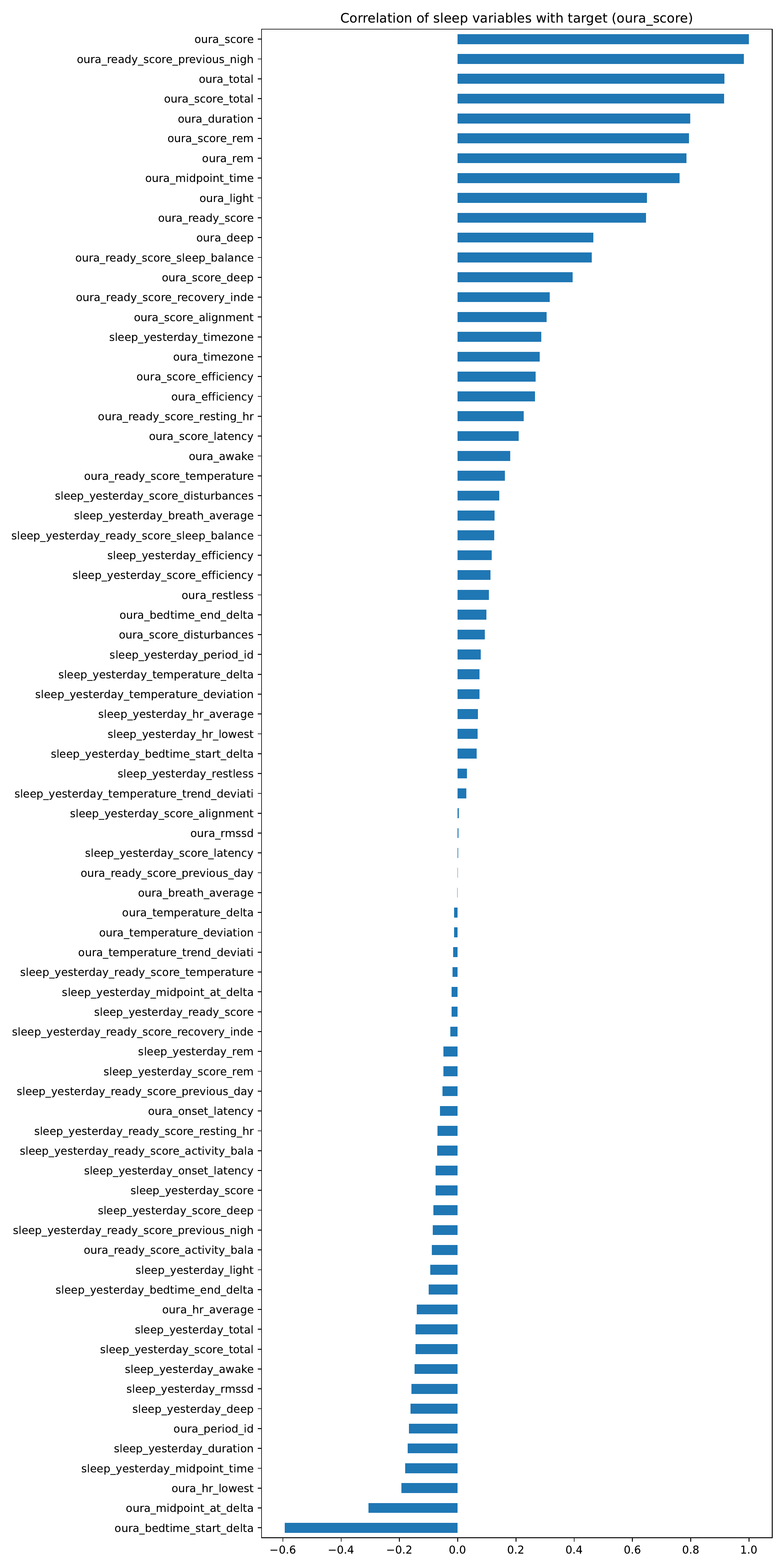}
    \end{subfigure}
    \hfill
    \begin{subfigure}[b]{0.49\textwidth}
        \includegraphics[width=\textwidth]{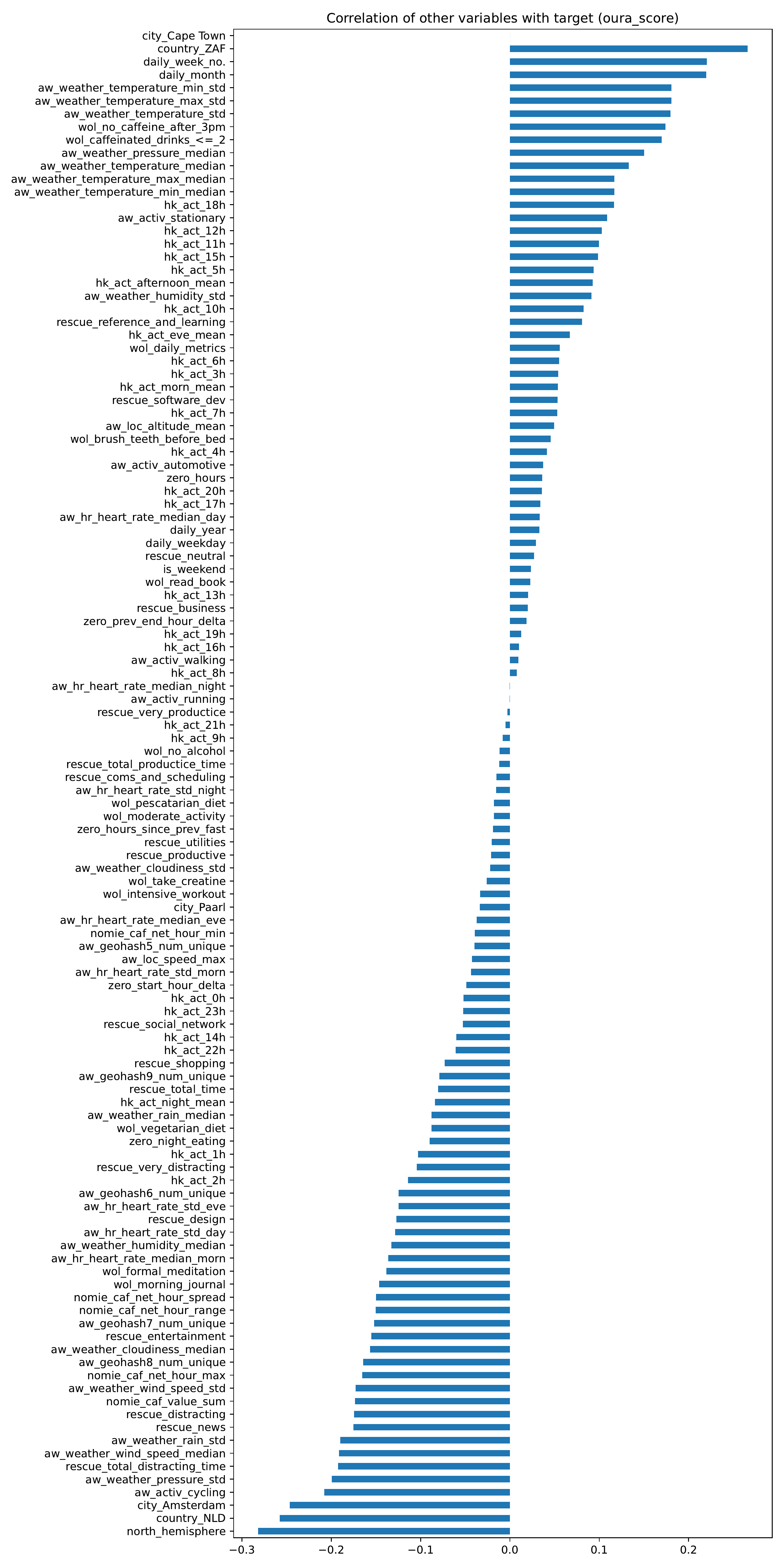}
    \end{subfigure}
    \caption{Pearson correlations between features and the target (\texttt{oura\_score}). The left subfigure shows features that originate from the Oura ring, i.e. sleep related. The right subfigure shows features that originate from other data sources. Recall from \S \ref{ssec:dataset_concatenation} that the prefix \texttt{sleep\_yesterday\_} is used for a 1-day lag on the \texttt{oura\_} features to make it easier to remember to remove same-day sleep features that constitute a data leak. So \texttt{sleep\_yesterday\_total} can be interpreted as \texttt{oura\_total} with a shift of 1 day.}
    \label{fig:02_correlation_scores_bar}
\end{figure}

At first, we consider only correlations of each feature to the target (Fig. \ref{fig:02_correlation_scores_bar}).

As expected, sleep-related features were more strongly correlated with the sleep score than the non-sleep features. Specifically, measures of sleep duration and efficiency had a strong positive correlation with the target ($0.5 \leq r \leq 0.9$). This is expected, knowing that sleep score is a weighted linear sum over several of these features. Curiously, a few sleep-related features had no direct correlation with sleep score. Notably, those related to respiration rate and temperature. 

The bedtime feature had a strong negative correlation with the target ($r \approx -0.6$), which is likely because it is considered as part of the ideal sleep window calculation that comprises $10\%$ of the sleep score. It was important to remove features like this from the dataset prior to training the model, in order to prevent data leaks. 

The non-sleep features had much weaker correlations with sleep score ($-0.3 \leq r \leq 0.3$), highlighting that there were no obvious candidate features. The strongest correlations fell into the location, weather, and caffeine categories. Notably, time-related features like the week number and month number showed a notable correlation to sleep quality ($r \approx 0.2$), indicating some trend or seasonality. This would also explain why weather and location features showed strong correlations with the target. Analysis of data stationarity is an essential step (\S \ref{ssec:stationarity}). 

\subsubsection{Hierarchical correlational clustering}
\begin{figure}[H]
    \centering
    \includegraphics[height=18cm]{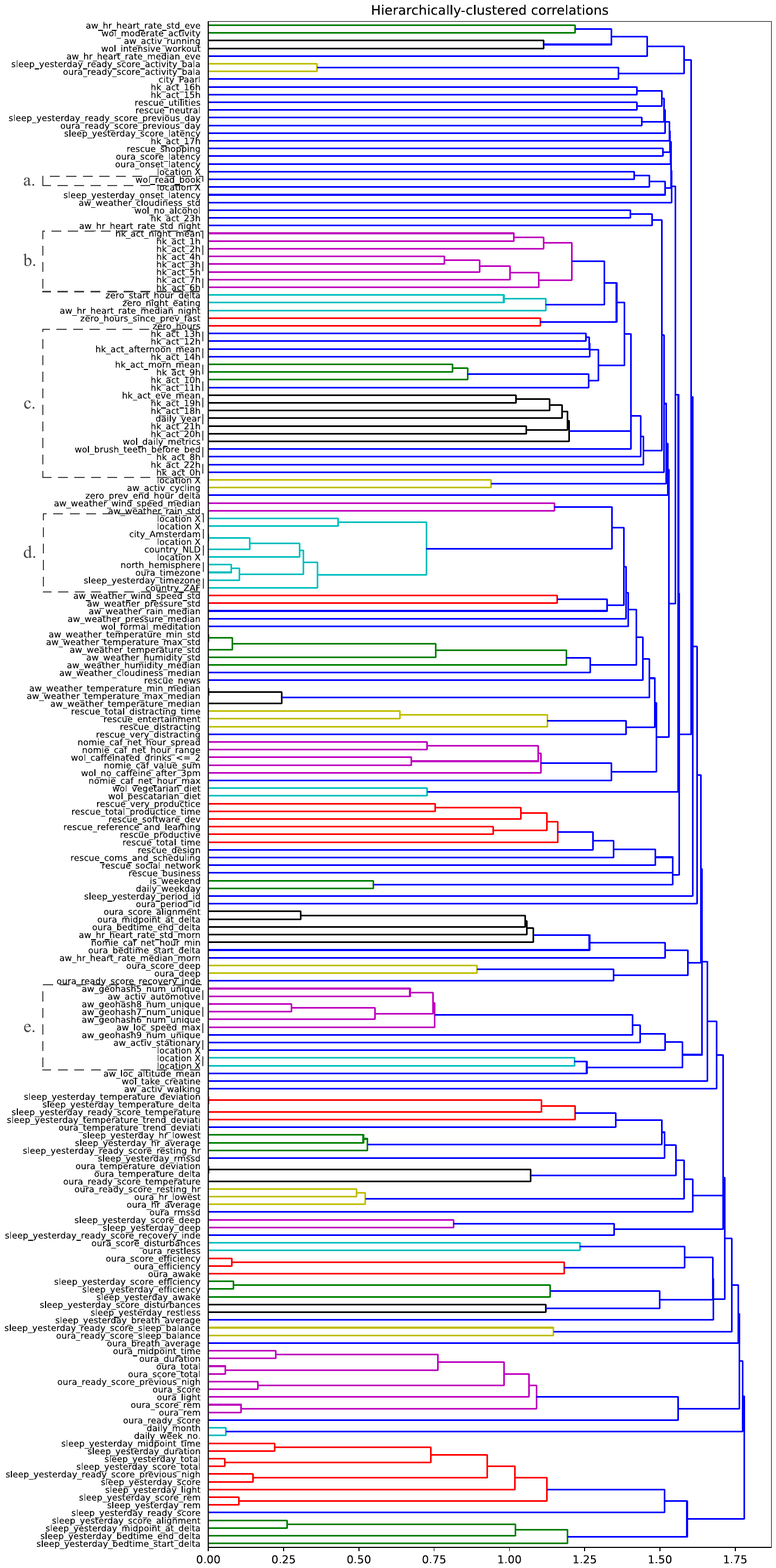}
    \caption{Dendrogram of hierarchical correlational clustering for all features with low missing data. Each of the $n$ features were clustered based on their distance in $n$-dimensional space to other features from the $n \times n$ correlation matrix. The dendrogram shows features on the vertical axis and pairwise Euclidean distance on the horizontal axis. Vertices indicate where clusters join into superclusters. The distance between clusters is represented by their vertical distance on the vertical axis. Colour-coding is used to illustrate primary clusters based on a 70\% similarity threshold. Note that some feature names are replaced with \texttt{location x} to protect private data.}
    \label{fig:02_correlation_hierarchy}
\end{figure}

It is also important to consider how features might correlate with each other. For small numbers of features, correlation matrices are ideal for such analysis. However, these graphics become difficult to interpret when there are more than a dozen features. Instead, this study made use of \textit{agglomerative clustering} of features by their correlations to produce a \textit{dendrogram} (Fig. \ref{fig:02_correlation_hierarchy}).

The dendrogram was generated as follows: First, the features were filtered to only consider those with less than 10\% of their values missing. Each of the $n$ features then had its absolute Pearson correlation to each other feature calculated, producing an $n \times n$ correlation matrix. Next, the pairwise Euclidean distances between each row of absolute correlations were calculated. Hierarchical clustering was performed on these pairwise distances using the Nearest-Point algorithm. These clusters were then plotted as a coloured dendrogram (Fig. \ref{fig:02_correlation_hierarchy}).

This hierarchical approach to correlations between features is immensely useful and offers more intuitive interpretations than a standard correlation matrix. By summarising the correlations between all $n \times n$ pairs of features into distances in $n$-dimensional space, the complexity of the relationships is interpreted for us. This highlights groups of features that are strongly correlated with one another but not with other clusters of features. Moreover, we can use the dendrogram representation to interpret just how correlated subsets of features are. This is immensely useful for validating (and refining) feature engineering in order to achieve the higher sample efficiency needed to model wide QS data. 

In Fig. \ref{fig:02_correlation_hierarchy} we can see that the hourly movement data (prefix \texttt{hk\_act\_}) clustered together in chunks of time and was highly correlated with aggregation features for this data. Specifically, we can see that movement between the hours of 1 and 7 was very highly correlated and therefore also highly correlated with their mean (\texttt{hk\_act\_night\_mean}) (Fig. \ref{fig:02_correlation_hierarchy}b). This tells us that most of the (linear) information about nighttime movement can be captured from a single feature that averages over the hourly features. We can also notice that there was low correlation between different time windows of movement features. The morning, afternoon, and evening features clustered within their time windows but not across them (Fig. \ref{fig:02_correlation_hierarchy}c). This tells us that the different windows of time throughout the day --- morning, afternoon, evening, night --- capture different information that might be relevant to our model. 

A final observation from the dendrogram is that location data formed a number of related, but strongly-separated clusters. Geohashes gave similar information and were clustered (Fig. \ref{fig:02_correlation_hierarchy}e), but were very far from the cluster of features relating to city, country, and hemisphere (Fig. \ref{fig:02_correlation_hierarchy}d). This suggests that two levels of resolution are important --- frequent changes in location at a resolution of metres, and infrequent changes in location at a resolution of kilometres.

\subsubsection{Autocorrelation}\label{sssec:autocorrelation}

\begin{figure}[H]
     \centering
     
     \begin{subfigure}[b]{0.49\textwidth}
         \centering
         \includegraphics[width=\textwidth]{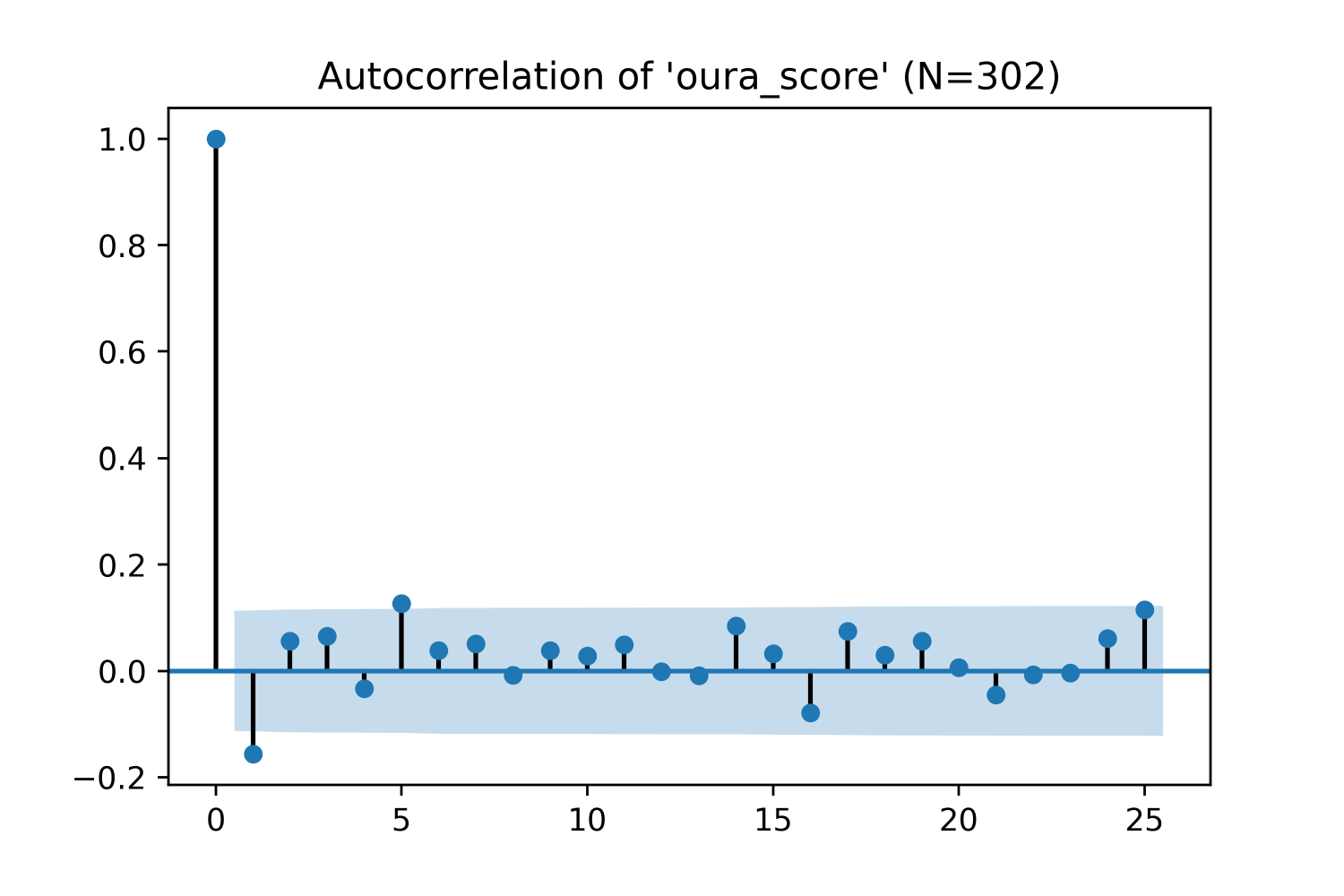}
         \caption{No autocorrelation.}
         \label{fig:autocorrelations_a}
     \end{subfigure}
     \hfill
     \begin{subfigure}[b]{0.49\textwidth}
         \centering
         \includegraphics[width=\textwidth]{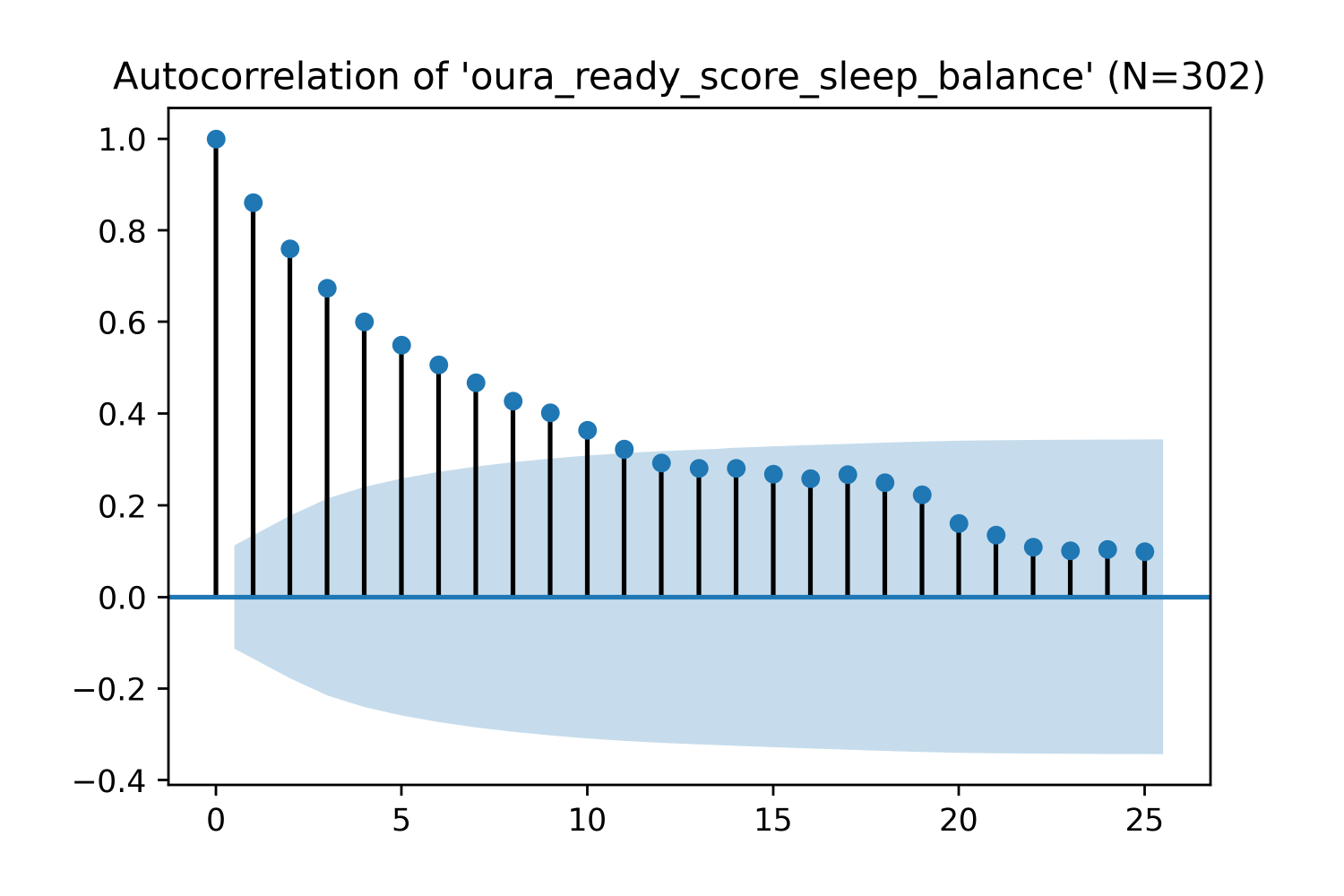}
         \caption{Trending autocorrelation.}
         \label{fig:autocorrelations_b}
     \end{subfigure}
     \hfill
     \begin{subfigure}[b]{0.49\textwidth}
         \centering
         \includegraphics[width=\textwidth]{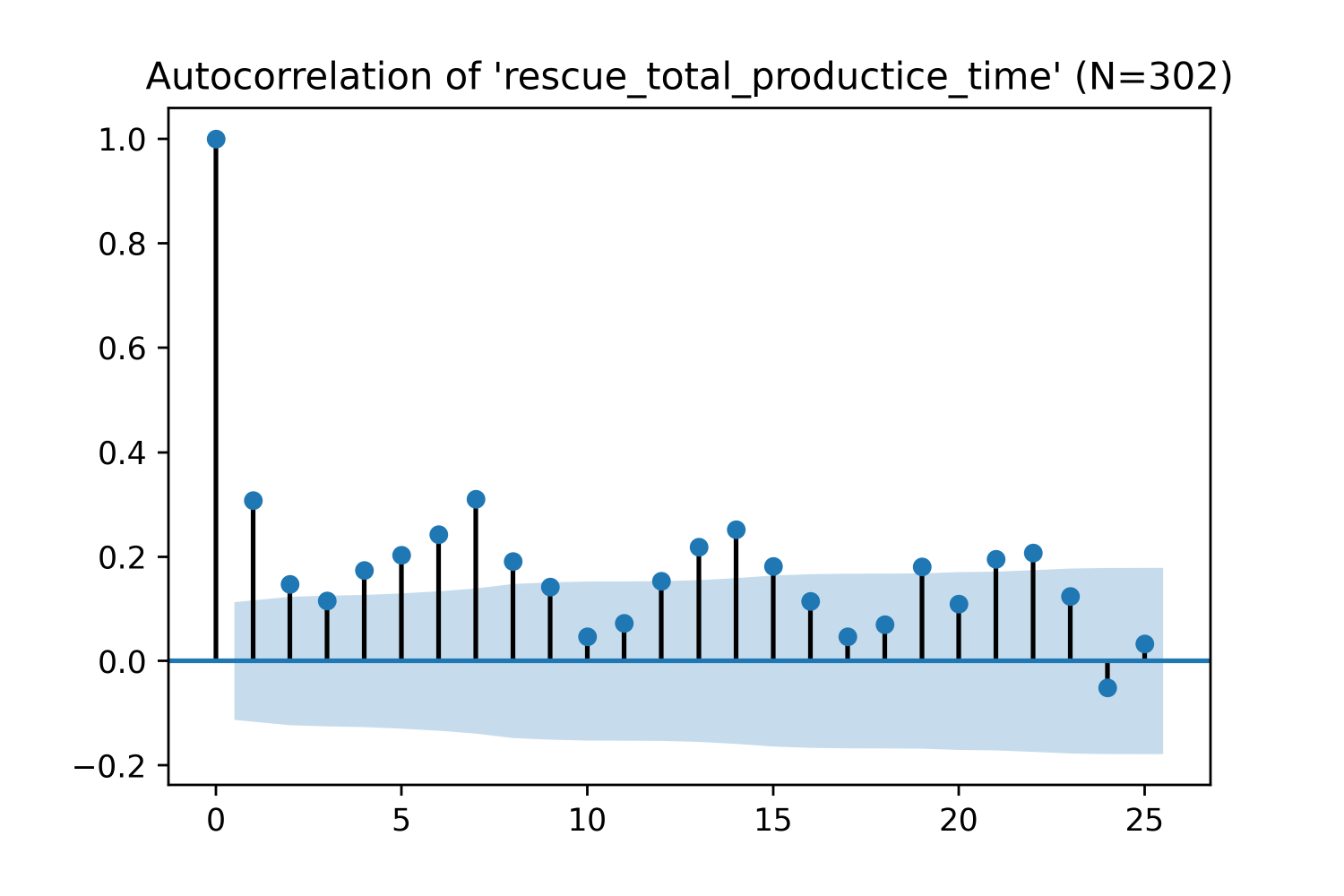}
         \caption{Periodic autocorrelation.}
         \label{fig:autocorrelations_c}
     \end{subfigure}
     \hfill
     \begin{subfigure}[b]{0.49\textwidth}
         \centering
         \includegraphics[width=\textwidth]{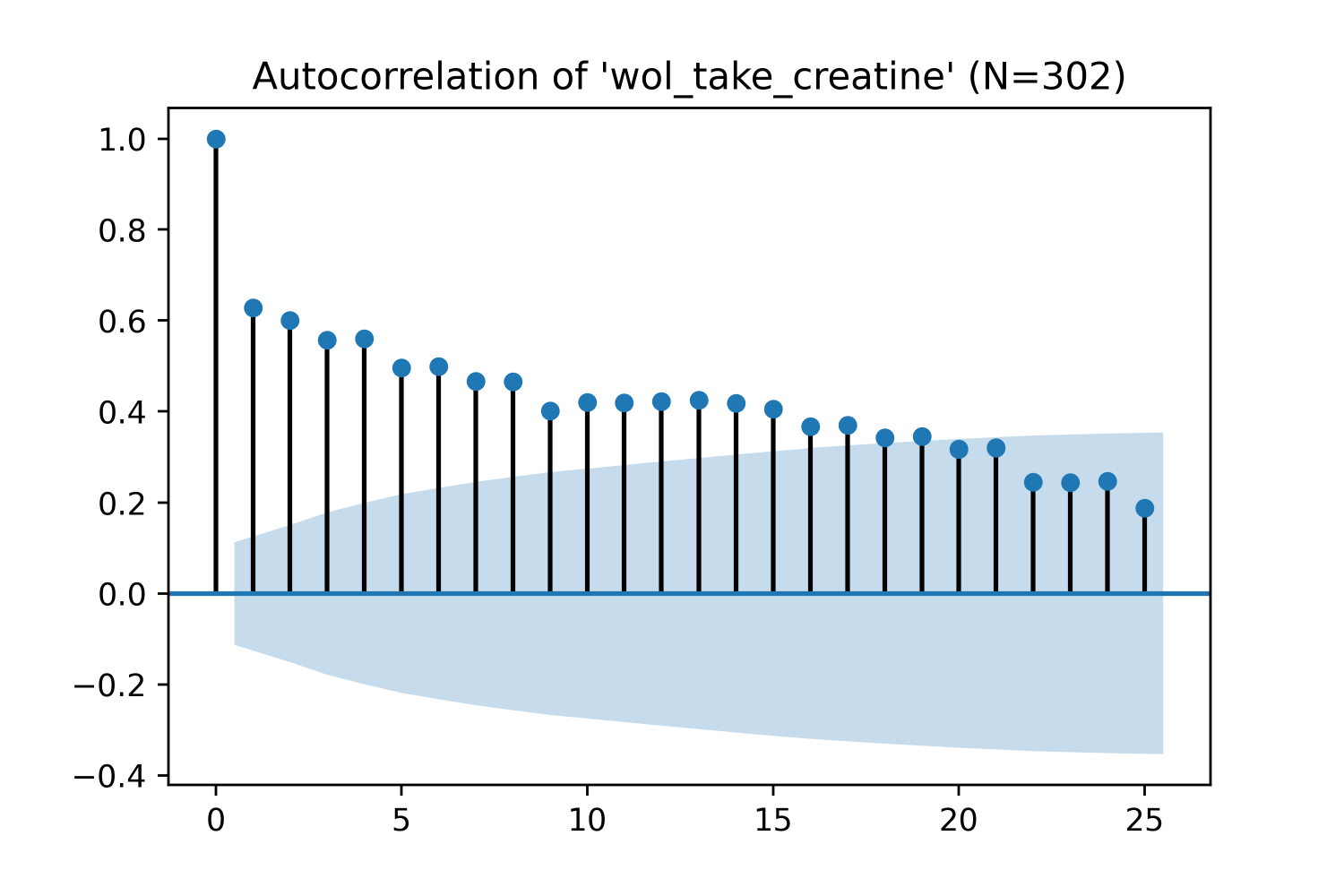}
         \caption{Trending autocorrelation.}
         \label{fig:autocorrelations_d}
     \end{subfigure}
     \hfill
    \caption{Notable autocorrelation patterns observed across features. Each plot shows the correlation on the vertical axis and the lag (in days) on the horizontal axis. The shaded regions represent a 95\% confidence interval for the null hypothesis at each lag level, so values that fall within this region are unlikely to be significant.}
    \label{fig:autocorrelations}
\end{figure}

Autocorrelation can be detected by computing the correlation between a feature and a copy of the feature where values are shifted (lagged) by 1 or more points. Autocorrelation gives us an indication of how well a feature predicts its own future values \cite{brockwell2016introduction}. This was immensely useful in this study because of the inherent time series nature of most QS data. Behavioural and biological patterns often have periodic variations or trends over time. For example, it is common for people to dramatically shift their sleeping patterns on weekends compared with weekdays. This can often be detected by spikes in autocorrelation at intervals of 6-7 days --- last week somewhat predicts this week. For this study, the autocorrelation of each feature was analysed for varying shifts (lags) up to 25 days. Four interesting patterns are presented in Fig. \ref{fig:autocorrelations}. 

We can see that the target feature (\texttt{oura\_score}) had a very low degree of autocorrelation (Fig. \ref{fig:autocorrelations_a}). This is extremely surprising, as it implies that previous sleep quality does not have a strong linear relationship to current sleep quality. This highlights how difficult the task of predicting sleep quality is. Interestingly, a lag of 1 day had a small negative correlation with the target ($r_1 \approx -0.2$). This suggests that sleep oscillates in quality on adjacent nights. We can also see that some features showed a clean trend of autocorrelation with respect to lag (Fig. \ref{fig:autocorrelations_b} and \ref{fig:autocorrelations_d}). This is to be expected of a number of features, such as habits and temporally-derived features like \texttt{sleep\_balance}. What is worth noting is the lag point at which the autocorrelation becomes statistically insignificant. This gives us a rough indication of how many days of history may be relevant for our model. 

Finally, we can see that some features  displayed prominent periodicity (Fig. \ref{fig:autocorrelations_c}). Productivity time is derived from time spent on my laptop working with specific software or websites that are marked as productive. We can see from the spikes in the plot that the periodicity had a peak-to-peak length of exactly 7 days. This makes a great deal of sense, as my productivity levels are highly influenced by the day of the week. 

\subsection{Stationarity}\label{ssec:stationarity}
A key principle of modelling time series data is that it should be stationary. Specifically, it should exhibit no periodicity or trends, leaving only irregular variations that we attempt to predict using other features \cite{Hoogendoorn2018}. However, most real-world data is not stationary. As we have already seen, many of the relevant features in this dataset are periodic. 

\begin{figure}[H]
     \centering
     \begin{subfigure}[b]{0.48\textwidth}
         \centering
         \includegraphics[width=\textwidth]{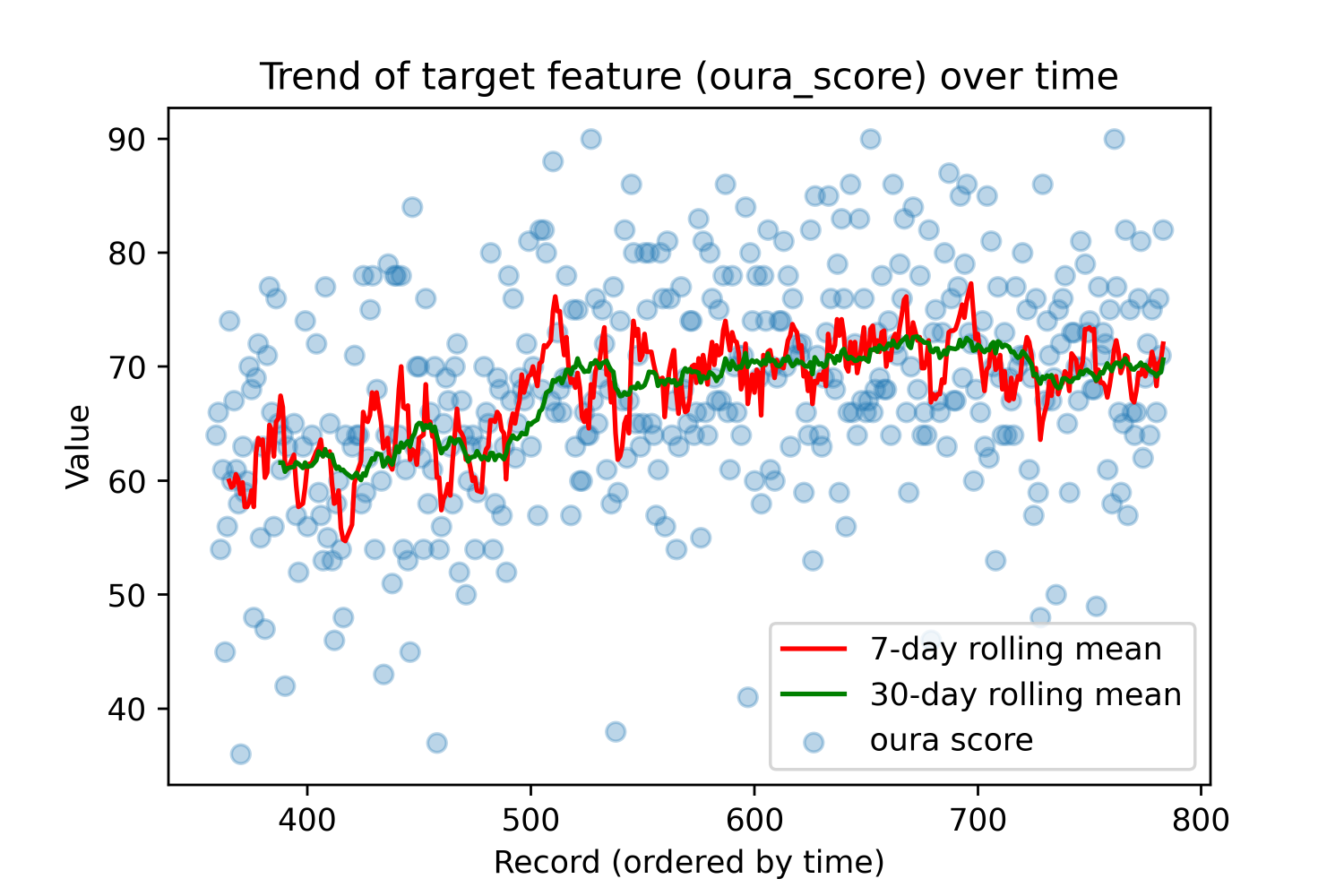}
         \caption{Trend of mean for target.}
         \label{fig:stationarity_a}
     \end{subfigure}
     \hfill
     \begin{subfigure}[b]{0.48\textwidth}
         \centering
         \includegraphics[width=\textwidth]{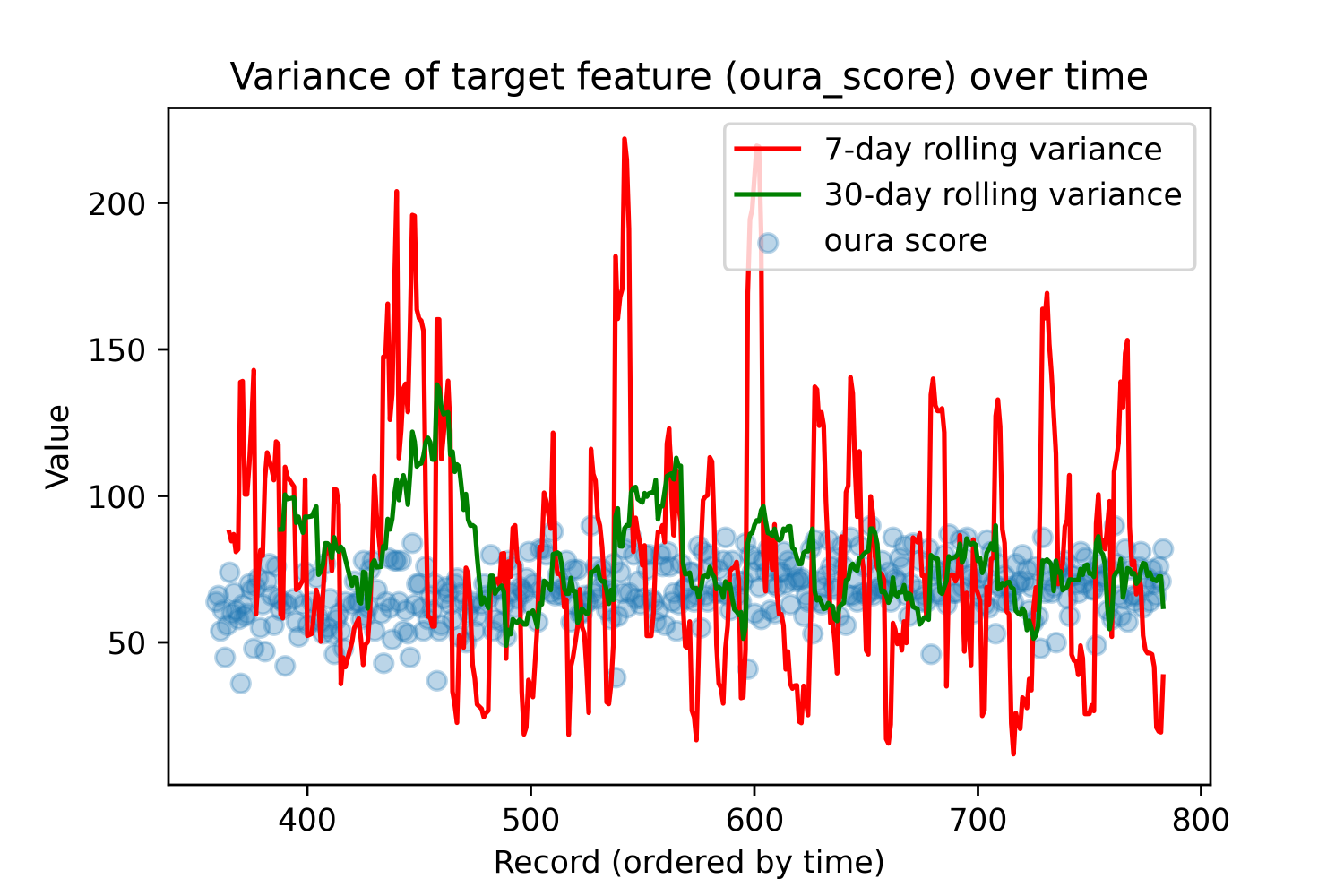}
         \caption{Trend of variance for target.}
         \label{fig:stationarity_b}
     \end{subfigure}
     \hfill
    \caption{Scatterplots of target features (\texttt{oura\_score}) over time (indexed observations). The left figure superimposes trends of the 7- and 30- day rolling mean. The right figure superimposes trends of the 7- and 30- day rolling variance.}
    \label{fig:stationarity}
\end{figure}

The augmented Dickey-Fuller test can be used to assess the stationarity of a sequence of data \cite{said1984testing}. This test was applied to each feature in the dataset. A total of 34 input features were found to be non-stationary ($p>0.05$). Many of these were explicitly temporal features like the year, month, or day. Many were naturally non-stationary data like weather patterns.

What we are most concerned with, however, is whether the target feature is stationary. If not, techniques like statistical differencing need to be applied before building a model \cite{Hoogendoorn2018}. Unfortunately, such applications make interpretation much more difficult. Fortunately, the target feature (\texttt{oura\_score}) was found to be stationary with reasonable confidence ($p<0.01$). This means that variations had no major periodic or trending patterns. This is illustrated visually with rolling averages of both the mean and variance in Fig. \ref{fig:stationarity}. This was very fortunate for this study, as the interpretation step is easier if we do not have to transform the target feature. This stationarity of the sleep features was likely a happy result of the pandemic conditions. The data included for modelling begins from October 2019. When including data back to December 2018, the target was less likely to be stationary ($p \approx 0.09$).

\section{Overcoming missing data}\label{sec:overcoming_missing_data}

\subsection{Theory of missing data}\label{ssec:theory_of_missing_data}
There are three main types of missing data identified in the statistics literature \cite{little2019statistical}. Understanding their different properties is integral to selecting robust imputation techniques for filling in the missing values \cite{beaulieu2018machine}.

\textbf{Missing completely at random (MCAR)}: Missing values are considered MCAR if the events that caused them are independent of the other features and occur entirely at random. This means that the nullity (missingness) of the values is unrelated to any of our study features and we can treat the data this \textit{is} present as a representative sample of the data as a whole. Unfortunately, most missing values are not MCAR \cite{beaulieu2018machine}.

\textbf{Missing at random (MAR)}: Despite the name, MAR occurs when the nullity is \textit{not} random, but can be fully accounted for by other features that have no missing values. For instance, I did not explicitly log the city where I slept every night, but the data is MAR because it is fully accounted for by the date index and geohash features, from which I can accurately infer the missing values for the city.

\textbf{Missing not at random (MNAR)}: Data is MNAR when the nullity of the value is related to the reason that it is missing. This makes the dataset as a whole biased. For instance, I am more likely to forget to log my mood when I am very happy or utterly miserable. The missing extreme values in the data are thus MNAR. 

\subsection{Analysis of missing data}
\begin{figure}[]
    \centering
    \includegraphics[width=\textwidth]{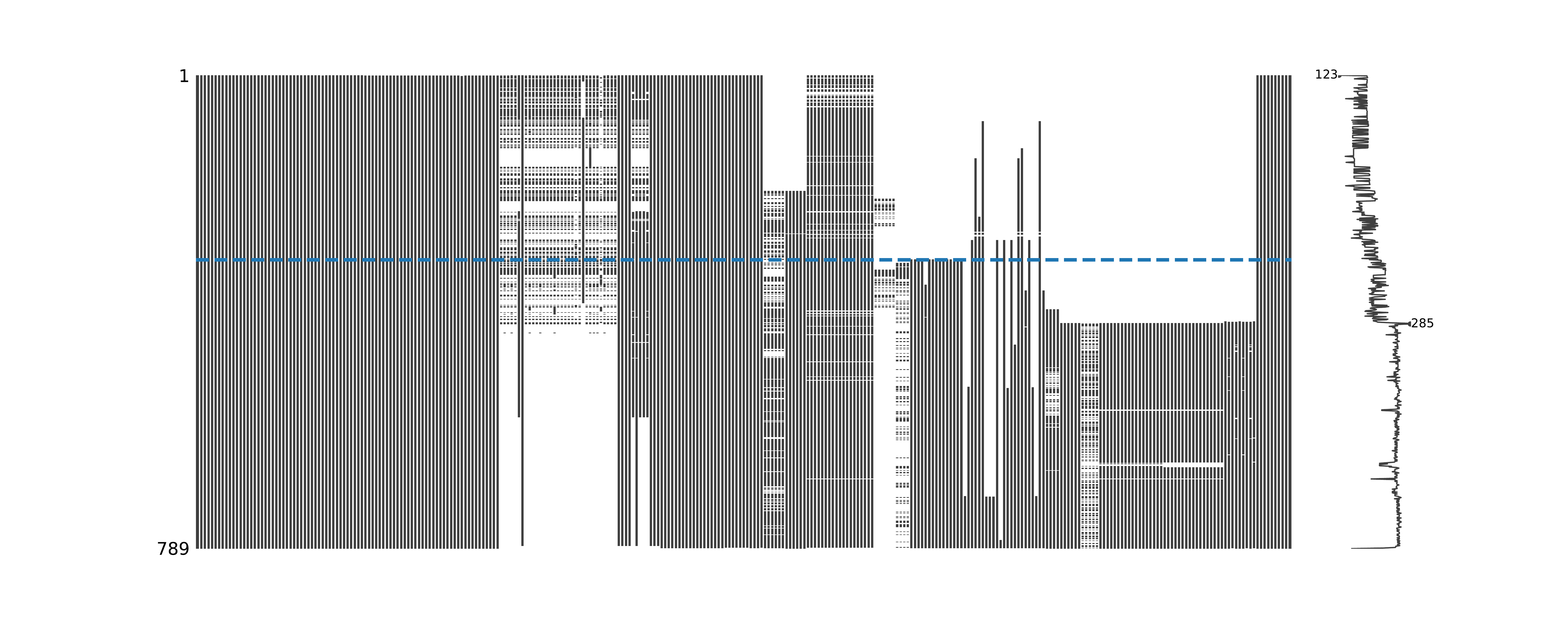}
    \caption{Missing data matrix for the entire raw dataset. The vertical axis represents rows in the dataset (corresponding to daily observations). The horizontal axis represents columns in the dataset (corresponding to features). The dashed blue line indicates the start of the period of time considered for this study. The sparkline on the right indicates the general shape of the data completeness, with the rows of minimum and maximum nullity labelled with a count of the non-missing values in the rows.}
    \label{fig:02_missing_unified_raw}
\end{figure}

\begin{figure}[]
    \centering
    \includegraphics[width=\textwidth]{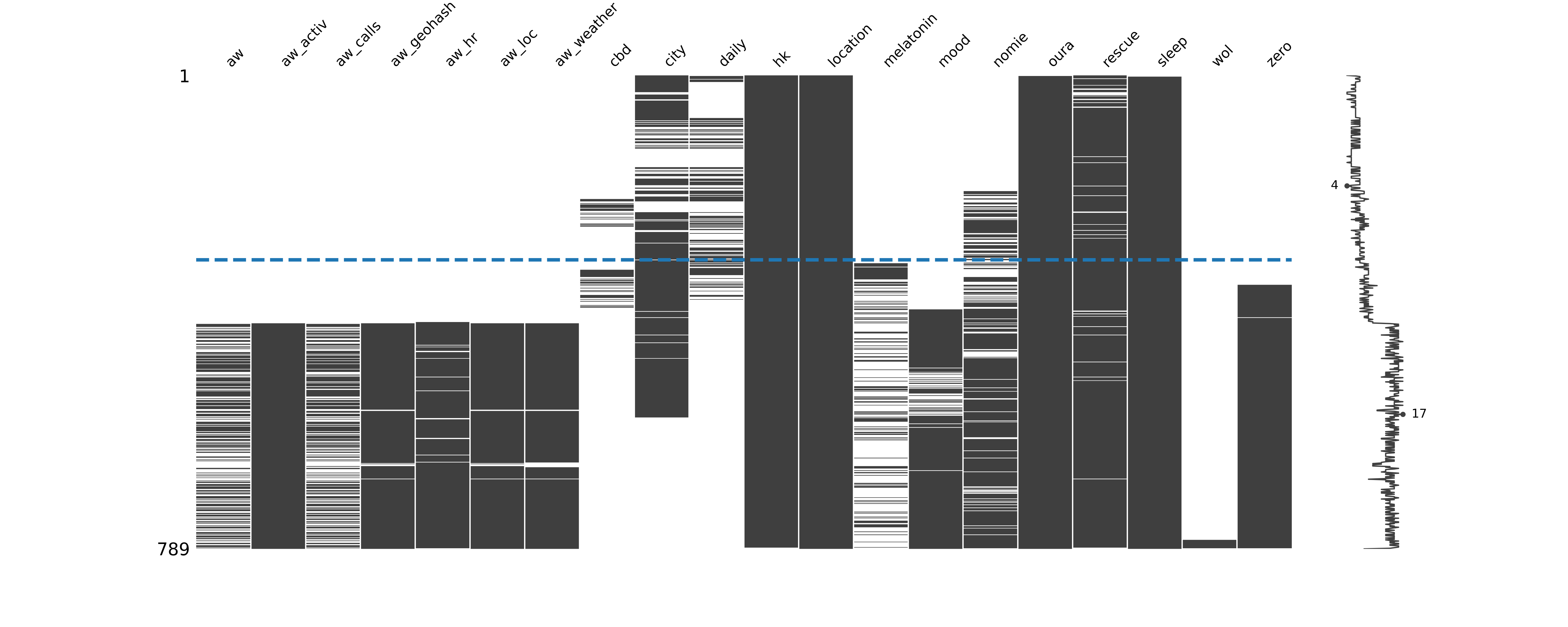}
    \caption{Missing data matrix for groups of features in the raw dataset. The vertical axis represents rows in the dataset (corresponding to daily observations). The horizontal axis represents columns in the dataset (corresponding to features). The dashed blue line indicates the start of the period of time considered for this study. The sparkline on the right indicates the general shape of the data completeness, with the rows of minimum and maximum nullity labelled with a count of the non-missing values in the rows. The column names are the prefixes of each source (see Table \ref{tab:sources}).}
    \label{fig:02_missing_simplified_raw_matrix}
\end{figure}

The absence of data points is a major factor in Quantified-Self (QS) projects \cite{Hoogendoorn2018}, especially when combining data from multiple sources. An upfront analysis of the quantity and distribution of missing values is essential before missing values can be rectified. Missing data matrices are an invaluable visualisation in this regard. They give an impression of the dataset in the same rows-as-observations and columns-as-features format that we are accustomed to, with shading to indicate where data is present. In Fig. \ref{fig:02_missing_unified_raw}, the entire dataset is rendered as on of these matrices using the superb \textit{missingno} library \cite{bilogur2018missingno}. It is important to note that this version of the dataset included more than the 15-month timeline of the final dataset. This was for illustrative purposes. Much of the top half of the dataset was ultimately discarded before modelling. In Fig. \ref{fig:02_missing_simplified_raw_matrix}, the columns are grouped by the source prefix (Table \ref{tab:sources}) in order to get a better understanding of which data sources are to blame for which missing data. When combining columns from the same source, missing values took precedence. In other words, the simplified matrix represents the worst-possible combination of the columns from that source, in terms of nullity. 

Anecdotally, these patterns are typical of multiple-source QS projects. New data sources are added over time, whilst others fall out of use. Some data is tracked very consistently (often automatically), whilst many of the manually-tracked sources have short sequences of missing values spread all over --- due to poor adherence to tracking protocols. 

We can see from the missing data matrices that the sources used changed dramatically over time. The AWARE (\texttt{aw}), Nomie, and Zero sources were only adopted later on. Fortunately for the focus of this study, the sleep data (\texttt{oura}) was complete and had no missing data, as it was automatically tracked and the ring was worn every night. By having a target feature free of missing data, this study was well-positioned to mitigate the missing data in the other features. 

\subsection{Approaches to handle missing data}
Most modelling techniques assume complete data, meaning that missing values pose a major problem \cite{burkov2019hundred}. There are three common strategies for dealing with missing data: omission, full analysis, and imputation \cite{little2019statistical, burkov2019hundred, kang2013prevention}.

\textbf{Omission} (dropping): The most common approach is to omit sections of the dataset that contain missing values. Typically, this involves discarding observations (rows) that contain one or more missing values. This leaves only complete observations, but discards much of the information in the original dataset. In QS datasets with many features from heterogeneous sources, it is often the case that the majority of observations contain at least one missing value. In such cases, dropping all these rows would discard almost all of the dataset. Some features (columns) have more missing values than others, so it is often desirable to discard the worst-offending features first, then discard the observations with missing values from the features that remain. This often results in retaining more overall information. In n-of-1 QS studies, we typically cannot afford to discard much data, so finding the right trade-off between retaining observations and features is a challenge.

\textbf{Full analysis}: Some learning algorithms are designed to use all available data and to tolerate missing values. Many of these are beyond the scope of this study, but the XGBoost implementation of gradient boosted Decision Trees is of interest. XGBoost is a well-engineered ensemble technique that learns default directions in the branches of the Decision Trees that are taken when missing values are encountered \cite{chen2016xgboost}. This leverages the values that are present to learn reasonable estimates for missing values. XGBoost is also a high-performing algorithm \cite{chen2016xgboost} and was thus a prime candidate from the outset of this study.

\textbf{Imputation}: Instead of omitting missing values, we can try to estimate them using imputation. This has the advantage of retaining the size and shape of the dataset, but the disadvantage of incorporating estimation error into the data --- resulting in additional noise for the model to overcome \cite{little2019statistical, kang2013prevention}. This study employed a variety of imputation strategies (\S \ref{ssec:imputation_strategies}) in combination and compared the resulting model performance. When our data is an ordered time series, we can use a specific kind of imputation called interpolation \cite{Hoogendoorn2018}. This fills the gaps between available samples by assuming some regular rate of change between samples. For most features in the dataset, the sample frequency was too low for interpolation to be viable. One notable exception was the heart rate data, which was sampled over a thousand times each day. This feature was first resampled at intervals of 1 minute to make it regular, then linear interpolation was applied to smooth over discontinuities. 

\subsection{Knowledge-based filling}\label{ssec:knowledge_based_filling}
Much of the data in observational n-of-1 studies is MAR or MNAR, both of which can be predicted with an accuracy much greater than random guessing \cite{beaulieu2018machine}. Furthermore, some QS studies have a unique kind of MCAR data that can be filled accurately, albeit not with prediction. We can refer to these as \textit{informative absences}. They are the result of the feature engineering step, but can be overcome due to intimate knowledge of the study protocol and adherence --- a perk unique to QS research. 

The informative absences resulted from event logs that were aggregated into daily summaries and, in most cases, could simply be filled with a suitable placeholder value. For instance, alcohol consumption was only logged (with a timestamp) when it occurred. Because I was both the subject and the researcher, I can be extremely confident that I almost never failed to log events like alcohol because I was conscious of my own adherence patterns. When the event logs were aggregated, there were very many days where no alcohol was consumed that simply had no data. Because I was confident that I adhered well to alcohol logging, I could simply fill this missing data with appropriate values indicating that no alcohol was consumed: $0.0$ for the sum features, and $-1.0$ for the temporal features. I refer to the manual filling of these kinds of MCAR data as \textit{knowledge-based} filling, as it relies on expert and domain knowledge to determine when and what values should be filled. We can see these informative absences in Fig. \ref{fig:02_missing_simplified_raw_matrix} for \texttt{melatonin}, \texttt{nomie}, and \texttt{aw\_calls}.

\subsection{Imputation strategies}\label{ssec:imputation_strategies}
After manually filling the informative absences with the knowledge-based approach, many MAR and MNAR values remained. These were imputed using one of several imputation strategies (see Table \ref{tab:imputation_strats}).

\subsubsection{Univariate imputation}
The most straightforward imputation strategy is univariate imputation. This is where missing values for the $j$-th feature are imputed based on the non-missing values for that $j$-th feature. In the case of continuous features, the mean or the median of the non-missing values is calculated for each feature. That value is then filled into the missing values for the respective feature. In this study, both mean and median univariate imputation were utilised (see Table \ref{tab:imputation_strats}).

\subsubsection{Multivariate imputation}
In the case of data that is MAR, the missing values for the $j$-th feature may be more accurately imputed when considering each observation $i$ individually and conditioning on information from the other features. For instance, a missing value relating to a particular day's activity levels could likely be imputed quite accurately based on non-missing values for weather and heart rate for that same day. This is the intuition behind multivariate imputation. Conceptually, each feature $\boldsymbol{x}_j$ in our data matrix $\boldsymbol{X}$ is modelled as a function $f_j$ of the other features $\boldsymbol{X}_{\neg j}$. A missing value at $x_{i,j}$ can then be imputed using the function $f_j(\boldsymbol{X}_{i, \neg j})$. In reality, the implementation of a specific imputation technique might look very different from this conceptual view. In this study, several approaches to multivariate imputation were explored (see Table \ref{tab:imputation_strats}). 

\subsubsection{Multiple imputation}
So far, we have considered single rounds of univariate or multivariate imputation. However, because many of the multivariate imputation techniques are stochastic, it is common to run multiple independent rounds of imputation on the entire dataset, in what is known as multiple imputation. These rounds can then be either (1) compared to select the one that minimises some cost function, or (2) can be aggregated into a single imputed dataset. The aim of these multiple rounds is to achieve a more accurate estimate of the imputed values. However, when the imputed dataset is then used for subsequent modelling, the uncertainty of these estimates is not considered, meaning there is a greater quantity of noise in the data that the model must overcome. 

To remedy this, some practitioners keep all versions of the dataset generated from the rounds of multiple imputation and concatenate them instead of aggregating them \cite{YouTubeMissingData}. If the original dataset was $m \times n$ in shape, the multiply-imputed dataset is $km \times n$ in shape, where $k$ is the number of rounds of multiple imputation. This effectively adds many observations that are slight variants of each other to the dataset, in the hope that the downstream model will make sense of the uncertainty in the imputed values. In this study, a number of multiple imputation strategies were employed (see Table \ref{tab:imputation_strats}). For the MICE technique, the imputation rounds were concatenated into a longer dataset to capture the uncertainty \cite{kang2013prevention}. 

\subsection{Baseline dataset with missing values}\label{ssec:baseline_dataset_withnans}
The unified dataset was filtered down to only numeric features from after \texttt{2019-10-05}, resulting in 482 observations and 271 features. Features which had more than 70\% of their values missing were removed, leaving 234 features. The remaining data showed a bimodal distribution of nullity. That is, 424 observations had fewer than 15 missing values across all 234 features, whilst the other 119 observations had more than 15 missing values across all the features. 

\begin{figure}[H]
     \centering
     \begin{subfigure}[b]{0.45\textwidth}
         \centering
          \includegraphics[width=\textwidth]{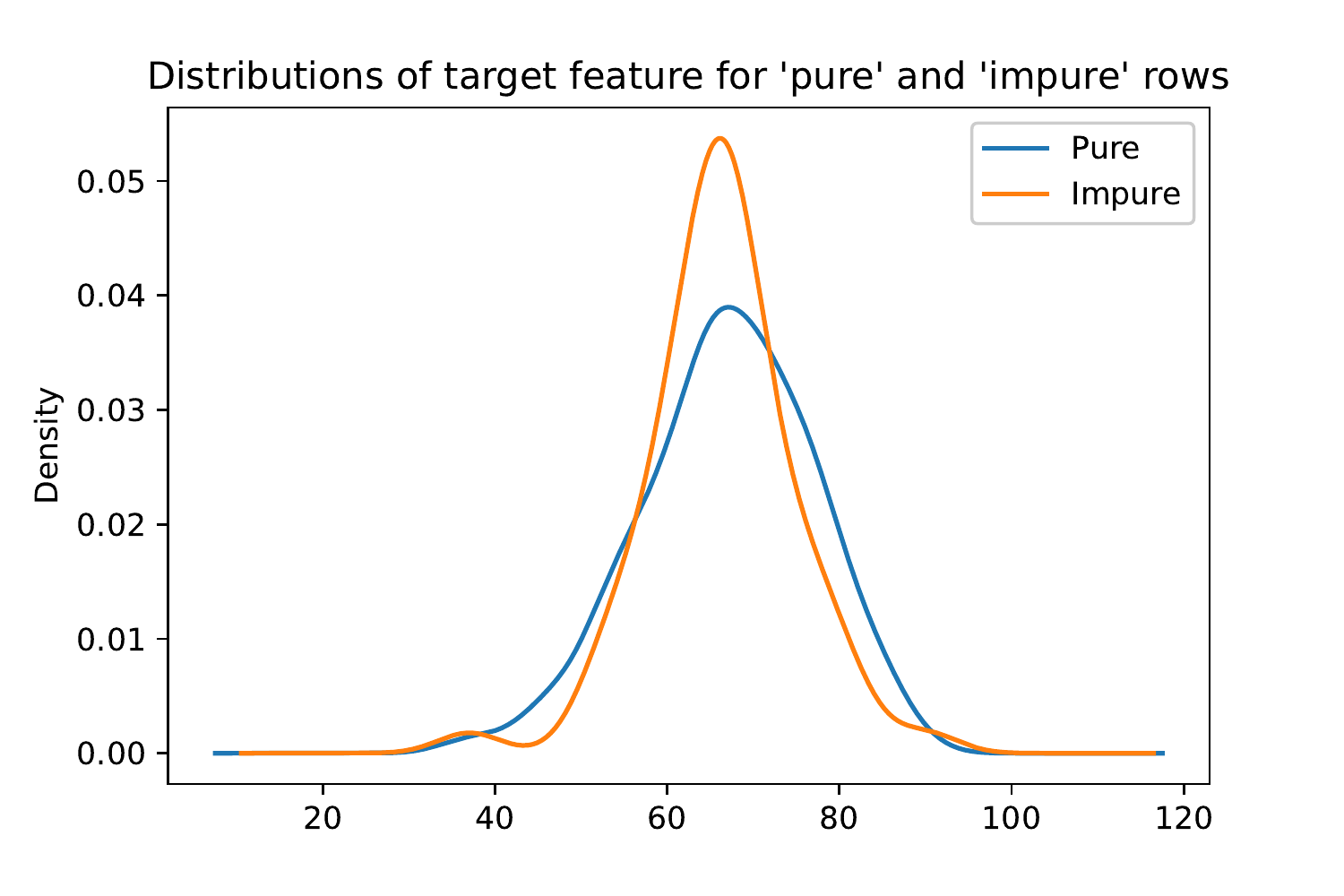}
            \caption{Comparison of target distributions.}
            \label{fig:03b_pure_impure_dist}
     \end{subfigure}
     \hfill
     \begin{subfigure}[b]{0.45\textwidth}
         \centering
         \includegraphics[width=\textwidth]{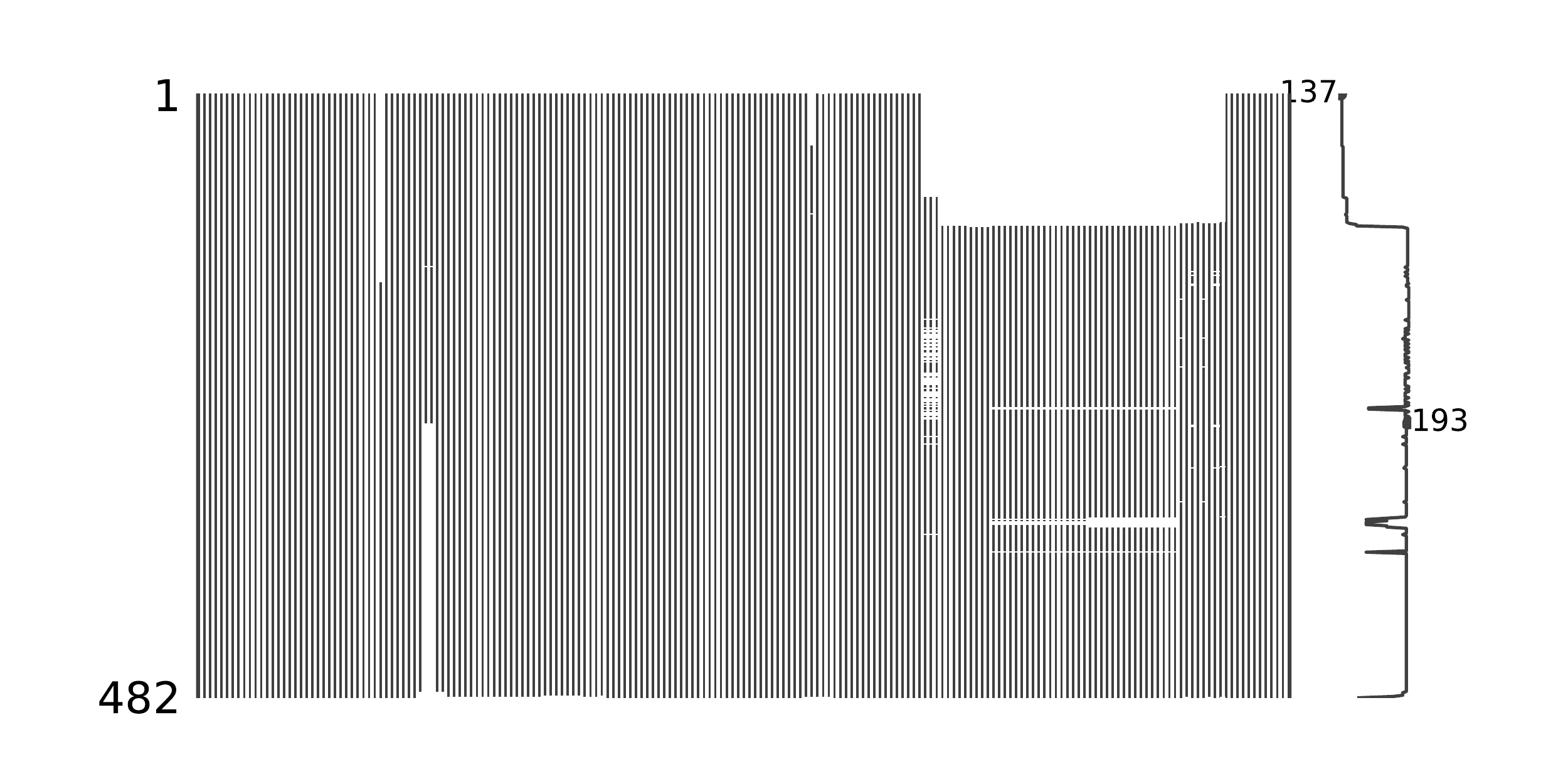}
         \caption{Missing data matrix.}
         \label{fig:pre_markov_with_nans.csv}
     \end{subfigure}
        \caption{Left: Comparison of target feature (sleep quality) for ``pure'' observations, which had no missing data in any of their features, and ``impure'' observations, which had some missing values. Right: missing data matrix for \texttt{with\_nans} dataset.}
        \label{fig:with_nans_subfigs}
\end{figure}

A subset of 61 observations were ``pure''. That is, they had no missing values in any of their features. An additional column (\texttt{is\_pure}) was appended to the dataset to label these observations. Only these observations were sampled for scoring model performance in later experiments. This had the advantage of isolating the effect of imputation techniques to only the training data, but came with two caveats. Firstly, there was a huge imbalance between pure and impure features (61:421). Secondly, the target distribution differed between these subsets (Fig. \ref{fig:03b_pure_impure_dist}). 

Data-leaking features with the \texttt{oura\_} prefix were removed. The resulting \texttt{with\_nans} dataset of 482 observations and 193 features (including the target) was used as the basis for all the imputations and as a non-imputed baseline. The missing values of this dataset are visualised in Fig. \ref{fig:pre_markov_with_nans.csv}.

\subsection{Quantifying imputation distance}\label{ssec:imputation_distance}
To quantify the change to a dataset from imputation, the imputation distance metric $D_{\text{imp.}}$ was defined. What follows is the derivation of the imputation distance metric from foundational principles of relative entropy.

The Jensen-Shannon distance (JSD) between two probability arrays $P$ and $Q$ is defined as:
\begin{equation}
    JSD(P\mid \mid Q) := \sqrt{\frac{D_{KL}(P \| M)+D_{KL}(Q \| M)}{2}}
\end{equation}
where $M = \frac{P + Q}{2}$ is the point-wise mean of the arrays $P$ and $Q$ and $D_{KL}$ is the Kullback-Leibler divergence (relative entropy).

\newpage
To calculate the JSD for two distributions of some feature, we first have to generate probability array $P$ and $Q$ for each distribution of the feature. We approximate this by constructing $1000$ equal-width bins over the range of each distribution and then counting how many of the observed values fall into those bins. These counts are divided by the number of observations to produce a probability array for each distribution.

The JSD has a number of desirable properties. Like KL-divergence, it is a principled information-theoretic measure that is non-negative and is zero only when the distributions are identical. Unlike KL-divergence, however, it is \textit{symmetric} and \textit{smoothed}, making it more robust \cite{lin1991divergence}. 

The JSD was used to compute the change in each feature $j \in \mathbb{R}^n$ from the original dataset (with missing values) $\boldsymbol{X}$ to the imputed dataset $\boldsymbol{X}'$. The average distance of an imputation $D_{\text{imp.}}$ could be estimated by taking the arithmetic mean over all $n$ features:
\begin{equation}
    D_{\text{imp.}}(\boldsymbol{X} \mid \mid \boldsymbol{X}') := \frac{1}{n} \sum_{j = 1}^{n} JSD\left[P(\boldsymbol{x}_j) \mid \mid Q(\boldsymbol{x}'_j)\right]
\end{equation}
The average imputation distances relative to the raw dataset were calculated for each imputation strategy (Table \ref{tab:imputation_strats}). 

\begin{table}[H]
\caption{Overview of the imputation strategies used in this study. The name is used to refer to the technique in the paper. The distance between the raw dataset and the imputed dataset was computed using the mean Jenson-Shannon (JSD) distance over all the features (\S \ref{ssec:imputation_distance}). A larger distance value indicates more dramatic differences between the distributions of the raw and imputed versions of the dataset. The implementation column refers to the Python class used to implement the technique, from either the\textit{Scikit-learn} \cite{pedregosa2011scikit} (*) or \textit{fancyimpute} \cite{fancyimpute} (\dag) libraries.}
\label{tab:imputation_strats}
\begin{tabularx}{\textwidth}{lllll}
\hline
\textbf{Name}   & \textbf{Distance}         & \textbf{Scope} & \textbf{Imputation}                          & \textbf{Implementation}         \\ \hline
    Univariate mean                              & 0.0923                     & Univariate     & Single              & SimpleImputer *          \\
    Univariate median                            & 0.0535                     & Univariate     & Single              & SimpleImputer *          \\
    Iterative Bayesian ridge (50)                & 0.0986                     & Multivariate   & Multiple            & IterativeImputer *       \\
    MICE (3) \cite{buuren2010mice}               & 0.2079                     & Multivariate   & Multiple            & IterativeImputer *       \\
    MICE (15) \cite{buuren2010mice}              & 0.2292                     & Multivariate   & Multiple            & IterativeImputer *       \\
    KNN (K=3)                                    & 0.0658                     & Multivariate   & Single              & KNN \dag                 \\
    SoftImpute \cite{mazumder2010spectral}       & 0.1319                     & Multivariate   & Multiple            & SoftImpute \dag          \\
    Iterative SVD \cite{troyanskaya2001missing}  & 0.2147                     & Multivariate   & Multiple            & IterativeSVD \dag        \\
    Matrix factorisation \cite{takacs2008matrix} & 0.1897                     & Multivariate   & Multiple            & MatrixFactorization \dag \\ \hline
\end{tabularx}
\end{table}

Univariate imputation of the median had the lowest imputation distance across all features. Univariate mean imputation had a much greater distance than univariate median imputation. This is likely because many of the features had non-normal distributions and the median is far less sensitive to outliers. Iterative SVD and MICE had the highest imputation distances. Greater numbers of concatenated MICE iterations (15 versus 3) resulted in a greater imputation distance.

\newpage
\section{Collapsing time with Markov unfolding}\label{sec:collapsing_markov_unfolding}
\subsection{Markov assumption}
Building a predictive model of sleep requires us to make the Markov assumption \cite{russel2013artificial} --- that the current night's sleep depends only on a finite fixed number of previous days' data. 

Imagine that we are trying to calculate a probability distribution $P$ over the possible sleep scores $y_t \in [0, 100]$ on day $t$, conditioned on all features from the past. Formally, we are trying to model:

\begin{equation}\label{eq:markov1}
    P(y_t \mid \boldsymbol{X}_{1:t-1})
\end{equation}
where $y_t$ is the current night's sleep score and $\boldsymbol{X}_{1:t-1}$ is the matrix of features for all days prior to day $t$. For simplicity, $\boldsymbol{X}_{1:t-1}$ includes the vector $\boldsymbol{y}_{1:t-1}$, the previous target values. 

Eq. \ref{eq:markov1} implies that the sleep on night $t$ depends on all past nights of sleep and all past days of other features. In other words, we require the full history. 

Now, if we make the Markov assumption:

\begin{equation}
    P(y_t \mid \boldsymbol{X}_{1:t-1}) = P(y_t \mid \boldsymbol{X}_{t-\tau:t-1})
\end{equation}

we assume that modelling only the past $\tau$ days produces as good an estimate of the current sleep as looking at the full history. This is a very useful assumption, as it allows us to reframe our time series prediction task to an independent-and-identically-distributed (i.i.d.) prediction task. This, in turn, makes many more techniques --- for both modelling and interpretation --- available to us. 

But how can we assume that the Markov property holds for this data? Clearly, if we set $\tau = t$ then the Markov property holds trivially, as we are using the full history. So there must be some $\tau < t$ at which the Markov property breaks down for our data. In reality, we do not care where the property first breaks down, but rather where the amount of information retained is sufficient for our prediction needs. We can estimate the optimal $\tau$ value by analysing our data for autocorrelation, then use it to construct an i.i.d. representation that incorporates the Markov property \cite{russel2013artificial}.

The autocorrelation analysis from Section \ref{sssec:autocorrelation}, alongside domain knowledge, highlighted that previous days' behaviour plays a role in the current characteristics and resulting sleep quality. For instance, I may have had a healthy and balanced day today, but if I was up late partying or revising for an exam the night before, then my sleep is likely to be affected. 

\newpage
\subsection{Markov unfolding}
We can implement this conversion from a time series to an i.i.d. dataset with a technique I dub \textit{Markov unfolding}. We begin with our $m \times n$ data matrix $\boldsymbol{X}$. The data in $\boldsymbol{X}_{t-\tau : t-1}$ (i.e. the past $\tau$ days) is ``unfolded'' into a row vector of length $\tau n$. This is repeated for each time point $t \in [\tau, m]$ and the resulting vectors are stacked onto $\boldsymbol{X}$ to produce a new dataset $\boldsymbol{X'}$ with shape $(m-\tau) \times (\tau + 1) n$. 

\begin{figure}[H]
    \centering
    \includegraphics[width=0.8\textwidth]{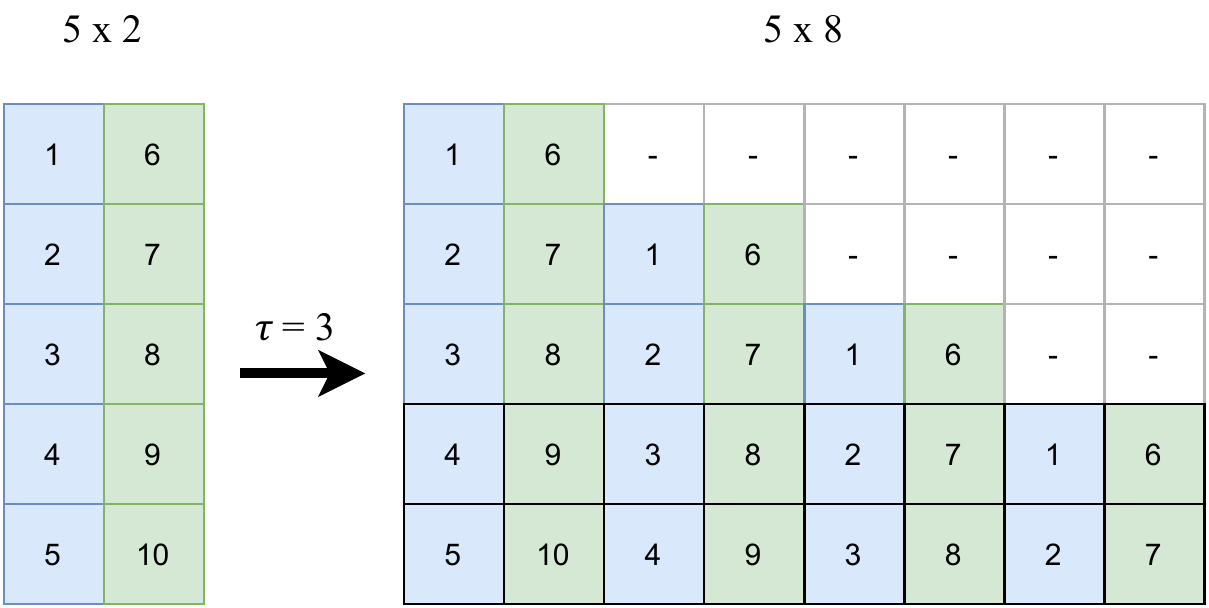}
    \caption{Illustrative example of Markov unfolding a $5 \times 2$ dataset into a $5 \times 8$ dataset using lag of $\tau = 3$. After removing the missing values, the final dataset is of shape $2 \times 8$ (outlined).}
    \label{fig:my_label}
\end{figure}

Intuitively, we are copying the columns of the original dataset, shifting them down by $1$ row, and stacking them as new columns on the right of the dataset. We repeat this with shifts of $2, 3, ..., \tau$; stacking the new columns each time. We have to throw away the first $\tau$ rows from the beginning of the dataset as we did not have enough previous states for them, resulting in missing values. We now have a dataset that is a few rows shorter, but $\tau +1$ times the width, than what we started with. We can now treat the rows (observations) in our dataset as totally independent, allowing us to use any supervised learning approach that assumes i.d.d. data.

The $\tau$ value was selected by analysing the autocorrelation of key features to set an upper limit (in this case, $\tau=7$). This is intuitive, as many cycles in behaviour occur at a weekly level, so allowing the models to look up to 7 days back is useful for capturing these patterns. For instance, behaviour around activity, working hours, alcohol consumption, and other lifestyle factors often differs dramatically on weekends. These patterns of behaviour (and their effect on sleep) are often a good predictor of the next week's behaviour, making them useful features to include. 

All the numeric features (including the target, \texttt{oura\_score}) were unfolded for $\tau$ days prior. This went some of the way to incorporating the Markov assumption into the time series, thus allowing each day to be treated as an independent observation. 

One major drawback of Markov unfolding is that it can result in low feature efficiency. In other words, too many features, with those from farther back (larger $\tau$) being only loosely-correlated to the current ones. This sometimes results in the ``curse of dimensionality,'' which can prevent the data from being learnable, given the limited number of observations \cite{vapnik2013nature, Hoogendoorn2018}. This is somewhat mitigated by the use of feature selection (\S \ref{sssec:rfe}).

\section{Model interpretation}\label{sec:model_interpretation}
Using predictive models to capture relationships in the data is integral to observational studies. Models can tell us much more than correlations between pairs of features, as they (1)  allow us to control for the effects of other features and (2) can be regularised so that they are desensitised to the noise in the data. There is often a trade-off between the quality and explainability of the model \cite{hernan2010causal}. In order to gain \textit{descriptive} value from a predictive model, we need techniques for interpretation.

There are a number of approaches to model interpretation \cite{molnar2019}. Two of the most relevant approaches for observational n-of-1 QS projects are (1) intrinsically-interpretable models and (2) model-agnostic techniques. 

\subsection{Interpreting model parameters}
Some learning algorithms --- like linear regression and its variants --- are \textit{intrinsically} interpretable, as their parameters explain the effect of each feature independently of the other features \cite{molnar2019}.  Consider the following formula, which describes a standard linear model:
\begin{equation}\label{eq:linear_model}
    \hat{y}_i = \beta_{\text{bias}} + \sum_{j=1}^n \beta_j x_{i,j} = \beta_{\text{bias}} + \beta_1 x_{i,1} + \beta_2 x_{i,2} + \cdots + \beta_n x_{i,n} 
\end{equation}where $\hat{y}_i$ is the predicted value for observation $i$, $(\beta_1, \ldots, \beta_n, \beta_{bias})$ are the internal model parameters, and $(x_{i, 1}, \ldots, x_{i, n})$ are the feature values for observation $i$. By analysing these internal model parameters (the $\beta$-parameters), we are able to interpret the magnitude and direction of the relationships between each feature and the target feature \cite{molnar2019}. 

\subsubsection{Regularised linear models}
One major challenge in n-of-1 QS studies is that datasets are wide --- there are more features than observations. In a linear model, each feature has a corresponding $\beta$-parameter. It is well known that having a large number $\beta$-parameters allows us to fit arbitrary functions. With our wide dataset, that means we are modelling the noise in the dataset along with the signal, which overfits our model to the data and makes any results spurious \cite{burkov2019hundred, hawkins2004overfitting}. One of the best techniques to resolve this is \textit{regularisation}, which penalises the model for having large $\beta$-parameters \cite{burkov2019hundred}. In Ridge regression, a weighted L2 norm of the $\beta$-parameters is added to the loss function. In Lasso regression, a weighted L1 norm is used instead. The mathematical properties of the L1 and L2 norm allow us to shape the effects of the regularisation in desirable ways. In particular, the L1 norm in Lasso regression favours \textit{sparse} $\beta$-values --- with values of 0 for features that have only a small effect on the prediction. This serves as built-in feature selection \cite{tibshirani1996lasso, burkov2019hundred}. Only features that are ``worth their weight'' will have non-zero $\beta$-parameters. Both Ridge and Lasso were included as candidate algorithms in this study. 

\subsubsection{Recursive feature elimination (RFE)}\label{sssec:rfe}
Lasso is a rare example of an intrinsically-interpretable model with built-in feature selection. For other linear models (and tree-based models), best performance can be achieved when pre-selecting the most important features, then fitting a (regularised) model on that subset. RFE \cite{chen2007enhanced} recursively trains some model using various subsets of the features \cite{guyon2002rfe}. In the past, feature selection techniques did not consider internal parameters, and instead tried all combinations of features to establish a ranking of their importance \cite{Hoogendoorn2018}. This has immense computational cost, as for $n$ features a total of $2^n -1$ models must be trained --- one for each subset of features. RFE makes this process more efficient by utilising internal parameters to help rank feature importance. For best results, the rankings can be averaged over $5$-fold cross-validation. These top features can then be fed to the model. The resulting parameters can be analysed as before. Additionally, the ranking of the features from cross-validated RFE is a good secondary check on which features are most explanatory. RFE was used for both purposes in this study. 

\subsection{Model-agnostic interpretation}
With the proliferation of neural networks, it is increasingly common for the best-fitting model to be a ``black box''. That is, its internal workings are inseparable, making them impossible to interpret in isolation. \textit{Model-agnostic} methods do not rely on internal model parameters. Instead, they rely on access to the dataset and the ability to test the model's predictions on various input-output pairs. By systematically varying the inputs, model-agnostic methods can interpret model behaviour and allow us to understand the underlying relationships. This is achieved by building a linear \textit{explanation model} $g(\boldsymbol{x}_i)$ which is an interpretable approximation of the black-box predictive model $f(\boldsymbol{x}_i)$ for any observation $\boldsymbol{x}_i$ in our dataset.

To construct the explanation model, we turn to the SHapley Additive exPlanations (SHAP) method \cite{SHAPpaper2017}. Because it is grounded in coalitional Game Theory, it boasts many desirable properties. It is also far more intuitive to interpret than other contemporary techniques, like local interpretable model-agnostic explanations (LIME) \cite{molnar2019}.

\subsubsection{Shapley values}
SHAP applies the concept of Shapley values with some algorithmic enhancements. Shapley values were originally demonstrated in Game Theory as a technique for fairly distributing payouts amongst players \cite{shapley1953value}. They can be used for model interpretability by reframing the problem: Let each feature be a player and let the total payout for each observation be proportional to the accuracy of the prediction. By evaluating all possible coalitions of features and repeating this for a number of examples, we can calculate Shapley values for each value that each feature takes on \cite{molnar2019}. These values represent the average contribution of a feature-value to the prediction. 

For a model with $n$ features, let the set of feature vectors be $N = \{\boldsymbol{x}_1, \boldsymbol{x}_2, \ldots, \boldsymbol{x}_n\}$, such that $\boldsymbol{x}_j$ is the column vector of values for feature $j$. The Shapley values of feature $j$ are calculated as:
\begin{equation}
    \phi_{j}(v)= \frac{1}{n!} \cdot \sum_{S \subseteq N\backslash\left\{\boldsymbol{x}_j\right\}} |S|! \cdot (n-|S|-1)! \cdot \left[v\left(S \cup\left\{\boldsymbol{x}_j\right\}\right)-v(S)\right]
\end{equation}
where $S \subseteq N\backslash\left\{\boldsymbol{x}_j\right\}$ denotes a subset $S$ of all features $N$ that do not include feature $j$, and $v: S \to \mathbb{R}$ is a payoff function that evaluates how predictive some subset $S$ of the features is, given dataset $\boldsymbol{X}$ and a trained model.

Intuitively, the Shapley values of a feature $j$ are how much each of its observed values shift the model prediction away from the mean prediction. 

Shapley values have been proven to satisfy a number of axioms in coalitional Game Theory \cite{shapley1953value}. This makes the technique preferable to earlier approaches like LIME, which fail to ensure that the predictive value is fairly distributed among the features. Moreover, Shapley values allows for contrastive explanations, which many other techniques (including LIME) do not \cite{molnar2019}. These are immensely useful for comparing different subsets of the data in terms of the relationships between features --- e.g. how does a day with terrible sleep compare to one with outstandingly good sleep?

Whilst extremely computationally expensive\footnote{In reality, most implementations of Shapley values for model interpretation rely on Monte-Carlo sampling to estimate the values efficiently \cite{molnar2019}.}, this approach allows much more nuanced interpretations of the relationships between features, as each feature is explained with a \textit{distribution} instead of a mere point estimate. So where interpreting $\beta$-parameters gave us an indication of each feature's \textit{global} importance, Shapley values give us their \textit{situational} importance --- based on their effects during each observation.

\subsubsection{SHAP}
SHapley Additive exPlanations (SHAP) \cite{SHAPpaper2017} inherits the desirable mathematical properties of the Shapley value, but builds on them in two key ways. Firstly, SHAP offers methods for global interpretations based on aggregations of Shapley values. Secondly, SHAP can utilise approximation methods that drastically speed up Shapley value computation on certain model architectures \cite{molnar2019}. 

SHAP builds a linear model $g$ that approximates the predictive model $f$ for any observation $\boldsymbol{x}_i$ in our dataset:
\begin{equation}
    f(\boldsymbol{x}_i) \approx g\left(\boldsymbol{x}_i\right)=\phi_{\text{bias}}+\sum_{j=1}^{n} \phi_{j} x_{i,j}
\end{equation}
where $\phi_j$ is the Shapley value of feature $j$ for observation $x_{i,j}$.

Essentially, SHAP has generated a linear model of the feature contributions for some black box model we gave it, for some specific instance. Now, we can interpret the $\phi$-parameters just as we interpret the $\beta$-parameters of any linear model (Eq. \ref{eq:linear_model}). Except that $\beta$-parameters are \textit{global} and thus apply over all observations in the data, whilst the $\phi$-parameters from SHAP apply to a single observation $\boldsymbol{x}_i \in \boldsymbol{X}$. 

\section{Experiments}\label{sec:experiments}
Four experiments were performed to evaluate the techniques presented in this study: (1) assessing the effectiveness of Markov unfolding, (2) comparing imputation techniques, (3) measuring overall predictive performance, (4) final model interpretation. The first three experiments made use of a grid-search over 3 parameters for 10 repeats. The fourth experiment leveraged the results of the prior experiments to train and interpret the most promising model.

\textbf{Preparing dataset variants}: Each of the 10 imputation variants of the dataset was loaded into the grid-search script, duplicated, then transformed with $7$-day Markov unfolding. This resulted in 20 dataset variants of varying sizes.  

\textbf{Learning algorithms}: A total of 7 different learning algorithms were included\footnote{Earlier experiments found that support vector algorithms had mediocre performance --- especially without hyperparameter optimisations --- and fully-connected neural networks were unable to converge due to the limited data, resulting in massive prediction error, as well as excessive training time. These were excluded for the final analysis.}. The first 6 were well-established linear and tree-based algorithms. Lasso and Ridge regression were included as regularised linear models. A basic Decision Tree regressor was included for reference, along with two ensembles of Decision Trees --- Random Forest (bagging) and XGBoost (boosting). The 7th algorithm was a ``Naïve regressor'' that always predicts the median value of the target feature from the training set. This served as a baseline for predictive value.

\newpage
\textbf{Feature selection}: The final datasets often had far more features (columns) than observations (rows), making them incredibly wide and difficult for algorithms to learn effectively \cite{fortmann2012understanding}. Initially, principal component analysis (PCA) was tested as a means of reducing the feature-space, but the number of components in PCA is limited by the number of observations, making this non-viable for these wide datasets. Recursive feature elimination (RFE) \cite{chen2007enhanced} with a Decision Tree base model was used to rank the features independently for each of the 20 dataset variants, based on a single random train-test split of the data. Each non-target feature in the datasets was scaled independently prior to feature selection to speed up convergence.

\textbf{Grid search}: The grid search performed 10 cross-validation repeats over each of the 20 dataset variants. In each repeat, the dataset was split into train- and test- sets with a custom splitting function: each call drew a sample of $40$ ``pure'' observations (\S \ref{ssec:baseline_dataset_withnans}), without repeats, for the test set and kept the remaining observations as the train set. This meant that models were only scored on observations that had complete data and had not undergone imputation --- which was essential for fair comparison. The train and test sets were scaled independently to avoid data leakage between them.

The grid search iterated over all algorithms and values of $n \in [2, 200]$, training each on the $n$-highest-ranked features in the train set and scoring the resulting model on the test set. Two of the dataset variants --- \texttt{with\_nans} and \texttt{pre\_markov\_with\_nans}--- still had missing values. In the case of XGBoost, which can tolerate missing values \cite{chen2016xgboost}, the scaled training set was given directly to the algorithm. For all other algorithms, the observations with missing values were dropped from the training set before scaling and training. 

\textbf{RMSE metric}: The prediction score was measured using the root mean squared error (RMSE) metric:

\begin{equation}
    \text{RMSE}(\boldsymbol{y}, \boldsymbol{\hat{y}}) = \sqrt{\frac{1}{m} \sum_{i=1}^{m}(y_i - \hat{y}_i)^2}
\end{equation}
where $\boldsymbol{y}$ were the true (target) values, $\boldsymbol{\hat{y}}$ were the predicted values, and $m=40$ was the number of observations in the test set. RMSE was selected because it is both (1) sensitive to outliers and (2) easily interpreted in the same units --- sleep score $\in [0, 100]$ --- as the original target feature.

\newpage
\section{Results and discussion}\label{sec:results_and_discussion}

For each algorithm, dataset variant, and number of features, the RMSE over all 10 repeats was summarised by the mean and standard deviation. 

\subsection{Effectiveness of Markov unfolding}
\begin{figure}[H]
    \centering
    \includegraphics[width=0.8\textwidth]{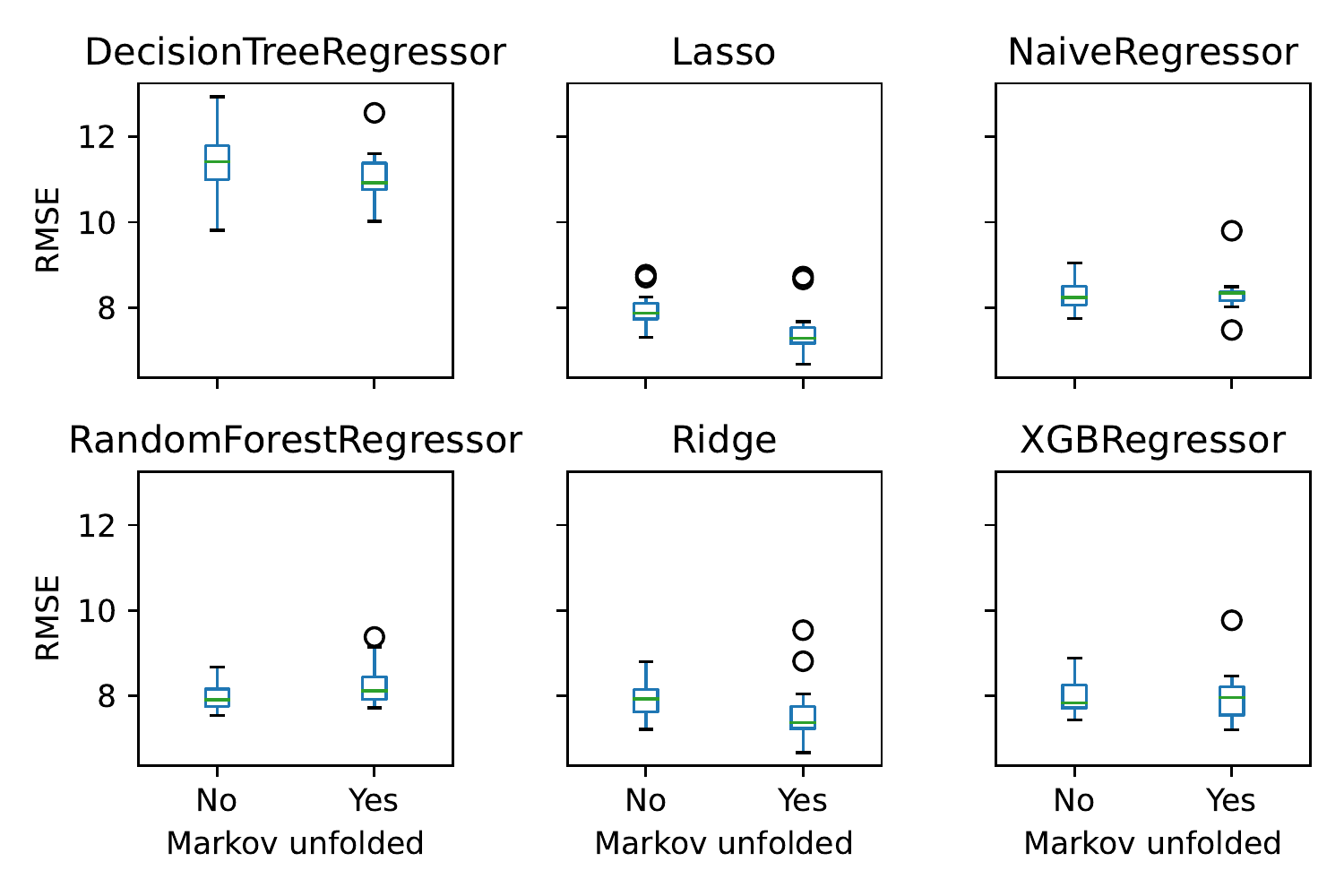}
    \caption{Comparison of prediction error (RMSE) across algorithms between the original dataset and the wider Markov-unfolded dataset. Results are distributions of mean RMSE scores for best-performing hyperparameter combinations for each dataset variant. Boxplots span the interquartile range (IQR) of the distribution, with a line indicating the median. The whiskers extend to the minimum and maximum values observed. Values beyond $1.5$ times the IQR are considered outliers and plotted as circles.}
    \label{fig:result_markov_unfolding_boxes}
\end{figure}

The best-performing configurations\footnote{The best configurations for each algorithm and dataset were selected by finding the number of features that produced the lowest mean RMSE for that algorithm and dataset.} for each model were collected and grouped by whether or not Markov unfolding was applied. Boxplots of the RMSE distributions are shown in Fig. \ref{fig:result_markov_unfolding_boxes}.

Markov unfolding did allow most algorithms to improve predictive performance, but the improvement was generally small. This was likely because the benefit of the additional information contained in the unfolded features traded off against the added burden of much higher dimensionality \cite{bishop2006pattern}.

\newpage
By inspection, it is clear that the differences between algorithms were greater than the differences between regular and unfolded datasets in terms of the resulting model error. Lasso yielded the lowest error and saw the greatest benefit from Markov unfolding ($-0.61$), but Ridge and XGBoost also saw small improvements. On average, the magnitude of the differences between the regular and Markov-unfolded variants of the datasets was small. This likely indicates limitations in the use of RFE to select features when the datasets are so wide\footnote{Manual inspection of the RFE feature rankings for each dataset revealed a great deal of variation in the top features.}.

\begin{figure}[H]
    \centering
    \includegraphics[width=\textwidth]{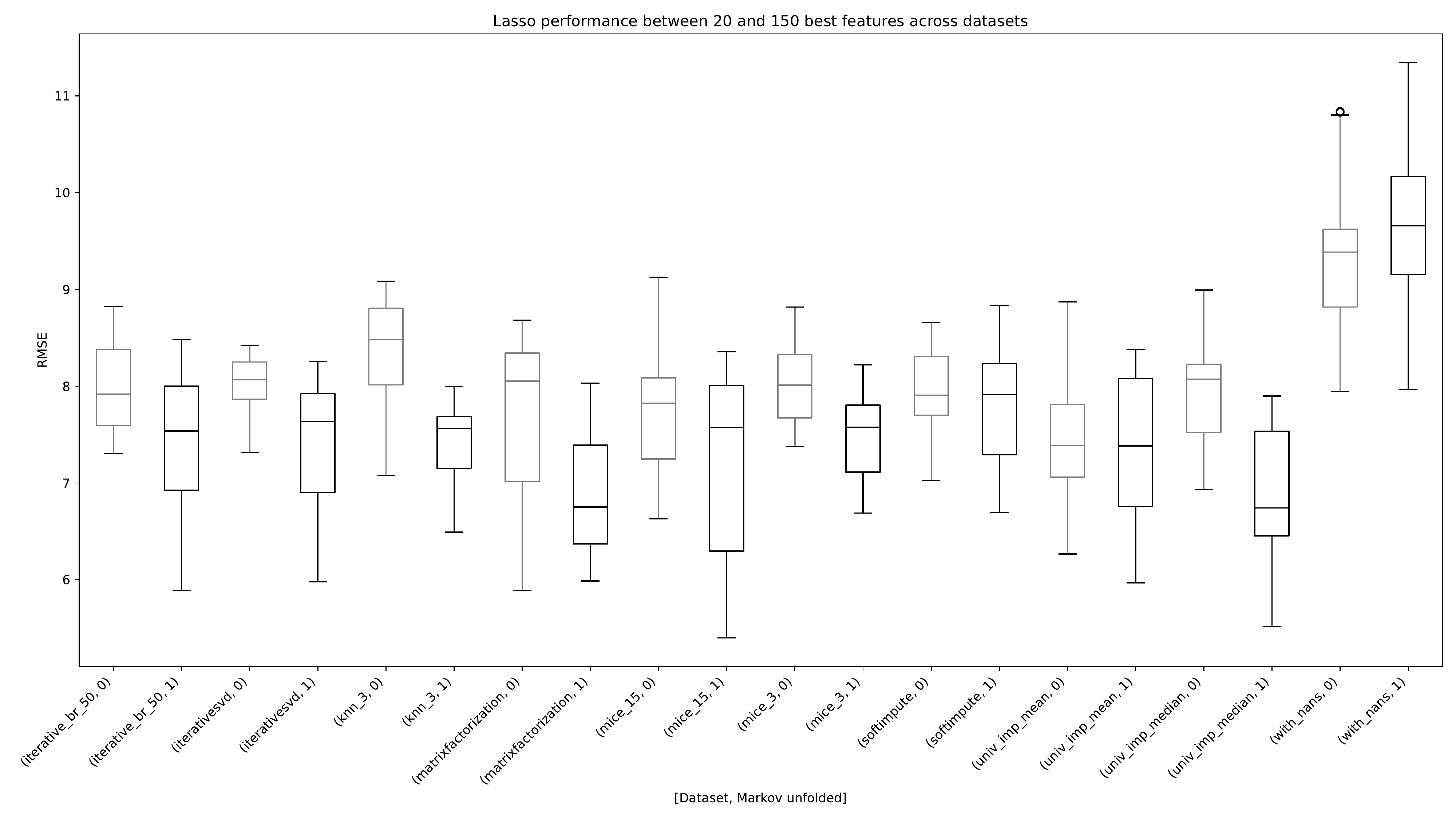}
    \caption{Distributions of prediction error over all repeats for Lasso model for between 50 and 150 features, across all dataset variants. The narrow version of each dataset is shown in grey. The Markov-unfolded version of each dataset is shown in black. Boxplots span the interquartile range (IQR) of the distribution, with a line indicating the median. The whiskers extend to the minimum and maximum values observed.}
    \label{fig:result_markov_lasso}
\end{figure}

Lasso was much more effective at leveraging the Markov unfolding than other algorithms, likely due to the sparse L1 regularisation serving as an additional layer of feature selection \cite{tibshirani1996lasso}. The effectiveness of Markov unfolding as a technique is best illustrated by comparing the performance of Lasso across all dataset variants paired with their Markov-unfolded versions. This is shown in Fig. \ref{fig:result_markov_lasso}.

Markov unfolding lowered the prediction error for Lasso across all dataset variants except the variant that underwent no imputation (and thus contained only 21 training observations). These results highlight the immense value of capturing historical data in each observation, provided the algorithm can overcome the increased dimensionality.

\newpage
\subsection{Comparison of imputation techniques}

\begin{figure}[H]
    \centering
    \includegraphics[width=\textwidth]{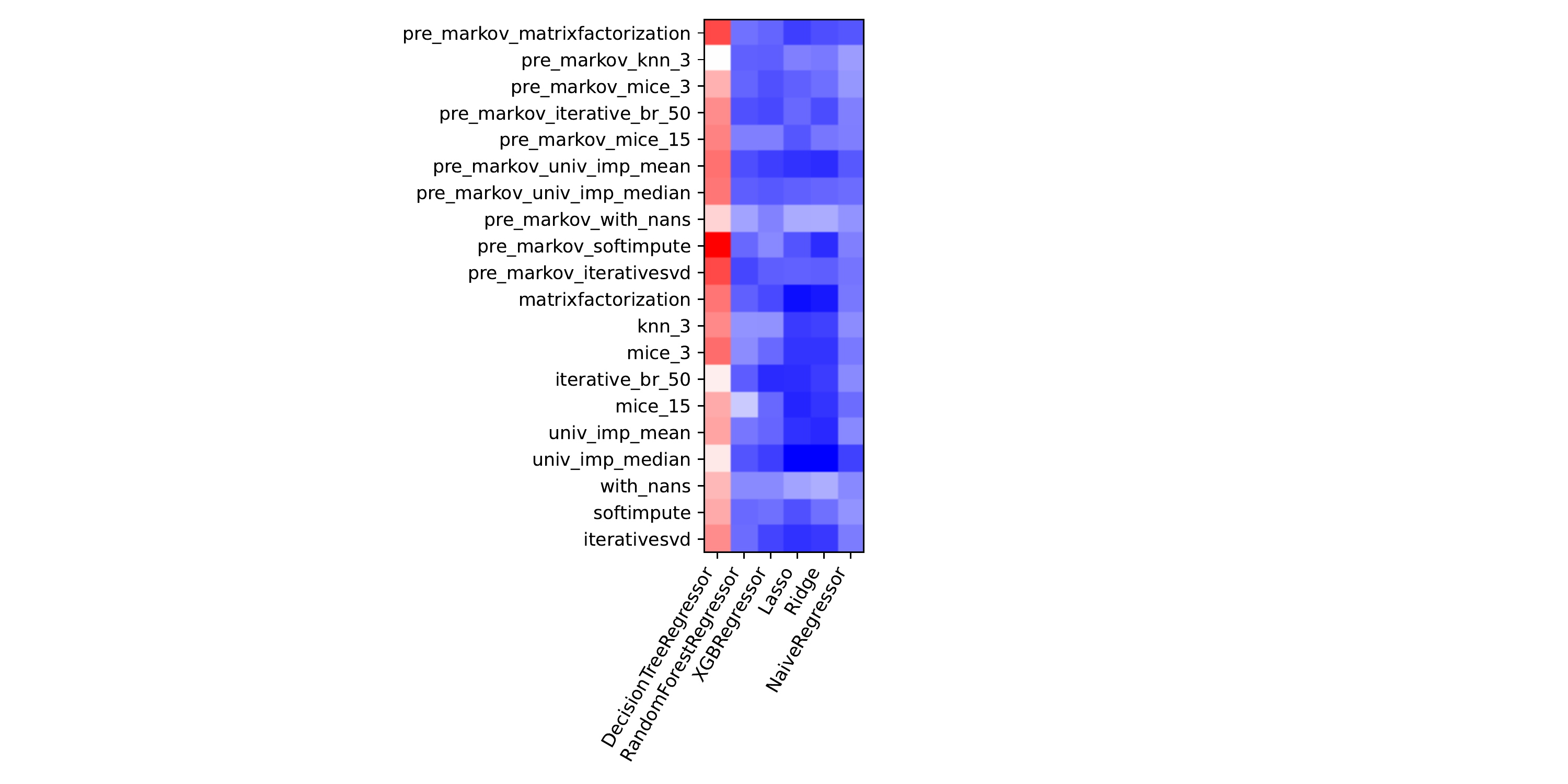}
    \caption{Heatmap of mean prediction error (RMSE) over datasets (rows) and algorithms (columns). Colour map: lower values are \textit{colder} (dark blue), medium values are near white, and higher values are \textit{hotter} (dark red).}
    \label{fig:result_rmse_heatmap}
\end{figure}

Regularised linear algorithms performed the best with imputed data, particularly for the matrix factorisation, KNN, MICE, iterative SVD, and univariate imputation strategies. Fig. \ref{fig:result_rmse_heatmap}, compares the best configurations for each algorithm-dataset pair in a heatmap of mean RMSE values over all 10 cross-validation iterations. This presentation highlights the poor results of the Decision Tree across all datasets, as well as the notable results for Lasso and Ridge when using matrix factorisation, MICE, and univariate median imputation. It is also clear that using no imputation (\texttt{with\_nans}) resulted in worse results for almost all the algorithms in both the Markov unfolded and pre-Markov treatments. 

There were weak correlations between imputation distance (\S \ref{ssec:imputation_distance}) and prediction error for the tree-based algorithms: Decision Tree ($r=0.415$), Random Forest ($r=0.367$), XGBoost ($r=0.499$). These are well above the (clearly spurious) correlation of the Naïve regressor ($r=-0.203$). The regularised linear models, however, both had negligible correlations: Lasso ($r=-0.276$), and Ridge ($r=0.107$). These correlations were for mean RMSE under the optimal number of features for each dataset.

\begin{figure}[H]
     \centering
     \begin{subfigure}[b]{0.48\textwidth}
         \centering
         \includegraphics[width=\textwidth]{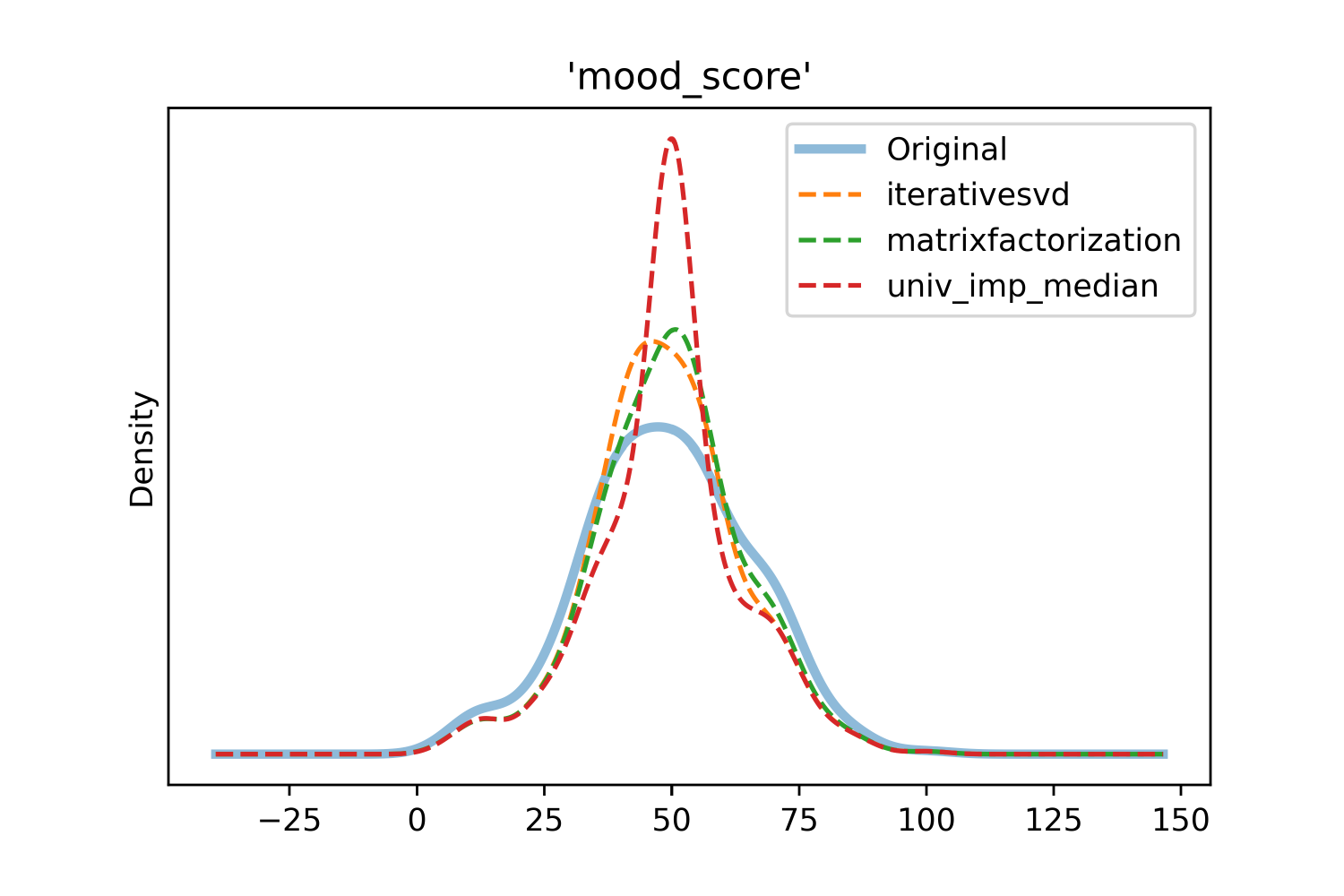}
         \caption{Gaussian shape.}
         \label{fig:comp_imp_mood_score}
     \end{subfigure}
     \hfill
     \begin{subfigure}[b]{0.48\textwidth}
         \centering
         \includegraphics[width=\textwidth]{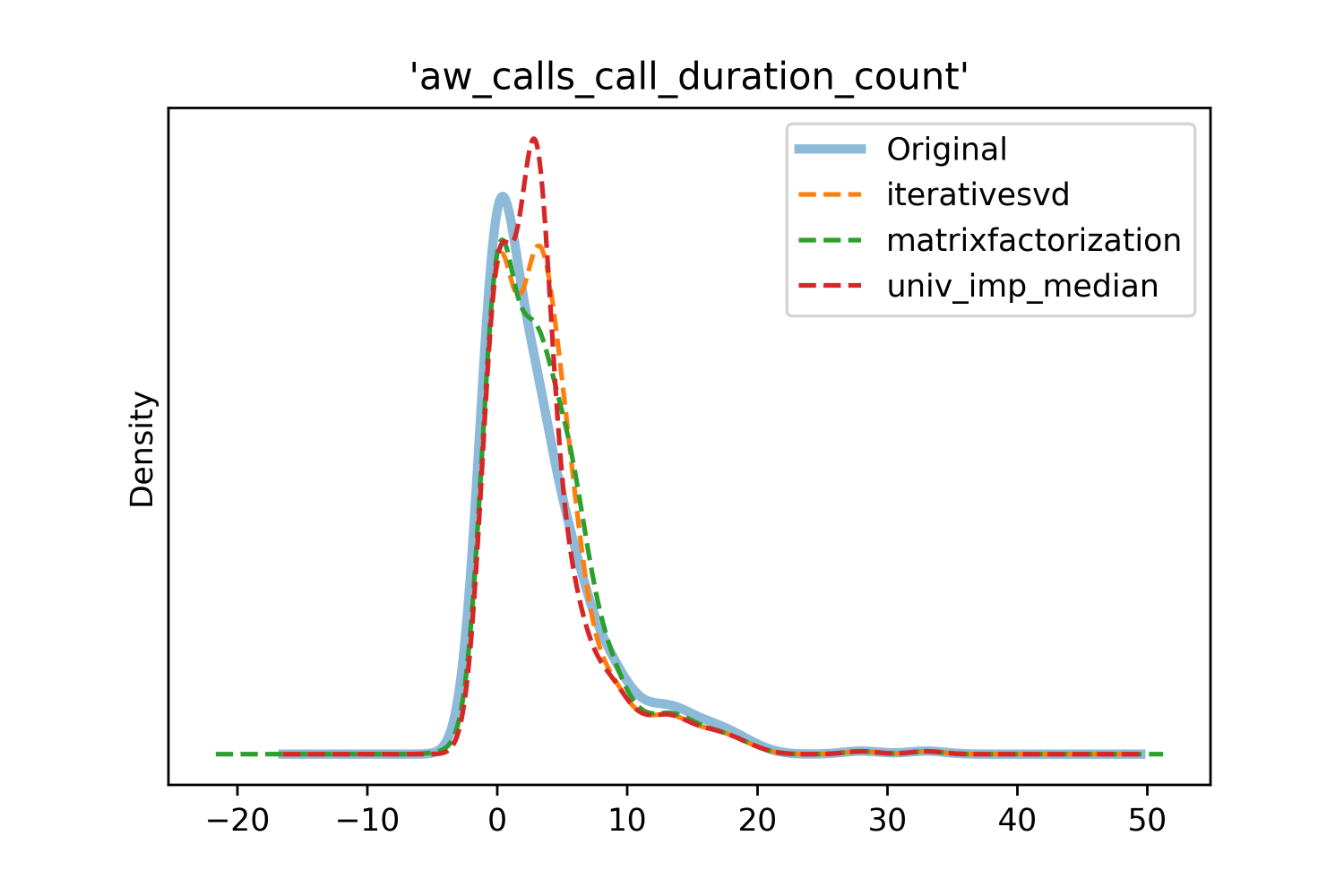}
         \caption{Log-normal shape.}
         \label{fig:comp_imp_aw_calls_call_duration_count}
     \end{subfigure}
     \hfill
    \begin{subfigure}[b]{0.48\textwidth}
         \centering
         \includegraphics[width=\textwidth]{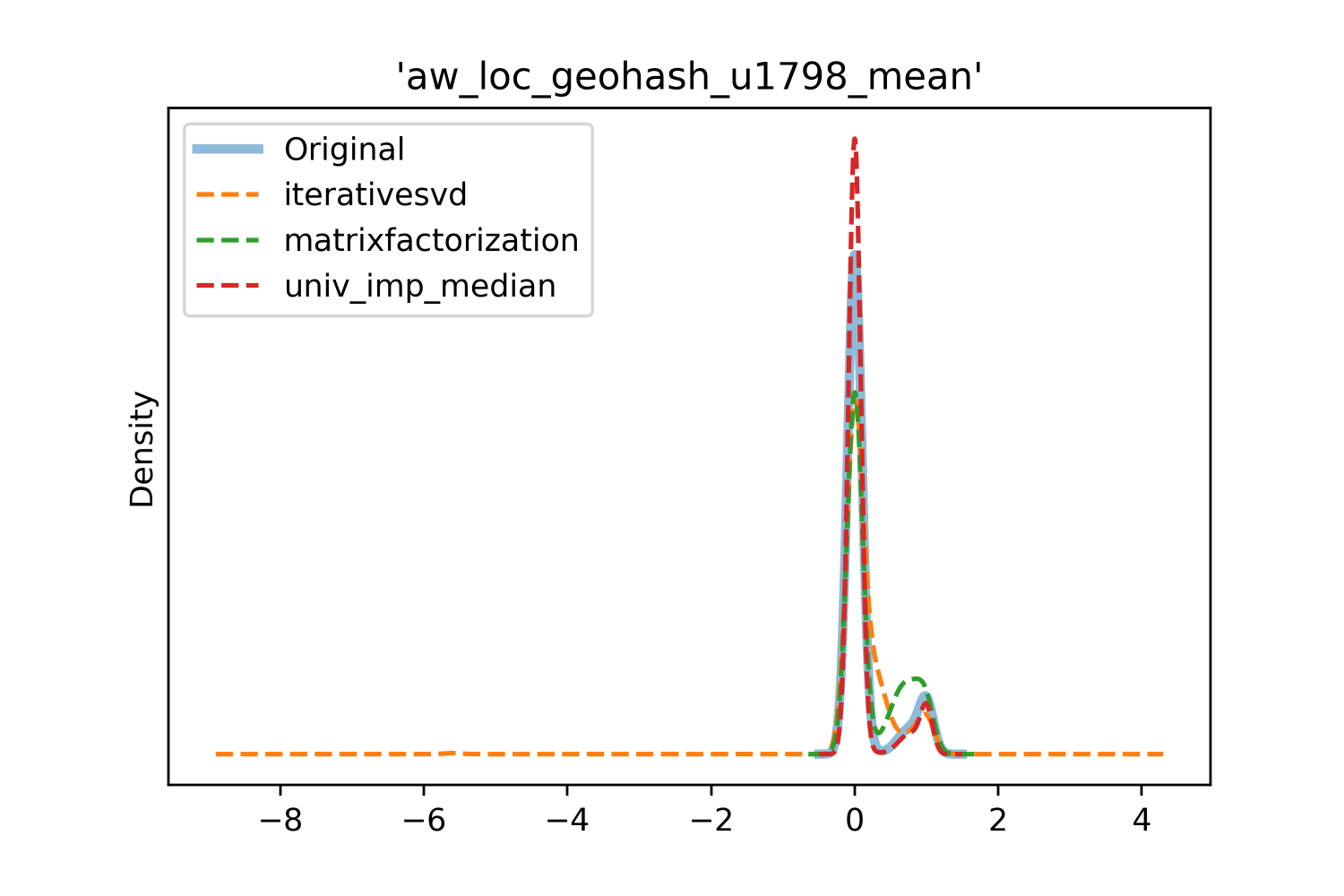}
         \caption{Bimodal shape.}
         \label{fig:comp_imp_aw_loc_geohash_u1798_mean}
     \end{subfigure}
     \hfill
     \begin{subfigure}[b]{0.48\textwidth}
         \centering
         \includegraphics[width=\textwidth]{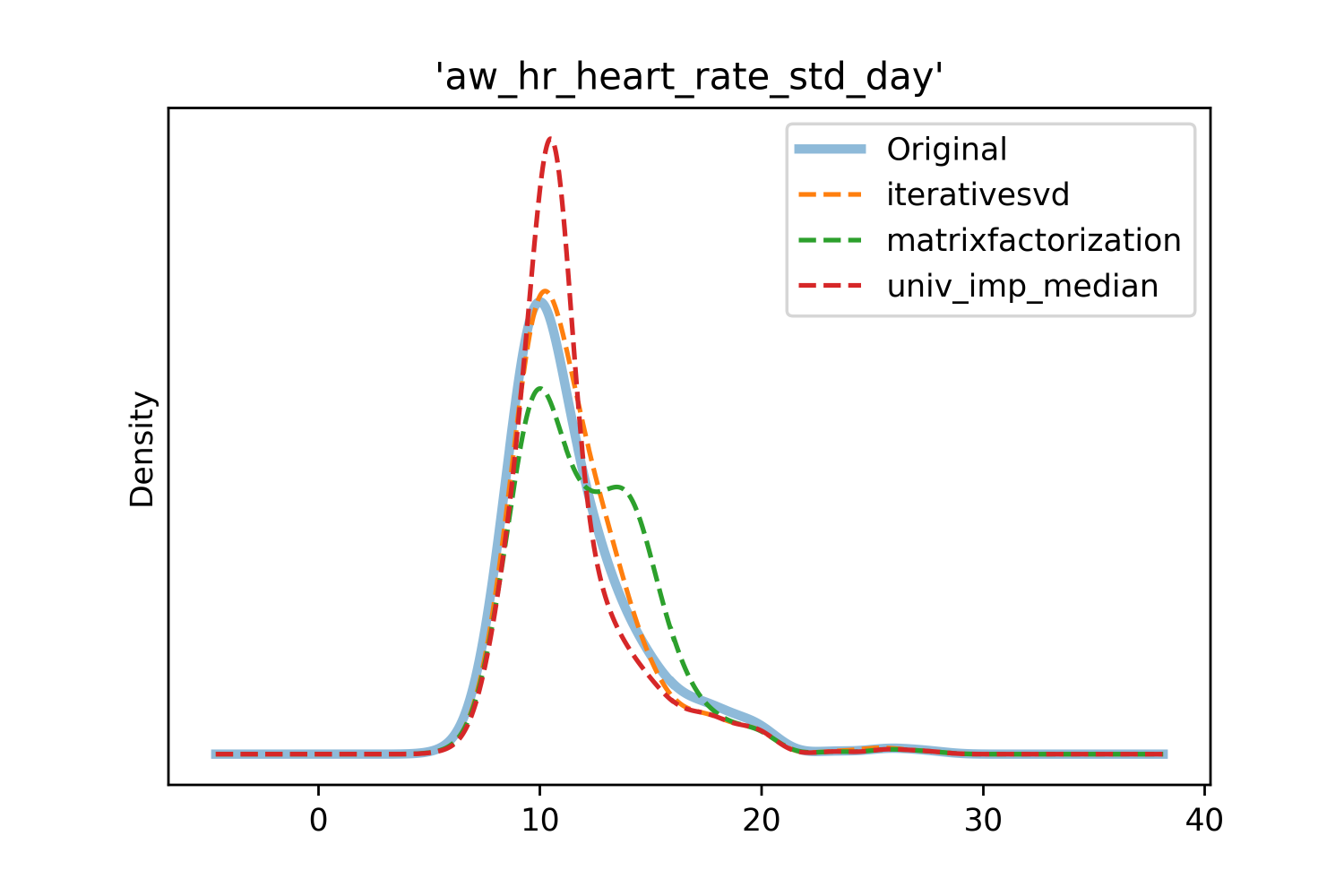}
         \caption{Hybrid shape.}
         \label{fig:comp_imp_aw_hr_heart_rate_std_day}
     \end{subfigure}
     \hfill
    \caption{Comparison of feature distributions for three notable imputation techniques on different types of underlying distributions. The original data distribution is shown with a solid line. The distribution after imputing missing values is shown with dashed lines, coloured differently for each imputation technique.}
    \label{fig:comp_imputations}
\end{figure}

Most of the imputed data was in the contiguous chunk of missing data from the AWARE source at the start of the project. This is visible in the top right of Fig. \ref{fig:pre_markov_with_nans.csv}. To better understand the different classes of imputation and why they produced such different imputation distance scores, it is helpful to look at some examples (Fig. \ref{fig:comp_imputations}).

Fig. \ref{fig:comp_imp_mood_score}: Median imputation copes best with a Gaussian-like distribution, as it minimises the average noise. However, the multivariate techniques appear to capture much more variance in the data, which gives the final model more information (if it is relevant) to capture underlying relationships. 

Fig. \ref{fig:comp_imp_aw_calls_call_duration_count}: A log-normal shape causes univariate imputation to overestimate most imputed values. Median imputation is much less sensitive to the long tail than mean imputation, which explains the lower imputation distance (and better predictive performance). Again, the multivariate imputation techniques appear to capture the underlying properties better. 

Fig. \ref{fig:comp_imp_aw_loc_geohash_u1798_mean}: The dramatic changes (in location and behaviour) caused by the global pandemic resulted in many features having bimodal distributions. These resulted in a great deal of error for the median imputation, as it fills values equivalent to the larger mode. In the case of location, this was precisely the wrong mode. Matrix factorisation did a superb job of inferring which values to fill based on other features, thus capturing the underlying distribution quite well. Iterative SVD tracked a similar curve, but also exhibited a failure mode of extremely long (and thin) tails. Note in this case that the feature was only defined on the continuous interval $[0, 1]$, meaning that the iterative SVD went way beyond plausible imputations for some values. Whilst infrequent, these likely resulted in its much larger imputation distance (and lower predictive performance). 

Fig. \ref{fig:comp_imp_aw_hr_heart_rate_std_day}: Some distributions were hybrids. The standard deviation of daytime heart rate, for instance, should typically follow a weak power law. But the pandemic changes would have resulted in two different log-normal shapes combining. In this case, the univariate median imputation was overly biased to the peak of the curve, whilst the multivariate imputation techniques appeared to capture more underlying structure. In this specific example, the difference between the iterative SVD and matrix factorisation methods is notable. 

\subsection{Predictive performance}
In general, Lasso had the best predictive results over the 10 cross-validation iterations, except for the outright-best configuration\footnote{Ridge regression on 5 features from the univariate median imputation dataset with Markov unfolding had RMSE of $6.67 \pm 0.59$.}. Most top results were Lasso regression on between 4 and 120 features with univariate median imputation and Markov unfolding (RMSE $6.68 \pm 0.62 - 6.82 \pm 0.68$). However, other top contenders were Lasso with the matrix factorisation imputation and Markov unfolding with 50 to 110 features (RMSE $6.84 \pm 0.59 - 6.85 \pm 0.59$). These had slightly higher error, but also slightly lower variance.

\begin{figure}[H]
     \centering
     \begin{subfigure}[b]{0.49\textwidth}
         \centering
         \includegraphics[width=\textwidth]{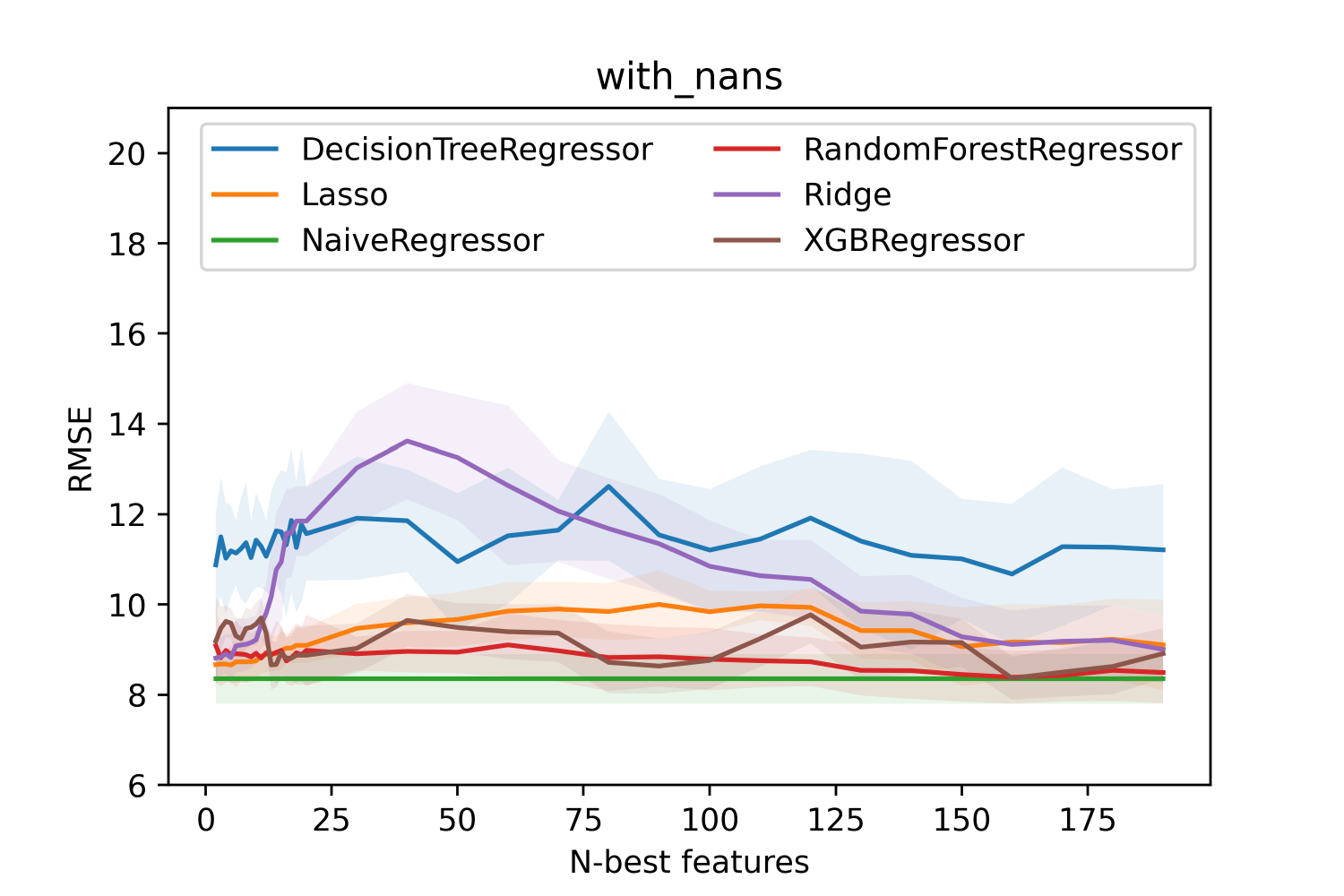}
         \caption{Dataset with missing values.}
         \label{fig:result_algs_with_nans}
     \end{subfigure}
     \hfill
     \begin{subfigure}[b]{0.49\textwidth}
         \centering
         \includegraphics[width=\textwidth]{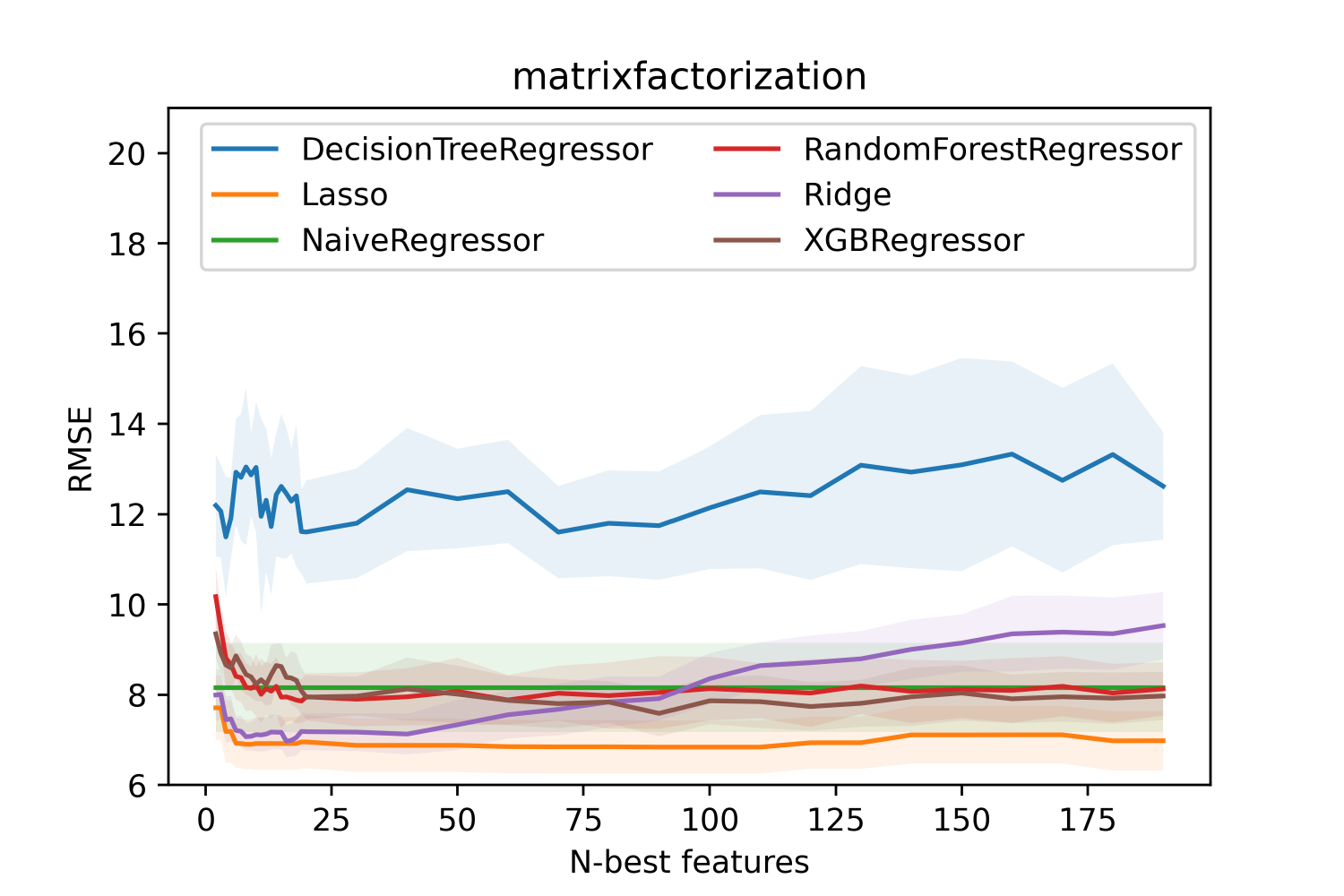}
         \caption{Dataset with matrix factorisation imputation.}
         \label{fig:result_algs_univ_imp_median}
     \end{subfigure}
     \hfill
        \caption{Comparison of model prediction error (RMSE) with respect to the number of features for a dataset with missing values (left) and a dataset where missing values were imputing using matrix factorisation (right). Coloured lines represent mean error over 10 independent resamplings. Shaded regions around each line indicate $\pm 1 \sigma$ of variance.}
        \label{fig:results_algs}
\end{figure}

Fig. \ref{fig:results_algs} compares the performance of all the algorithms with respect to the number of features for two different datasets. All algorithms other than the Decision Tree had much lower error when using an imputed dataset. Recall that only non-imputed observations were used to score the algorithms, allowing a direct comparison. Most imputation techniques resulted in more stable results and overall lower error. When there were missing values in the dataset, the median of the target (i.e. Naïve regressor) was a better predictor than any of the learning algorithms. When imputation was applied, however, the algorithms could better capture the relationships in the data and outperformed the naïve median estimate by a notable magnitude. On the whole, imputation provided the algorithms with vastly more training examples (and variance), resulting in a dramatic improvement in model quality.

\newpage
The number of features (which directly relates to the free parameters of the models) had a fairly mild influence on most algorithms. With the exception of Ridge, most algorithms actually reduced their out-of-sample error as the features were increased, instead of overfitting. Random Forest, Lasso, and XGBoost showed almost no trends for these error curves on the imputed dataset (Fig. \ref{fig:result_algs_univ_imp_median}). This was likely due to regularising effects built into these three algorithms. Lasso uses L1 regularisation on $\beta$-parameters \cite{tibshirani1996lasso}, Random Forest is a bagged ensemble --- which is known to reduce variance \cite{buhlmann2002analyzing} --- and XGBoost has mild L2 regularisation by default \cite{chen2016xgboost}. 

XGBoost also uses default directions in trees to tolerate missing data: when a value is missing, trees automatically follow the paths that maximise gain \cite{chen2016xgboost}. We can compare the built-in tolerance to explicit imputation for XGBoost by comparing the performance across Figures \ref{fig:result_algs_with_nans} and \ref{fig:result_algs_univ_imp_median}. Close inspection shows that XGBoost performs much better with even a simple explicit imputation than when using its own built-in tolerance for missing values. This reiterates the value of explicit imputation for n-of-1 QS projects. 

\subsection{Model interpretation}

A matrix-factorisation-imputed dataset with Markov unfolding was the dataset variant that yielded the most accurate and consistent predictions\footnote{Univatiate median imputation and matrix factorisation imputation had similar mean error on Lasso, but matrix factorisation was more consistent (lower $\sigma$). This is clear in Fig. \ref{fig:result_markov_lasso}.} with the Lasso algorithm. This combination was selected for the final model interpretation.

\begin{figure}[H]
    \centering
    \includegraphics[width=0.5\textwidth]{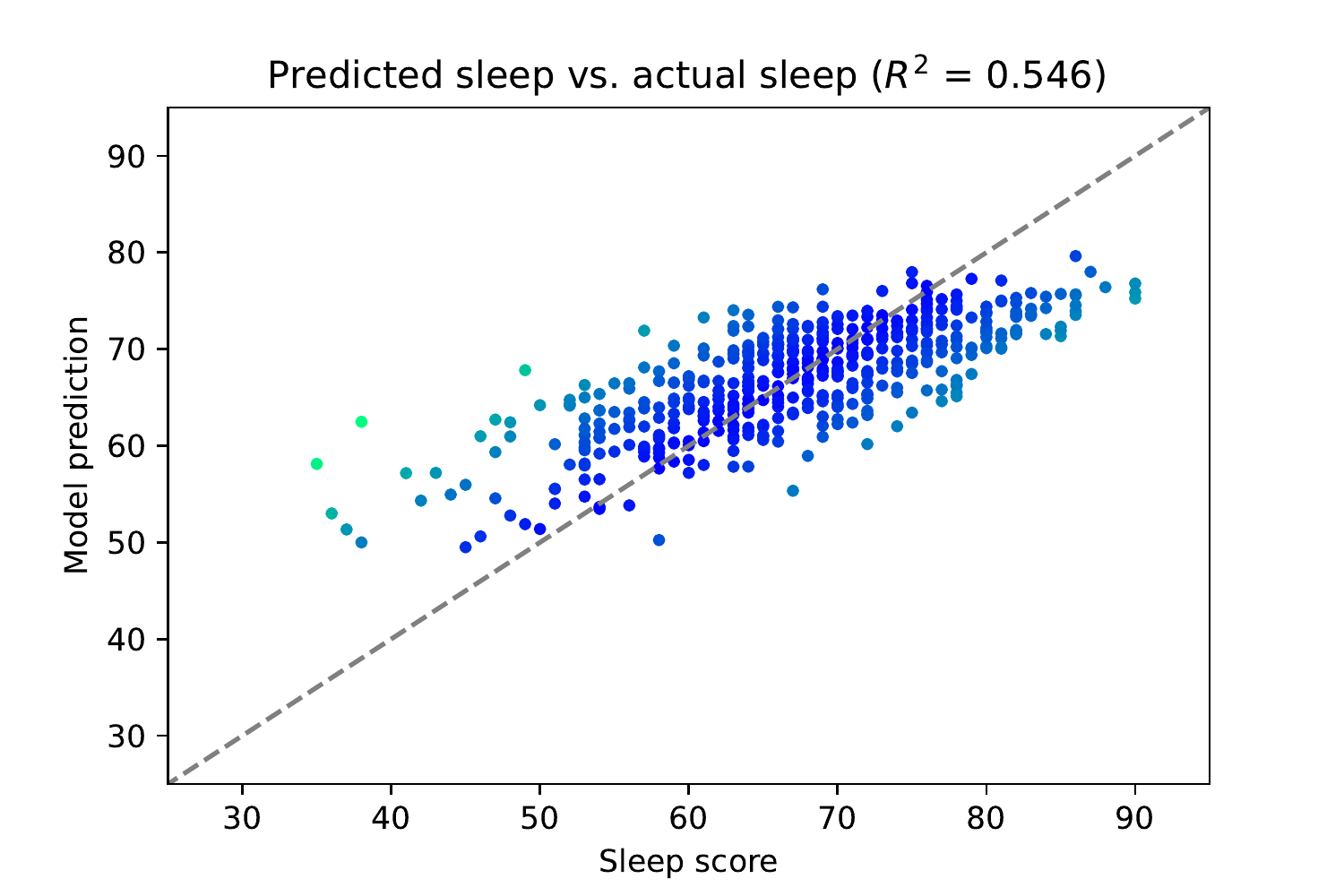}
    \caption{Calibration curve for the best algorithm-dataset pairing after 5-fold cross-validation. The vertical axis is the predicted sleep score. The horizontal axis is the true sleep score. The points are coloured by their residuals (prediction error), with lighter colours being higher residuals.}
    \label{fig:interp_calibration}
\end{figure}

5-fold cross-validation of Lasso on all of the observations in the final dataset variant yielded an $R^2$-score of $0.546$, with points of greatest error being near the left tail of the distribution. In other words, nights with very bad sleep were predicted as better than they were (see Fig. \ref{fig:interp_calibration}). A regularisation constant $\alpha=0.567$ was found optimal. 

\subsubsection{Interpreting cross-validated RFE}
RFE with 5-fold cross-validation on the Lasso base revealed that 36 features yielded the best average performance on the selected dataset variant. The 36 top features are listed in the Appendices (Table \ref{tab:rfe_top_features}). 

\subsubsection{Interpreting cross-validated Lasso}
$\beta$-parameters are sensitive to a number of factors (particularly with regularisation) \cite{Commonpitfallsininterpretationofcoe}. To obtain robust estimates, 20 rounds of 5-fold cross-validation were performed. In each round, 5 different subsets of the final dataset variant were used to fit a Lasso model ($\alpha = 0.567$). This process resulted in 100 versions of the Lasso model trained on slightly different subsets of 80\% of the final dataset variant.

\begin{figure}[H]
    \centering
    \includegraphics[width=0.9\textwidth]{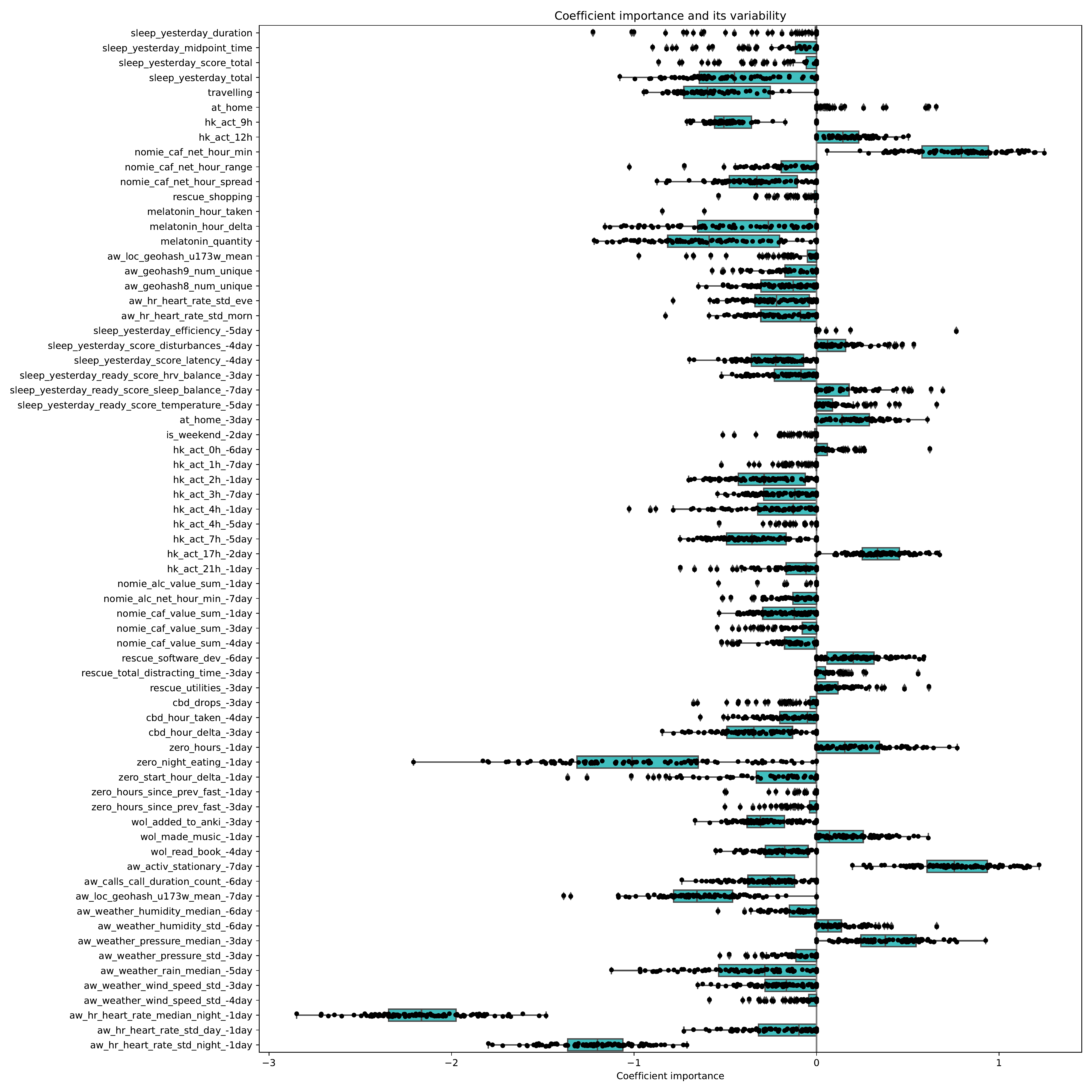}
    \caption{Overview of $\beta$-parameter variability for features after 20 rounds of 5-fold cross-validation on the best algorithm-dataset pair. Only the features which had non-negligible $\beta$ parameters (at least one $\geq 0.5$) are shown. Each point is a $\beta$-parameter from one of the 100 fittings for that feature. Boxplots are superimposed to illustrate the median, interquartile ranges, and non-outlier extremes of each distribution.}
    \label{fig:interp_coeff_importance_boxes}
\end{figure}

The 100 estimates of each $\beta$-parameter were used to construct distributions over all features and assess the robustness of the coefficient magnitudes. Because Lasso produces sparse $\beta$-parameters, those features which never yielded a coefficient with a magnitude greater than $0.5$ were ignored. The distributions for the remaining 69 features are shown in Fig. \ref{fig:interp_coeff_importance_boxes}.

Because all the features were scaled prior to fitting, we can directly compare the $\beta$-parameter distributions to get an indication of how relevant each feature is. As we see in Fig. \ref{fig:interp_coeff_importance_boxes}, many of the $\beta$-parameters varied wildly in magnitude as a result of the subset of the data they were fit on. Perhaps the factors that influence sleep varied depending on circumstances --- such as which country I was in at the time, or what time of year it was. But it is more likely that these wide ranges indicate the very high level of noise in the dataset. The relatively low $R^2$-score makes this a strong possibility.

Whilst the parameters varied greatly in magnitude, the direction was very consistent. Most of the features had $\beta$-parameters anchored at zero that varied into the positive or negative direction exclusively. This means that although those features were not always deemed relevant by the model, when they \textit{were} deemed relevant, it was always in the same direction. A few of the features had $\beta$-parameters with distributions centred well away from zero. These are strong candidates for the most-important features, as they were deemed strongly-relevant by the Lasso model regardless of the subset of the data they were fit on. 

\subsubsection{Interpreting model with SHAP}
\begin{figure}[H]
    \centering
    \includegraphics[width=\textwidth]{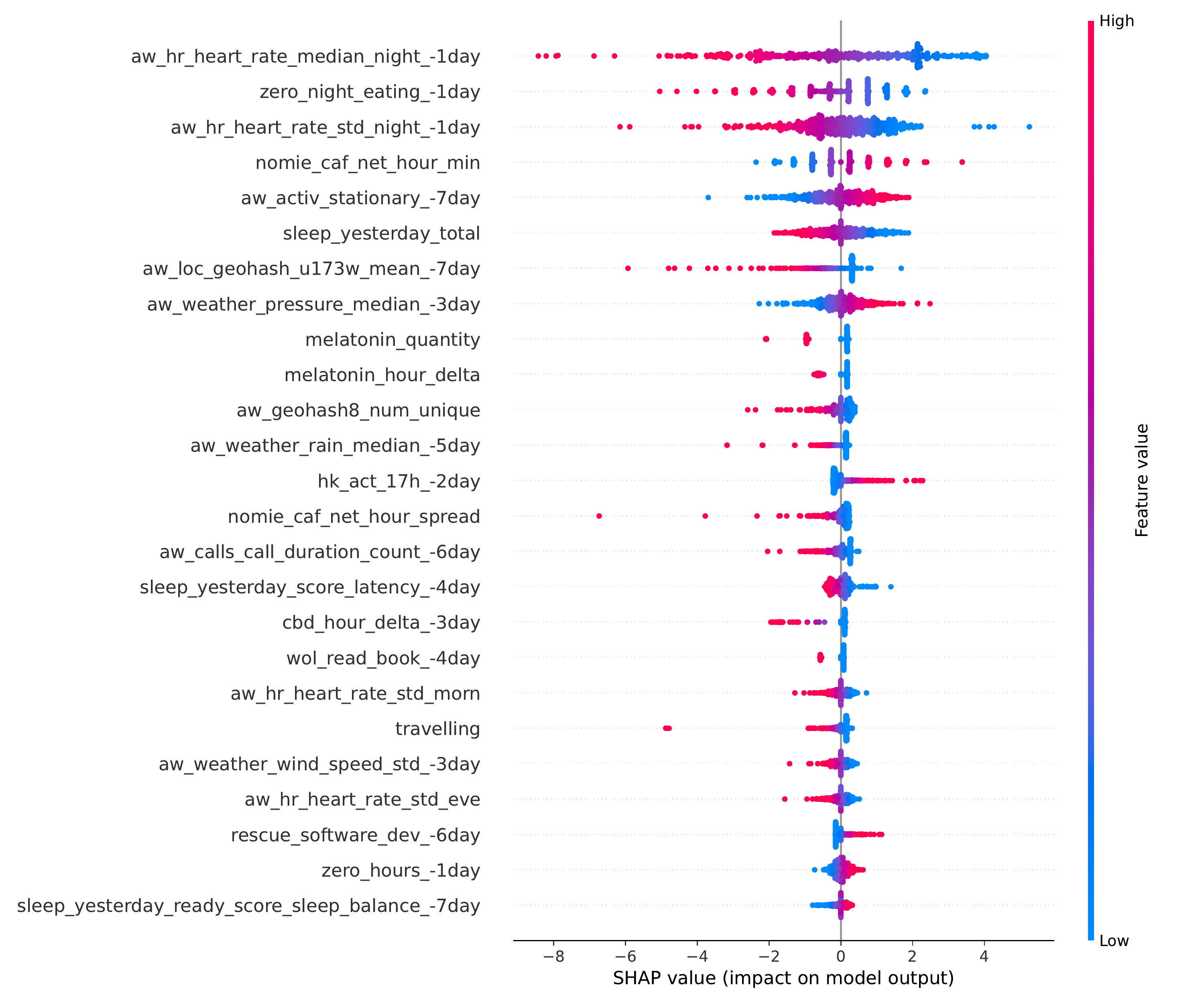}
    \caption{Beeswarm plot for interpreting the final Lasso model using SHAP. Each row is an explanatory feature, in descending order of importance. Each dot represents an observation for that feature. The colour of the dot indicates the observation's value for that feature, with low values being cooler and high values being warmer. The horizontal axis is the SHAP value for each observation, indicating the magnitude and direction of its impact on the final model prediction.}
    \label{fig:interp_shap_summary}
\end{figure}

SHAP analysis was applied to the optimised Lasso model trained on the final dataset variant. Using the cluster method\footnote{Running SHAP on a ``black box'' model is computationally expensive --- especially when there are thousands of features to consider. In the case of linear or tree-based models, the computation can be massively accelerated by allowing SHAP to inspect internal parameters \cite{SHAPpaper2017}. Alternatively, a weighted $k$-means clustering can be applied to the dataset prior to a ``black box'' SHAP analysis. This method was used, with $k=30$. This effectively summarised the data as 30 observations with features that were a weighted average of the points that comprised the cluster. Whilst this was primarily used to reduce the computation time for SHAP, it also served to smooth out a great deal of noise from the dataset and thus provide more robust results.}, SHAP computed the Shapley values for each feature-observation pair and used those to automatically rank the relevant features. These results are presented in a beeswarm plot in Fig. \ref{fig:interp_shap_summary}.

We can see that the majority of features had an negative impact on sleep score, as low (blue) values gather on the right side whilst high (red) values gather on the left. One interpretation of this is a \textit{subtractive} model of sleep quality, where many factors can detract from a baseline of good sleep. For instance, fewer hours of nighttime eating the previous day (\texttt{zero\_night\_eating\_-1day}) pushed the predicted sleep quality up (high SHAP value), whilst more hours of nighttime eating pulled the predicted sleep quality down (low SHAP value). 

Some features (like \texttt{aw\_hr\_heart\_rate\_std\_morn} and \texttt{zero\_hours\_-1day}) appear to be centred at their mean, with Gaussian-like distributions, whilst others (like \texttt{melatonin\_quantity} and \texttt{wol\_read\_book\_-4day}) clearly reflect their discrete distributions. 

A common pattern in the beeswarm plot is a strongly-skewed distribution of values on one tail. For example, \texttt{nomie\_caf\_net\_hour\_spread} measures the variance in the hours when caffeine was consumed throughout the day. A low or medium spread results in a minor increase in sleep quality. However, a high spread results in anywhere from a small to very large decrease in sleep quality. Whilst this does reflect the underlying feature distribution (which is positively skewed), it shows that extreme values can have a large range of effect sizes on the final prediction. Finally, the colour gradients on almost all features change very smoothly with respect to the SHAP value. This reflects the inherently linear nature of the Lasso model used. However, SHAP can be used to interpret any black-box model \cite{SHAPpaper2017} and could (in principle) illustrate non-linear relationships captured in the data. 

\begin{figure}[H]
     \centering
     \begin{subfigure}[b]{0.49\textwidth}
         \centering
         \includegraphics[width=\textwidth]{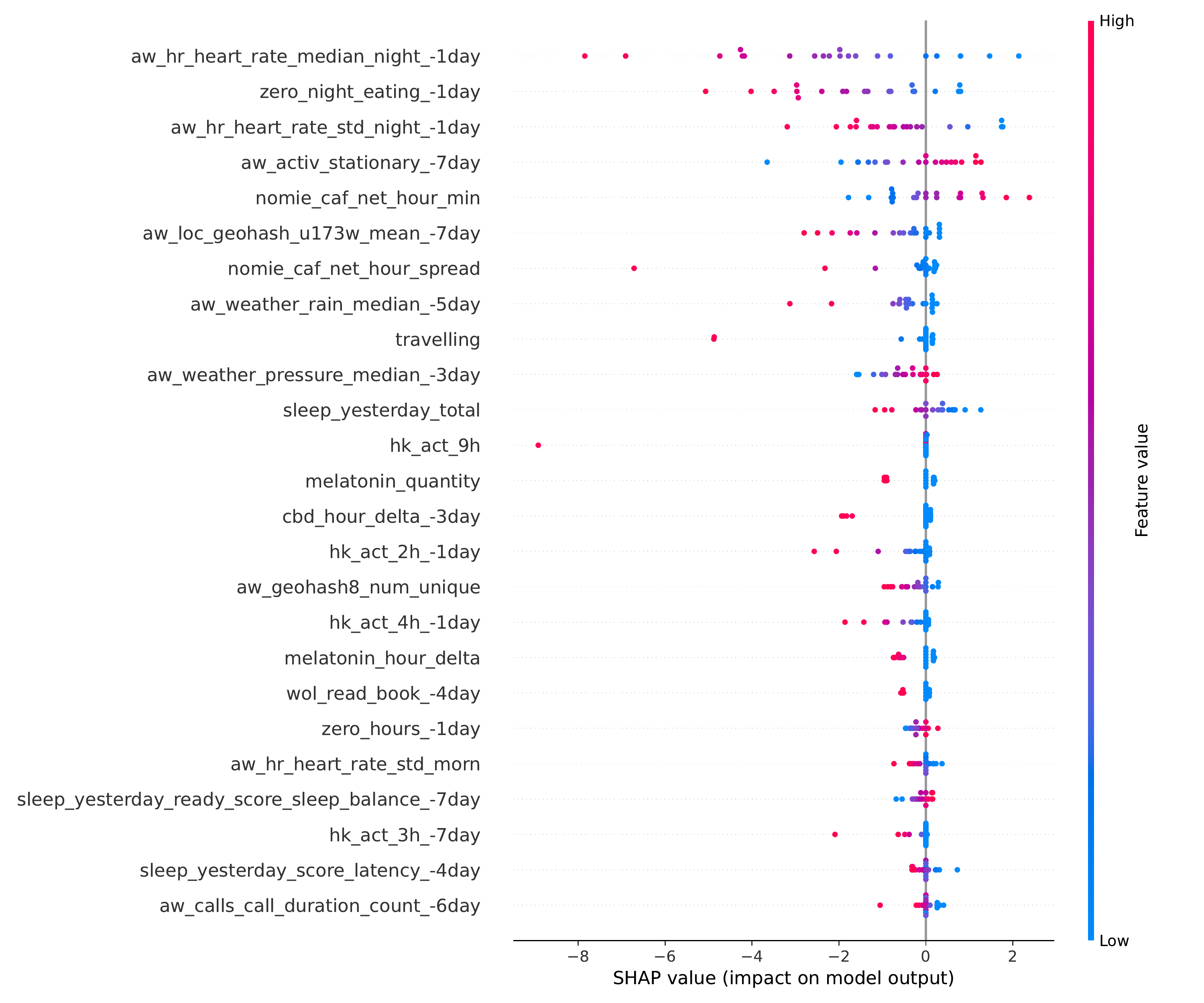}
         \caption{Sleep score $< 50$}
         \label{fig:interp_shap_low_values}
     \end{subfigure}
     \hfill
     \begin{subfigure}[b]{0.49\textwidth}
         \centering
         \includegraphics[width=\textwidth]{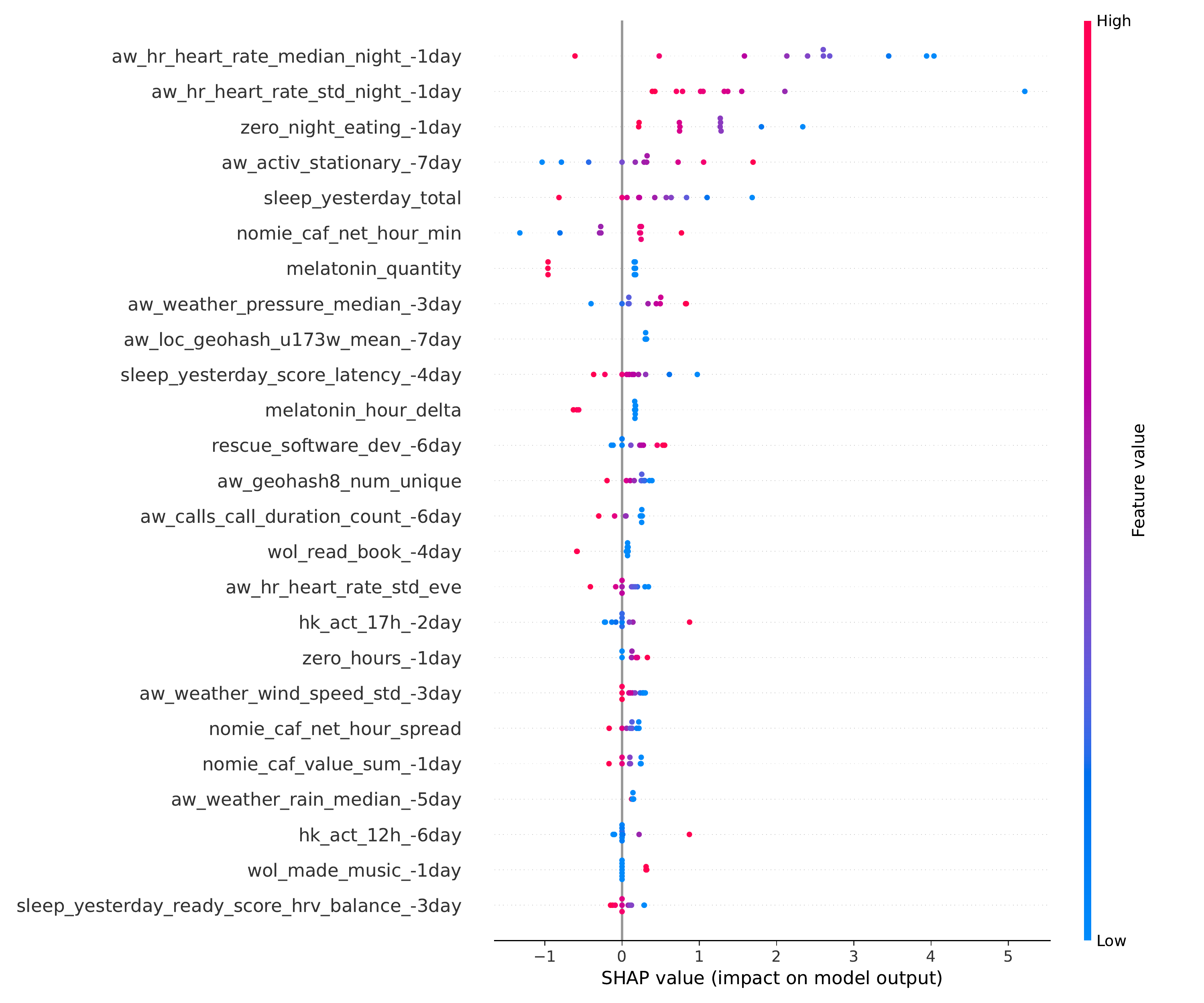}
         \caption{Sleep score $> 85$}
         \label{fig:interp_shap_high_values}
     \end{subfigure}
    \caption{Contrastive beeswarm plots for 21 nights with terrible sleep quality (left) and 11 nights with superb sleep quality (right).}
    \label{fig:interp_shap_values}
\end{figure}

\newpage
One of the strengths of SHAP is its ability to perform contrastive explanations. We can therefore compare the beeswarm plots of the worst (score $<50$) and best (score $>85$) nights in the dataset (see Fig. \ref{fig:interp_shap_values}).

The main difference to note is the reordering of the features. This indicates that some features contributed far more on one extreme of the sleep quality distribution than on the other. For instance, \texttt{travelling} appears ninth in the low-scoring nights (\ref{fig:interp_shap_low_values}) but not at all in the high-scoring nights (\ref{fig:interp_shap_high_values}). Moreover, we can see from the distribution of points that not all nights of poor sleep were due to \texttt{travelling}, but that it was to blame for poor sleep quality when it did occur. 

\subsubsection{Combined interpretation}
The top features from the RFE-based selection, the cross-validated Lasso $\beta$-parameters, and SHAP interpretation were combined to find standout features. These can inform future investigations and n-of-1 experiments. By combining the top features from all three interpretation strategies, 16 features were found to be consistent and notable. They are listed along with their full descriptions and relevant importance measures in Table \ref{tab:top_features_described}. 

\begin{table}[H]
\caption{Combined set of model interpretations, showing the 16 most predictive features (roughly ordered) and a written description of what each feature means. The $r$ column indicates the Pearson correlation coefficient between the feature and the target in the dataset. The $\beta$ column indicates the associated $\beta$-parameter in the final Lasso model. The $|S|$ column indicates the mean absolute Shapley values (over all observations) for each feature after interpretation with SHAP.}
\footnotesize\addtolength{\tabcolsep}{-2pt}
\begin{tabular}{lrrrl}
\hline
\textbf{Feature name}                     & $r$   & $\beta$ & $|S|$ & \textbf{Description}                              \\ \hline
aw\_hr\_heart\_rate\_median\_night\_-1day & -0.46 & -2.17   & 1.84   & Median heart rate in the evening, 1 day prior     \\
aw\_hr\_heart\_rate\_std\_night\_-1day    & -0.33 & -1.21   & 0.94   & Std. heart rate in the evening, 1 day prior       \\
zero\_night\_eating\_-1day                & -0.44 & -0.96   & 0.99   & Hours of eating after sunset, 1 day prior         \\
nomie\_caf\_net\_hour\_min                & -0.02 & 0.77    & 0.71   & Hour of first caffeine consumption                \\
aw\_activ\_stationary\_-7day              & 0.20  & 0.75    & 0.67   & Time spent stationary, 7 days prior   \\
aw\_loc\_geohash\_u173w\_mean\_-7day      & -0.29 & -0.63   & 0.44   & Time in location u173w, 7 days prior  \\
melatonin\_quantity                       & -0.21 & -0.54   & 0.47   & Quantity of melatonin consumed                    \\
travelling                                & -0.21 & -0.49   & 0.17   & Whether or not I was travelling long distances    \\
sleep\_yesterday\_total                   & -0.04 & -0.39   & 0.56   & Total time spent sleeping the previous day        \\
aw\_weather\_pressure\_median\_-3day      & 0.21  & 0.39    & 0.43   & Median weather pressure, 3 days prior             \\
melatonin\_hour\_delta                    & -0.20 & -0.37   & 0.30   & The hour melatonin was consumed                   \\
hk\_act\_17h\_-2day                       & 0.16  & 0.35    & 0.23   & Activity levels at 17h00, 2 days prior            \\
aw\_geohash8\_num\_unique                 & -0.16 & -0.17   & 0.25   & The no. unique level-8 geohashes visited \\
cbd\_hour\_delta\_-3day                   & -0.25 & -0.32   & 0.21   & The hour CBD oil was consumed, 3 days prior       \\
wol\_read\_book\_-4day                    & -0.21 & -0.18   & 0.22   & Whether I read a book, 4 days prior               \\
zero\_hours\_-1day                        & 0.35  & 0.20    & 0.16   & Hours of intermittent fasting, 1 day prior        \\ \hline
\end{tabular}
\label{tab:top_features_described}
\end{table}

We can see from the suffixes in Table \ref{tab:top_features_described} that 11 of the 16 features were lagged values generated through Markov unfolding\footnote{Recall that the prefix \texttt{sleep\_yesterday\_} is used for a 1-day lag on the \texttt{oura\_} features to make it easier to remember to remove same-day sleep features that constitute a data leak. So \texttt{sleep\_yesterday\_total} can be interpreted as \texttt{oura\_sleep\_total\_-1day}}. A total of 5 have a 1-day lag, 2 have a 7-day lag, and the remaining 4 have 2-4 day lags. This suggests that history is an important factor in sleep quality. In particular, the previous day and last week on the same day have notable impact. This justifies the need for techniques like Markov unfolding which incorporate history into each observation. It also hints that much of the value of Markov unfolding comes from capturing information from 1 day and one week prior.

It is interesting to compare the original correlations ($r$-values), the Lasso $\beta$-parameters, and the Shapley values for each of these features (see Table \ref{tab:top_features_described}). All but one of the $r$-values have the same positivity as the $\beta$-parameters. The exception is \texttt{nomie\_caf\_net\_hour\_min}, which had almost no correlation ($r = -0.02$). This tells us that correlation analysis correctly identified the direction that each feature ultimately affected sleep quality. However, the magnitudes of these correlations are fairly small and would not have been sufficient to select these 16 features from the full set, based only on correlation (Fig. \ref{fig:02_correlation_scores_bar}). Most notably, \texttt{nomie\_caf\_net\_hour\_min} and \texttt{sleep\_yesterday\_total} are important features but displayed negligible correlation to sleep quality. The magnitudes of the $\beta$-parameters and the Shapley values correspond closely, except for the \texttt{travelling} feature, which SHAP deemed as far less important on average. This is likely a result of SHAP (correctly) down-weighting the travelling feature, as it only has a non-zero value for a handful of observations. 

This indicates that modelling the data helped isolate which features most impacted sleep quality in a way that simple correlation analysis would not have been capable of. Overall, the results of model interpretation track the direction of simple correlation, but not the magnitude. 

\subsection{Feature explanation}
Some of the final 16 features are expected. Travelling, eating windows, previous sleep, and melatonin are all known to affect sleep quality \cite{st2016effects, lee2008sleep, bihari2012factors, walker2017we}. But the direction of some of these effects was unexpected. For instance, melatonin consumption is associated with a \textit{decrease} in sleep quality, when research suggests it should have be increase \cite{van2010use, dijk1997melatonin}. The previous night's sleep quantity also has an unexpected \textit{decrease} on the present night's sleep quality. That implies that there is some kind of trade-off between sleep \textit{quality} and sleep \textit{quantity} on successive nights. These unexpected directions might just be artefacts of the study design, but could also indicate idiosyncrasies in my sleep patterns. Through the process of modelling and interpreting, such unexpected connections are highlighted for future n-of-1 studies. 

On the other hand, some of the final 16 features were very much unexpected. Specifically, the lag in the features. It is not intuitive that pleasure reading and barometric pressure from many days prior could affect the current night's sleep. It is also strange that location changes from an entire week prior are still predictive of the current night's sleep. One explanation is that these features merely correlate with other (unmeasured) variables that influence sleep quality, and that this is being captured by the model \cite{hernan2010causal}. Another (complementary) explanation is that the sparsity effect of the Lasso model \cite{tibshirani1996lasso} caused it to discard all but one feature from each set of highly-correlated features. For instance, \texttt{wol\_read\_book} is highly correlated with a number of Oura-based features, as well as computer usage, location, alcohol consumption, and afternoon activity (Fig. \ref{fig:02_correlation_hierarchy}a). Given that the Lasso model already includes correlated features for the current day and previous day, it gains no predictive value by including \texttt{wol\_read\_book} with a lag of 0 or 1. In fact, the L1 regularisation would mean that this extra feature imposes a penalty on the model \cite{tibshirani1996lasso}. But, including \texttt{wol\_read\_book\_-4day} adds information from \textit{earlier} days. And because this reading feature is highly correlated with many other predictive features, it captures a great deal of predictive information from earlier days.

This line of reasoning suggests that it is not specifically these 16 features that are most-important for understanding sleep quality. Instead, it is clusters of features that are correlated with these 16 that are of importance. A thorough analysis (which is beyond the scope of this paper) would use results from Table \ref{tab:top_features_described} and the dendrogram from Fig. \ref{fig:02_correlation_hierarchy} together to build a comprehensive understanding of the relationships affecting sleep quality\footnote{Indeed, SHAP does include hierarchical interpretation techniques, but they rely on iteratively fitting XGBoost models on different subsets of the features, which is far too computationally expensive for large numbers of features.}.

The 16 final features are a much reduced subset of the 308 initial features that were explored throughout this observational study, yet they explain the majority of the variation in sleep quality (when controlling for all the other features). The sum of the SHAP scores for the 16 final features is $8.59$, while the sum of the SHAP scores for the other 1505 Markov-unfolded features is $2.31$.

The next step is to design a series of \textit{interventional} n-of-1 QS experiments to explore whether each of these factors is causal and what the effect sizes are. The design and implementation of such studies is outside the scope of this paper, but is well-covered in other literature \cite{kravitz2014design, grantwhite, choe2014understanding}. 

\section{Limitations and future work}\label{sec:limitations_and_future_work}

\subsection{Ground truth}
The lack of accurate ``ground truth'' was a theme at all levels. For instance, the Oura ring --- whilst a leader in wearable sleep trackers --- is still vastly less accurate than research-grade polysomnography equipment. Because of the opaque nature of sleep \cite{walker2017we}, even polysomnography is only an estimation \cite{kaplan2017gold}. All of this added a great deal of noise to the target. This made it harder to assess the quality of predictive models. Those models also had the challenge of noisy sensor data and biased event logs as features. 

Analysis of the underlying distributions of features in the dataset revealed some potential issues. Whilst many biological and environmental factors produce near-normal distributions \cite{lyon2014normal}, they were often very skewed or fat-tailed in this dataset. There were also a number of strongly-bimodal distributions caused by the dramatic changes during the first wave of the pandemic. These distributions are ``dodgy'' because they deviate \cite{taleb2020statistical} from the perfectly-Gaussian assumptions made by many statistical techniques and learning algorithms \cite{burkov2019hundred, bishop2006pattern}. 

The real-world nature of the dataset meant that there was no ground truth against which to directly evaluate the feature engineering, hierarchical correlation, imputation, Markov unfolding, and model interpretation techniques. This reduced the power of the evidence presented by the results of this study. Future work could apply the methods of this study to ``easier'' datasets and modelling problems to evaluate them in isolation. 

\subsection{Generalisability}
Because this was an n-of-1 study, most results are unlikely to generalise to other individuals. Moreover, many factors --- the specifics of the tracking tools used, environment, personal behaviour, social factors, pandemic consequences, etc. --- are likely to vary over time. This means that many of the results of this study may not even generalise to \textit{me} going forward. This is definitely true of specific findings, like the effect of melatonin on my sleep quality. But it may also be true of the results about different techniques used. Whilst Lasso and matrix-factorisation-imputation worked well on this specific dataset, that may not generalise to other n-of-1 QS datasets. Indeed, those who replicate these approaches should also replicate the variety of techniques used. By casting a wide ``net'' of possible techniques, it is more likely to find ones that apply best to the specific dataset being modelled. 

\subsection{Markov unfolding}
Markov unfolding was only notably effective in the regularised linear algorithms. This may be because the current implementation increased the dimensionality too much --- outweighing the value of the historical information. The linear algorithms, especially Lasso, likely dealt with this better because of their regularisation helping prioritise the most important features \cite{burkov2019hundred, tibshirani1996lasso}. The predominance of 1-day and 7-day lagged features in the final model interpretation suggests that there may be more sophisticated ways to apply Markov unfolding to maximise the historical information whilst minimising the increase in dimensionality. For example, future work could explore the use of rolling aggregations for the 2-6 day lags. Instead of feature $j$ becoming 8 features, it would only become 5: $j$, $j_{-1}$, $j_{-7}$, $j_{\mu}$, and $j_{\sigma}$. This captures the information-rich previous day, the cyclical weekly patterns of seven days prior, and a rolling distribution of ``the last few days''. This would also help make features more intuitive and models more interpretable. 

\subsection{Imputation}
There were three limitations to the imputation strategy of this paper. 

Firstly, no ground-truth of error levels across imputation techniques existed for this dataset. The custom distance metric was used as a proxy for the magnitude of change caused by imputation, but this did not directly help select the best-suited imputation algorithm or hyperparameters. Future work can first evaluate imputation techniques on \textit{synthetic} missing values from a sample of the same dataset. The error of the re-imputed values could then be used to select and optimise the imputation method for the dataset. This was beyond the scope of this study, as generating the synthetic missing values in a way that generalises to \textit{real} missing data is a complex topic --- MAR, MNAR, and MCAR effects (\S \ref{ssec:theory_of_missing_data}) all need to be synthesised. Moreover, the fact that there were only 61 observations in the data that did not need imputation would have meant a very biased sample and may have invalidated the synthetic approach on this specific dataset. 

Secondly, although all features used in imputation were numeric, a subset had an underlying binary structure. In principle, the imputation techniques used should only apply to continuous features \cite{little2019statistical}. In practise, there were sufficiently few binary features and sufficiently few missing values within them that this was \textit{probably} negligible. Future work could investigate automatic filtering methods for excluding such features. 

Thirdly, a large number of the imputed values were from a contiguous ``chunk'' in the AWARE-sourced features from prior to my use of AWARE. This chunk can be clearly seen in the top right of Fig. \ref{fig:pre_markov_with_nans.csv}. It is likely that imputing these values (especially with simpler univariate techniques) resulted in a great deal of additional noise for those features. That may have affected the results in a number of different ways. A key trade-off in this study was between number of observations and number/quality of features. The decision to impute over this large missing chunk was part of that trade-off. Future work could investigate the effects of imputing large chunks of contiguous values on final model interpretation. 

\subsection{Interpretation}
One of the major advantages of SHAP over $\beta$-value analysis is that it assigns Shapley values to each feature for each observation, allowing for a nuanced distributional analysis (e.g. Fig. \ref{fig:interp_shap_values}). By using a linear final model (Lasso), non-linear feature effects were not able to show up in the SHAP results. It is possible that there were no relevant non-linearities to the most predictive features, but by constraining the scope of this study to a single (linear) model, it became impossible to know. Future work could compare SHAP interpretations over different models and develop techniques for reconciling multiple interpretations.

\subsection{Hierarchical clustering}
The hierarchical clustering of feature correlations offers many advantages to projects of this nature. Whilst it was used to aid model interpretation, it was under-utilised in feature selection. This was mainly due to it being developed towards the end of this study, when it was too late to overhaul the data pipeline. Future work could (1) validate the hierarchical correlational method's efficacy in finding related features, and (2) explore the use of hierarchical correlation for feature selection. The hierarchical correlation of features may offer a better trade-off between accuracy and performance for automated feature selection than RFE or exhaustive search. This is especially relevant to n-of-1 QS projects where most features are weakly correlated to most other features and very weakly correlated with the target \cite{Hoogendoorn2018, swan2013quantified}. 

\section{Conclusions}

This paper presented a case study in how to conduct observational n-of-1 Quantified-Self (QS) research, combining relevant techniques from statistics and machine learning to obtain robust and interpretable results.

Several methods were presented for combining heterogeneous data sources and engineering day-level features from different data types and frequencies, including manually-tracked event logs and automatically-sampled weather and geo-spatial data. The resulting dataset was thoroughly analysed --- for outliers, normality, (auto)correlations, stationarity, and missing data --- and cleaned accordingly. A notable inclusion was the use of hierarchical clustering of the correlation matrix to identify groups of correlated features.

The missing data was organised and filled using a combination of knowledge-based and statistical techniques. The latter included several different imputation methods that have been presented in the literature. Regularised linear algorithms (Lasso and Ridge) performed the best with imputed data, particularly for the matrix factorisation, KNN, MICE, iterative SVD, and univariate imputation strategies. The use of imputation saved hundreds of observations from being discarded and improved overall performance of all the algorithms except the plain Decision Tree.

To collapse the time series into a collection of independent observations, the Markov unfolding technique was presented. This added lagged copies of features to each observation to incorporate values from recent history for each engineered feature. Markov unfolding improved the predictive performance of Lasso dramatically, but for other algorithms the improvement was less pronounced. This was likely because the benefit of the additional information traded off against the added burden of much higher dimensionality. Lasso likely performed best because L1 regularisation allowed it to effectively sift through the greater number of features. The paper suggests ways Markov unfolding could be made more feature efficient in future work.

From the extensive grid-search, a low-error, low-variance model and dataset combination was selected --- Lasso regression on a Markov-unfolded version of the dataset which had undergone matrix factorisation imputation. The final model was interpreted in two key ways: (1) by inspecting the internal $\beta$-parameters, and (2) using the SHAP framework, which builds local explanatory models for each observation. By repeatedly re-training the Lasso model on different subsets of the dataset, distributions of $\beta$-parameters were generated. Features with consistently-large $\beta$-coefficients were deemed \textit{globally} important. This was combined with SHAP's \textit{situational} assessment of the importance of each feature with respect to each observation --- which allowed contrastive analysis of extreme examples of sleep quality. These two interpretation techniques were combined to produce a list of the 16 most-predictive features.

By comparing the list of predictive features to the hierarchy of correlated features, it became apparent that Lasso's ability to learn sparse parameters helped it to find features that were in different correlational clusters, thus making the best use of all the information in the dataset without overfitting. Unfortunately, this made it more complex to infer a \textit{descriptive} model of sleep behaviour.
 
Overall, predictive modelling helped detect relationships in the dataset that simple univariate analysis would not have found. By identifying the factors that most affect sleep, this study showed that it is possible to use an \textit{observational} study to greatly reduce the number of features that need to be considered in \textit{interventional} n-of-1 QS research.

\section{Practicalities}
\subsection{Code and data availability}\label{ssec:code_data_availability}
The Python 3 code for methods presented in this study is \href{https://github.com/gianlucatruda/quantified-sleep}{publicly available} as a collection of Jupyter notebooks. These show outputs and documentation inline with the code, making it easier to understand the flow of this study in a sequential order. The code is available under a GNU General Public License (GPLv3), allowing modification and re-use. For the sake of personal data privacy, some outputs and code are redacted from the published notebooks. 

The dataset from this study will not be made available, as it contains extensive personal information. My current ongoing research into generative differential privacy hopes to find long-term solutions to this trade-off between data privacy and open science. If a version of the dataset does become available in future, it will be made available via the project repository (\href{https://github.com/gianlucatruda/quantified-sleep}{github.com/gianlucatruda/quantified-sleep}).

\subsection{Replicating this work}

Other QS enthusiasts are encouraged to apply these techniques to their future projects and report on their findings. Much of the progress in this niche of n-of-1 QS occurs outside of the formal scientific literature. One of the goals of this paper was to help bridge the gap between community rules-of-thumb and end-to-end studies. 

As mentioned, the dataset is not available for privacy reasons. However, much of the code in the Jupyter notebooks should generalise to other n-of-1 QS projects, provided the sources are similar. Those wishing to replicate the methodology of this study on their own data will need to write their own \textit{ingesters}. These are simple Python functions that take in an exported data file (e.g. CSV, JSON, etc.) and return a Pandas dataframe with observations as rows and features as columns. Most of the code was written in a way that minimises the number of changes required to adapt to different sources. Indeed, during the course of this study, I added and updated sources a number of times, finding that much of the code was robust to these changes. Current details of the structure of the code and how to best modify it for your own projects can be found in the \href{https://github.com/gianlucatruda/quantified-sleep}{repository documentation}.

\bibliography{bibliography}

\appendix

\section{Appendices} 

\begin{figure}[H]
    \centering
    \includegraphics[width=0.5\textwidth]{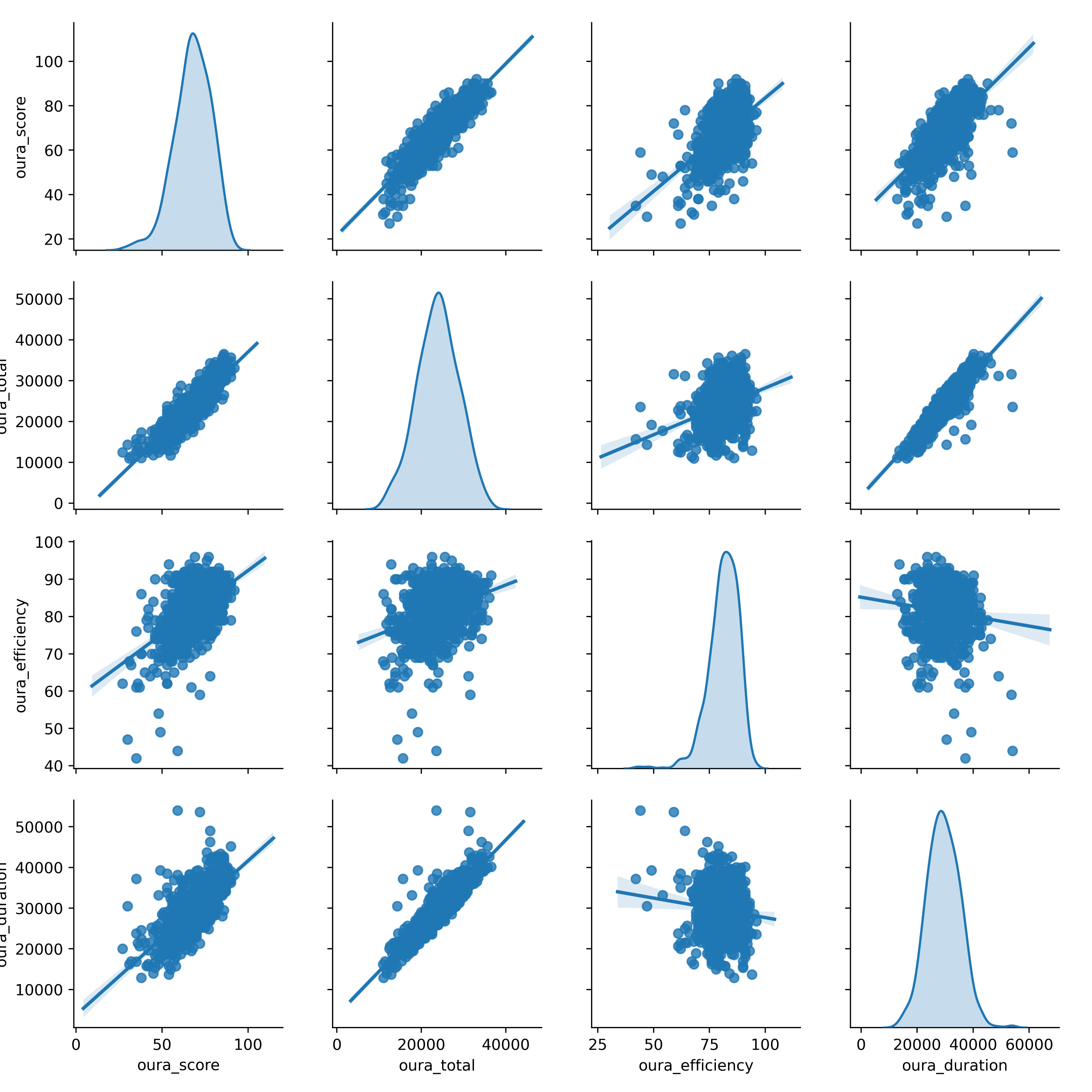}
    \caption{Pairplot matrix over 4 of the most important sleep features. In each row, a feature comprises the vertical axis. In each column a feature comprises the horizontal axis. In cell $i,j$, a scatterplot and regression line is shown for $i$ vs $j$. In cell $i, i$, the histogram of feature $i$ is shown instead of a redundant scatterplot.}
    \label{fig:01b_oura_sleep_pairplots}
\end{figure}

\begin{table}[H]
\tiny
\begin{tabular}{l}
\hline
\textbf{Feature name}                     \\ \hline
sleep\_yesterday\_midpoint\_time          \\
sleep\_yesterday\_rmssd                   \\
sleep\_yesterday\_score                   \\
sleep\_yesterday\_score\_alignment        \\
sleep\_yesterday\_score\_deep             \\
sleep\_yesterday\_score\_disturbances     \\
sleep\_yesterday\_score\_efficiency       \\
sleep\_yesterday\_score\_latency          \\
travelling                                \\
hk\_act\_9h                               \\
nomie\_caf\_net\_hour\_min                \\
nomie\_caf\_net\_hour\_range              \\
nomie\_caf\_net\_hour\_spread             \\
melatonin\_quantity                       \\
aw\_geohash8\_num\_unique                 \\
hk\_act\_2h\_-1day                        \\
hk\_act\_4h\_-1day                        \\
hk\_act\_7h\_-5day                        \\
hk\_act\_17h\_-2day                       \\
hk\_act\_night\_mean\_-1day               \\
nomie\_caf\_value\_sum\_-4day             \\
cbd\_hour\_delta\_-3day                   \\
zero\_hours\_-1day                        \\
zero\_night\_eating\_-1day                \\
wol\_intensive\_workout\_-2day            \\
wol\_read\_book\_-4day                    \\
aw\_activ\_stationary\_-7day              \\
aw\_loc\_geohash\_u173w\_mean\_-7day      \\
aw\_weather\_pressure\_median\_-3day      \\
aw\_weather\_rain\_median\_-5day          \\
aw\_weather\_wind\_speed\_median\_-5day   \\
aw\_weather\_wind\_speed\_std\_-3day      \\
aw\_hr\_heart\_rate\_median\_night\_-1day \\
aw\_hr\_heart\_rate\_std\_day\_-1day      \\
aw\_hr\_heart\_rate\_std\_morn\_-7day     \\
aw\_hr\_heart\_rate\_std\_night\_-1day    \\ \hline
\end{tabular}
\caption{The 36 most-predictive features (in arbitrary order) after $5$-fold cross-validated recursive feature elimination (RFE) for a Lasso regression model on a markov-unfolded-matrix-factorised variant of the dataset.}
\label{tab:rfe_top_features}
\end{table}

\end{document}